\documentclass[12pt]{iopart}
\pdfoutput=1
\usepackage[T1]{fontenc}
\usepackage[latin1]{inputenc}
\usepackage{iopams}
\usepackage{graphicx,epsfig}
\usepackage{subfigure}
\usepackage{wrapfig}
\usepackage{ae}
\usepackage[breaklinks=true]{hyperref}

\begin{document}
\newcommand{\rub}{$^{87}$Rb}
\newcommand{\pot}{$^{40}$K}
\newcommand{\he}{$^4$He}

\title[Heteronuclear quantum gas mixtures]{Heteronuclear quantum gas mixtures}
\author{
C. Ospelkaus\footnote{Present address: NIST Boulder; MS847; 325 Broadway; Boulder, CO 80305; USA},
S. Ospelkaus\footnote{Present address: JILA, University of Colorado; Campus Box 0440; Boulder, CO 80309; USA}
}
\address{Institut f\"ur Laser-Physik, Luruper Chaussee 149, 22761 Hamburg, Germany}
\ead{christian.ospelkaus@nist.gov, silke.ospelkaus@jila.colorado.edu}
\begin{abstract}
In this PhD tutorial article, we present experiments with quantum degenerate mixtures of fermionic and bosonic atoms in 3-dimensional optical lattices. This heteronuclear quantum gas mixture offers a wide range of possibilities for quantum simulation, implementation of condensed matter Hamiltonians, quantum chemistry and ultimately dense and quantum degenerate dipolar molecular samples.

We show how quantum degenerate mixtures of \pot\ and \rub\ are created in the experiment. We analyze stages of evaporative cooling and show how a dynamic mean field collapse occurs during the final stage of the evaporation~\cite{Ospelkaus2006b} as a result of attractive interactions. The particle numbers observed in our experiment have only been limited by this mean field collapse, resulting in an excellent starting point for our experiments.

We explore magnetic field induced Feshbach resonances and demonstrate tuning of interactions~\cite{Ospelkaus2006c} between \pot\ and \rub\ by means of heteronuclear Feshbach resonances. We observe both stable attractively and repulsively interacting mixtures. We analyze the mean field energy of the condensate and find qualitative agreement with a simple model. By making the interaction strong and attractive, we induce a mean field collapse of the mixture. For strong and repulsive interactions, we observe phase separation of the mixture.

When loaded into a 3D optical lattice, a whole zoo of novel quantum phases has been predicted for Fermi-Bose mixtures. We present the first realization of Fermi-Bose mixtures in 3D optical lattices as a novel quantum many body system~\cite{Ospelkaus2006e}. We study the phase coherence of the bosonic cloud in the 3D optical lattice as a function of the amount of fermionic atoms simultaneously trapped in the lattice. We observe a loss of phase coherence at much lower lattice depth than for a pure bosonic cloud and discuss possible theoretical scenarios including adiabatic processes, mean field Fermi-Bose Hubbard scenarios, and disorder-enhanced localization scenarios.

After considering this many-body limit of mixtures in lattices, we show how fermionic heteronuclear Feshbach molecules can be created in the optical lattice~\cite{Ospelkaus2006d} as a crucial step towards all ground state dense dipolar molecular samples. We develop rf association as a novel molecule association technique, measure the binding energy, lifetime, and association efficiency of the molecules. We develop a simple theoretical single channel model of the molecules trapped in the lattice~\cite{two_particles_hh} which gives excellent quantitative agreement with the experimental data.
\end{abstract}
\submitto{\JPB}
\maketitle

\section{Introduction}

Nature distinguishes between two fundamental types of particles: fermions and bosons, depending on the spin of the particle. Particles with half-integer spin  are called fermions and obey Fermi-Dirac statistics. As a result of Fermi-Dirac statistics, in a system of indistinguishable particles, at most one particle can occupy a given quantum state.  Particles with integer spin  are called bosons and obey Bose-Einstein statistics. Any single-particle eigenstate of a physical system can be occupied by an arbitrary number of bosons.

While all of the basic building blocks of an atom are fermions, an atom as a whole has bosonic or fermionic character depending on its total angular momentum. For a gas of atoms confined in an external potential, the quantum statistical properties of the atoms become important at ultralow temperatures where the thermal deBroglie wavelength of the constituents is on the order of the interparticle separation. For bosonic atoms, this leads to the onset of Bose-Einstein condensation as observed for the first time in 1995 in a gas of $^{87}$Rb at JILA~\cite{Anderson1995a}, in  $^{23}$Na at MIT~\cite{Davis1995a} and for $^{7}$Li at Rice University~\cite{Bradley1995a} in the special case of attractive interactions. For fermionic atoms, the onset of degeneracy is less spectacular due to the absence of a phase transition. In an ultracold spin-polarized fermionic gas, the appearance of a macroscopic Fermi sea has first been demonstrated at JILA in 1999~\cite{DeMarco1999b}.

While the pioneering work on quantum degenerate gases  revealed important quantum phenomena such as interference, superfluidity and nonlinear atom optics, recent years have seen spectacular progress in the realization of novel strongly correlated systems with ultracold quantum matter. Strong correlations are observed either when the interactions between the constituents become very strong (e.\ g.\ at Feshbach resonances) or when strong confinement imposes stringent boundary conditions (e.\ g.\ in the periodic potential of an optical lattice).

Atomic systems offer a high degree of control of both the external confinement and the interactions between the constituents. The latter has become possible with the advent of Feshbach resonances~\cite{Courteille1998a, Inouye1998a} which allow control of $s$-wave and even higher order scattering between atoms by means of external fields. Feshbach resonances have been the key to a series of ground breaking experiments. For two-component fermionic gases, they have allowed the exploration of the BCS-BEC crossover and demonstrated that fermions and bosons are not as far from one another as it may seem: A Bose-Einstein condensate of diatomic molecules made from two fermionic atoms can be continuously transformed into a BCS state of atomic Cooper pairs~\cite{Regal2004a,Bartenstein2004a,Zwierlein2004a,Bourdel2004a,Kinast2004a, Zwierlein2005a}.

The ``confinement-induced'' approach to strongly correlated phases was first proposed for a gas of repulsively interacting bosons in 1998 by D. Jaksch and coworkers~\cite{Jaksch1998a}. In particular, it was demonstrated that bosonic atoms loaded into an optical lattice are an ideal model system  for the simulation of the Bose Hubbard Hamiltonian (see~\cite{Fisher1989a}) known from condensed matter physics, and it was predicted that the phase transition from a superfluid to a Mott-insulating state can be induced by merely increasing the lattice depth.  The theoretical prediction, together with the experimental demonstration of~\cite{Greiner2002a}, highlighted the potential of ultracold atoms in optical lattices for simulations of quantum many-body systems and for the realization of strongly-correlated systems. Optical lattices have since been the key to the observation of intriguing phenomena such  as  a Tonks-Girardeau gas of atoms in a one-dimensional geometry~\cite{Paredes2004a, Toshiya2004a} or a Kosterlitz-Thouless transition in a 1D lattice (2D geometry)~\cite{Hadzibabic2006a}. Two-body bound states (molecules) in homonuclear systems have been engineered~\cite{Stoeferle2006a,Thalhammer2006a} and  evidence for fermionic superfluidity has been reported in a cloud of $^6$Li loaded into a 3D optical lattice~\cite{Chin2006a}.

As a completely new area in the field of ultracold quantum gases, multicomponent quantum gases in 3D optical lattices have recently attracted a lot of attention. In the case of mixtures of fermionic and bosonic atoms, the different quantum-statistical behavior of the components gives rise to  fundamentally novel quantum many-body phases. In the extreme case of pairing of fermions with one or more bosons, a whole zoo of new quantum phases of these ``composite fermions'' has been predicted~\cite{Lewenstein2004a}. Fermi-Bose mixtures in 3D optical lattices may exhibit fermionic pairing which is mediated by the presence of bosonic atoms in full analogy to solid state superconductivity, and there are interesting connections to high-$T_C$ superconductivity~\cite{Heiselberg2000a,Albus2004a, Wang2005a}. Even before such ``atom pairs'' form, Fermi-Bose correlations are predicted to become manifest in polaron-related physics of fermions dressed by a bosonic cloud~\cite{Mathey2004a} and quantum percolation~\cite{Sanpera2004a}. These phenomena are connected to disorder induced localization scenarios. In reduced dimensionality, phenomena such as  charge-density waves~\cite{Mathey2004a, Wang2005a} and supersolids~\cite{Buchler2003a} are predicted to occur.

Any atomic Fermi-Bose mixture is necessarily also a special case of a heteronuclear system, i.\ e.\ a two-component system where both nuclei have a different decomposition. In some cases, such as $^6$Li -- $^7$Li, the resulting mass difference is small, whereas in others, such as $^6$Li -- $^{133}$Cs, it is very large. The mass difference, as we shall see, gives rise to interesting features already in the case of harmonic trapping of a mixture, but it ultimately opens up an interesting and highly promising approach to dipolar gases, quantum computation and simulation: if the components of such a heteronuclear mixture are brought together to form a molecule in its internal (rovibrational and electronic) ground state, the mass difference gives rise to a permanent molecular dipole moment. The resulting dipolar interaction, together with the high density and ultracold temperature of the initial atomic samples, would pave the way for novel quantum gases with dipolar interactions and quantum computation with polar molecules~\cite{DeMille2002a}. In more exotic heteronuclear systems, these techniques could be used for precision measurements~\cite{Sandars1967,Kozlov1995a,Hudson2006}.

From the experimental point of view, there has been an impressive series of experiments on homonuclear systems  which we have partly mentioned above, but experiments on heteronuclear systems in lattices, which Fermi-Bose mixtures are a special case of, have been scarce. By the end of 2005, the only experiment with Fermi-Bose  mixtures in optical lattices has been reported  by Ott and coworkers at LENS~\cite{Ott2004b}. In these experiments,  the ``insulating'' behavior of a trapped ideal Fermi gas in a 1D lattice has been compared to  collisionally induced transport of fermionic atoms in the presence of  a bosonic cloud.

In this tutorial article, we present experiments performed at the University of Hamburg as part of our PhD theses~\cite{SilkeThesis,OspelkausC2006a}. The article is organized as follows: we start our discussion by giving a cartoon picture of harmonically trapped Fermi-Bose mixtures and review a simple Thomas-Fermi model for the density distributions. As a function of the heteronuclear interaction, we identify regimes of stable mixtures with attractive and repulsive interaction as well as regimes of collapse and phase separation for a large heteronuclear interaction strength~\cite{Molmer1998a}.

We show how we create large harmonically trapped Fermi-Bose mixtures of \pot\ and \rub\ in a magnetic trap. We analyze stages of evaporative cooling and identify signatures of a dynamical mean-field collapse of the mixture as a result of attractive interactions. We show that large particle numbers of $7\cdot 10^5$ \pot\ and $1.2\cdot 10^6$ \rub\ atoms can be achieved~\cite{Ospelkaus2006b}, only limited by the aforementioned mean field collapse.

We show how heteronuclear interactions can be tailored by means of Feshbach resonances. The tunability of interactions gives access to the full phase diagram of the harmonically trapped mixture and allows us to produce both stable attractive and repulsive mixtures and to observe phase separation and collapse~\cite{Ospelkaus2006c}.

As a novel quantum many-body system, we discuss properties of Fermi-Bose mixtures trapped in 3D optical lattices. We show how already a small admixture of fermionic atoms reduces the bosonic coherence  at much lower lattice depths than for the pure bosonic (superfluid -- Mott insulator transition) case~\cite{Ospelkaus2006e}. We discuss various theoretical scenarios, from mean field models over thermodynamic processes and disorder-enhanced scenarios.

Combining the ability to load the mixture into a 3D lattice and tunability of interactions, we demonstrate creation of heteronuclear Feshbach molecules in a 3D optical lattice~\cite{Ospelkaus2006d,two_particles_hh}. When combined with coherent Raman de-excitation schemes for the molecular ro-vibrational manifold, this constitutes a key step towards the production of all-ground state samples of dense, polar molecules.
\section{Fermions and bosons in a harmonic trap}

We will start our discussion with a basic introduction to density distributions of trapped interacting Fermi-Bose mixtures. This will help us identify regimes of stable mixtures as well as interaction-induced instabilities and give us a basic understanding of the phase diagram. Harmonically trapped interacting Fermi-Bose mixtures have first been considered in a paper by M\o{}lmer in 1998 \cite{Molmer1998a}, see also ref.~\cite{roth_structure_and_stability}. For the bosons, we will neglect the influence of the thermal cloud and consider a Bose-Einstein Condensate in the Thomas-Fermi approximation. For the fermions, we assume the Thomas-Fermi approximation at $T=0$. In this case, the bosonic and fermionic density distributions $n_B(\vec r)$ and $n_F(\vec r)$ are given by the self-consistent solution to the following set of equations:
\begin{eqnarray}
\label{eq:thomas_fermi_coupled}
n_B(\vec r) & = &
\frac{1}{g_{BB}}\mathrm{max}\left[ \mu_B-g_{FB}\cdot n_F(\vec r) - V_B(\vec r),0\right] \\
n_F(\vec r) & = &
\frac{(2m_F)^{3/2}}{6\pi^2\hbar^3}\mathrm{max}\left[\mu_F-V_F(\vec r)-g_{FB}\cdot n_B(\vec r),0
\right]^{3/2} \nonumber
\end{eqnarray}
The Bose and Bose-Fermi interaction strengths are related to the corresponding scattering lengths according to 
\begin{eqnarray}
g_{BB} & = & 2\pi\hbar^2a_{BB}/\mu_{BB}\\
g_{FB} & = & 2\pi\hbar^2a_{FB}/\mu_{FB}
\end{eqnarray}
where $\mu_{BB}$ and $\mu_{FB}$ are the reduced masses associated with the two-boson and the boson-fermion case. The bosonic and fermionic chemical potentials $\mu_B$ and $\mu_F$ are  fixed by the particle numbers according to:
\begin{eqnarray}
\label{eq:thomas_fermi_coupled_N}
N_B & = & \int_0^\infty n_B(\mu_B,\vec x)\,d^3x \\
N_F & = & \int_0^\infty n_F(\mu_F,\vec x)\,d^3x\nonumber
\end{eqnarray}
$V_B(\vec r)$ and $V_F(\vec r)$ is the external trapping potential acting on bosons and fermions, respectively. In this treatment, the two components influence each other's density distributions through a mean-field potential given by 
\begin{eqnarray}
\label{eq:hetnucpotFonB}
U_{F\rightarrow B}(\vec r) & = & g_{FB}\cdot n_F(\vec r)\\
\label{eq:hetnucpotBonF}
U_{B\rightarrow F}(\vec r) & = & g_{FB}\cdot n_B(\vec r)
\end{eqnarray}
acting on bosons ($U_{F\rightarrow B}$) and fermions ($U_{B\rightarrow F}$), respectively. Let us assume a harmonic trap with trapping frequencies $\omega_{x,y,z}^B$ for the boson and $\omega_{x,y,z}^F$ for the fermion. For simplicity, let us assume that the trapping potential is harmonic and the same for both bosons and fermions. This situation is given for example for magnetic trapping of stretched states. For a far enough detuned dipole trap, this would also be valid. In this situation, the trap frequencies are related according to $\omega^F /  \omega^B = \sqrt{m_B/m_K}$. We can introduce rescaled coordinates
\begin{equation}
\tilde x_i = \sqrt{\frac{m_F}{2}}\omega_{i}^F\cdot x_i \quad ,
\end{equation}
which have the dimension of the square root of an energy: $\tilde r^2=V_B=V_F$; the coupled Thomas-Fermi problem becomes:
\begin{eqnarray}
\label{eq:thomas_fermi_coupled_trans1}
n_B(\tilde r) & = &
\frac{1}{g_{BB}}\mathrm{max}\left[ \mu_B-g_{FB}\cdot n_F(\tilde r) - \tilde r^2,0\right] \\
\label{eq:thomas_fermi_coupled_trans2}
n_F(\tilde r) & = &
\frac{(2m_F)^{3/2}}{6\pi^2\hbar^3}\mathrm{max}\left[\mu_F-\tilde r^2-g_{FB}\cdot n_B(\tilde r),
0\right]^{3/2}
\end{eqnarray}
Note that  the problem only depends on $\tilde r = \sqrt{\tilde x_1^2+ \tilde x_2^2+ \tilde x_3^2}$. In the new set of coordinates, equations~(\ref{eq:thomas_fermi_coupled_N}) become
\begin{eqnarray}
\label{eq:thomas_fermi_coupled_N_trans1}
N_B & = & \left(\frac{2}{m_F}\right)^{3/2}\frac1{(\bar \omega_F)^3}\int_0^\infty 4\pi^2 \tilde
r^2\cdot n_B(\tilde r)\,d\tilde r \\
\label{eq:thomas_fermi_coupled_N_trans2}
N_F & = & \left(\frac{2}{m_F}\right)^{3/2}\frac1{(\bar \omega_F)^3}\int_0^\infty 4\pi^2 \tilde
r^2\cdot n_F(\tilde r)\,d\tilde r \quad ,
\end{eqnarray}
where we have used the spherical symmetry of the problem in the new coordinates. As one can see, the above problem can be formulated solely in terms of the potential energy $\tilde r^2$ of the mixture, and the trapping potential only enters the calculation through the geometric mean trapping frequency $\bar\omega_F=\sqrt[3]{\omega_1\omega_2\omega_3}$. As we shall see, this is particularly relevant when comparing mixtures produced in different experiments and different harmonic traps.

\begin{figure}
  \centering
  \subfigure[Non-interacting mixture]{\includegraphics[width=0.47\columnwidth]{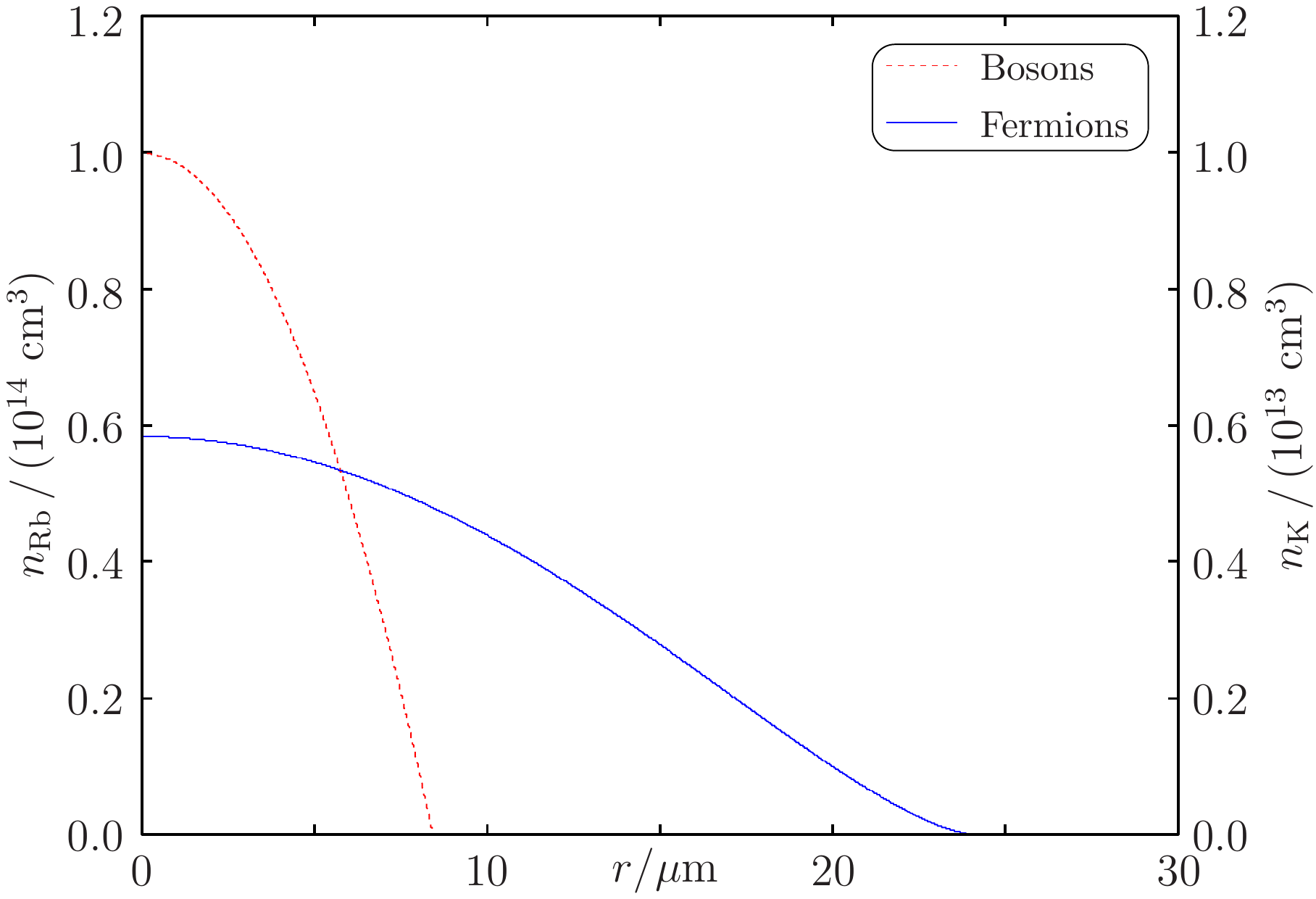}\label{fig:non_int}}
  \subfigure[weak attractive interaction ($g_{FB}/g_{BB}=-4$)]{\includegraphics[width=0.47\columnwidth]{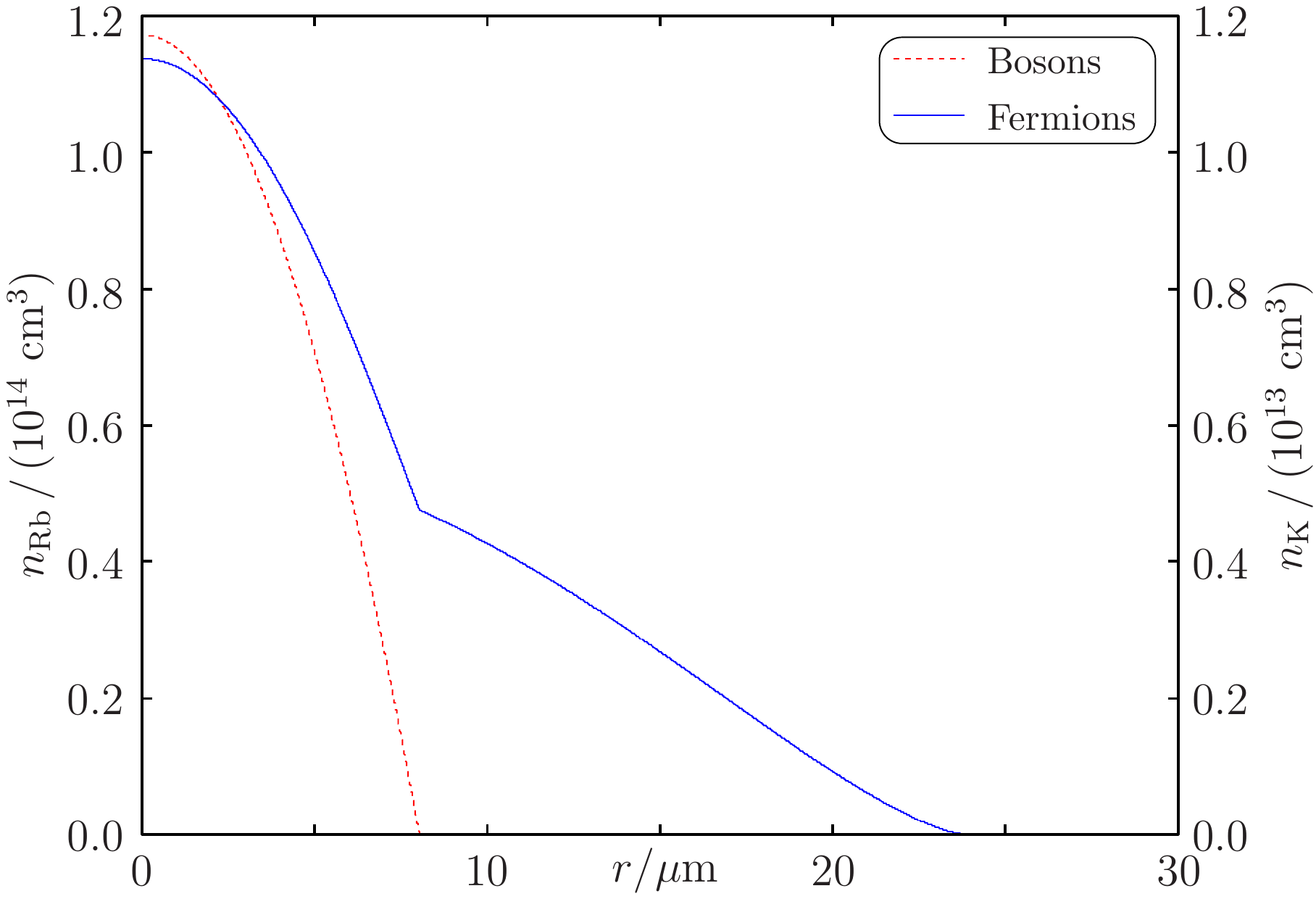}\label{fig:att_int}}
  \subfigure[weak repulsive interaction ($g_{FB}/g_{BB}=+4$)]{\includegraphics[width=0.47\columnwidth]{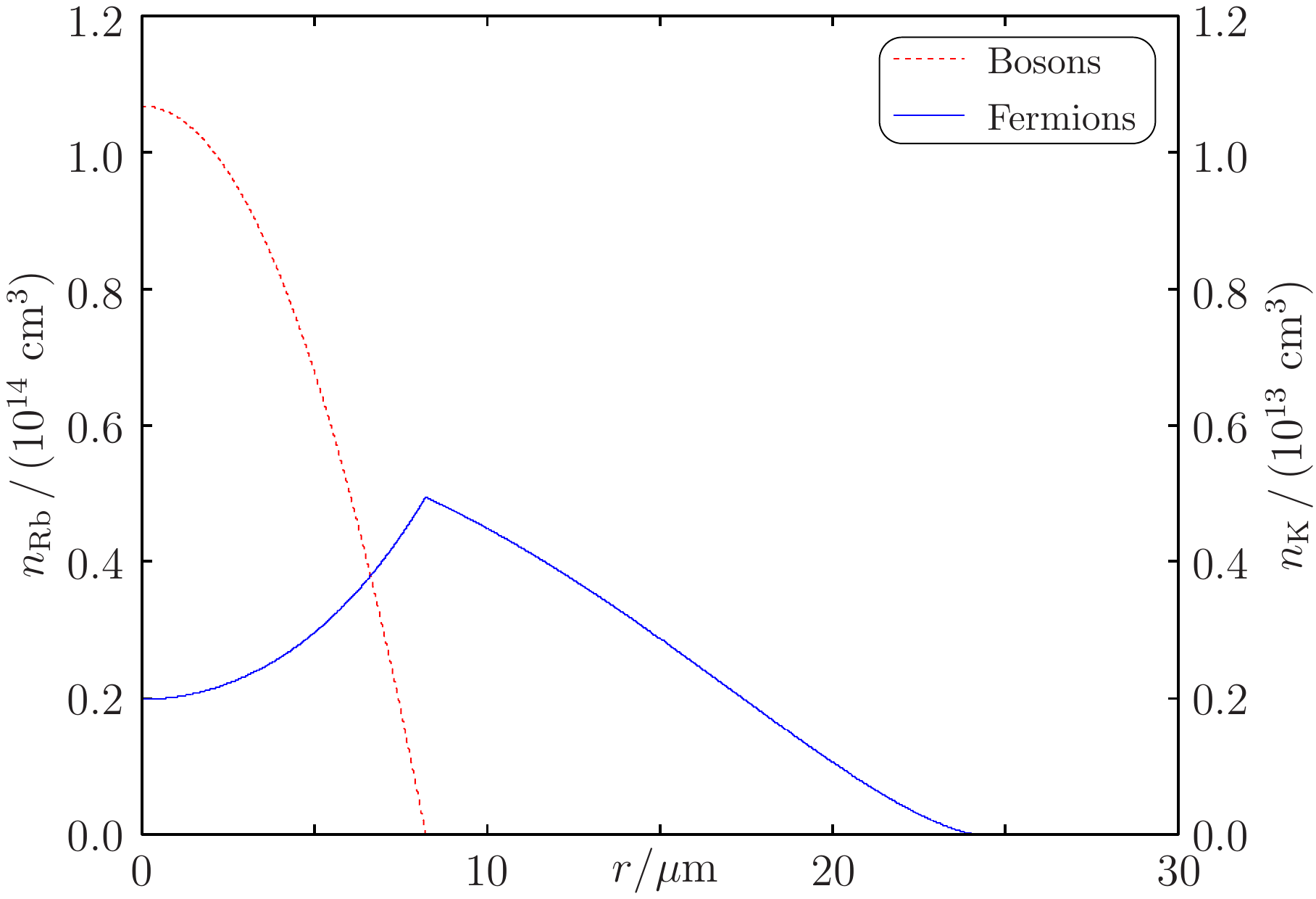}\label{fig:rep_int}}
  \subfigure[phase separation ($g_{FB}/g_{BB}=+8$
)]{\includegraphics[width=0.47\columnwidth]{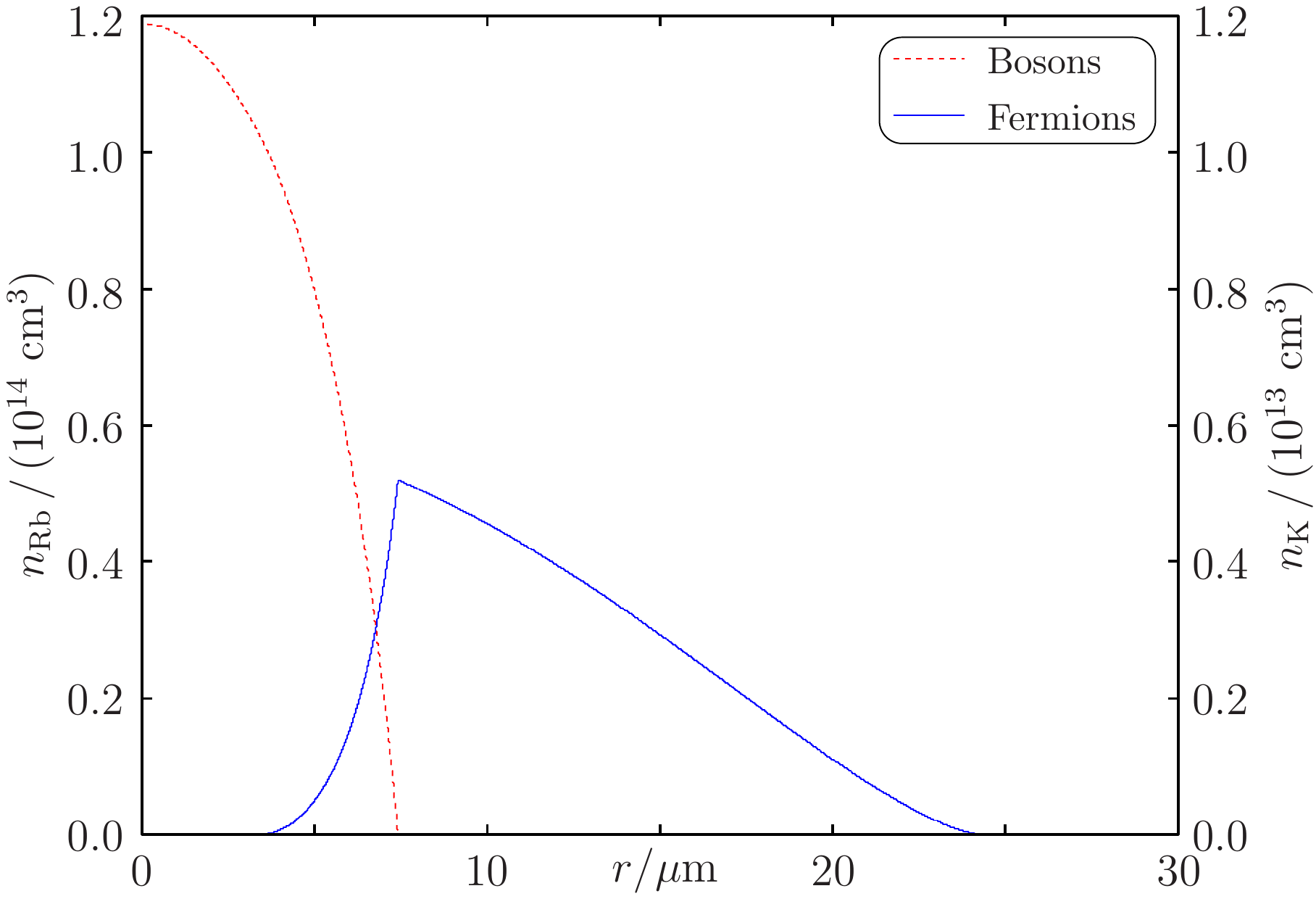}\label{fig:phase_sep}}
  \caption{Density distributions for trapped Fermi-Bose mixtures calculated in the Thomas-Fermi model. Parameters are $N_B=N_F=10^5$, and $\bar\omega_{\mathrm{B}}=2\pi\cdot50$~Hz. Also, $a_{BB}=98.98a_0$~\cite{Kempen2002a}, corresponding to \rub\ in $\left|F=2, m_F=2\right>$.}
  \label{fig:densitydist}
\end{figure}

In the following, we are going to analyze solutions to this coupled mean field scenario for various heteronuclear interaction strength combinations. Typically, equations~(\ref{eq:thomas_fermi_coupled_trans1}) and (\ref{eq:thomas_fermi_coupled_trans2}) are solved in an iterative scheme by initializing both bosonic and fermionic distributions with corresponding non-interacting distributions. In a first step, because the bosonic density is usually much higher than the fermionic density, equation~(\ref{eq:thomas_fermi_coupled_trans2}) is used to update the fermionic density, taking into account the bosonic mean field potential. The fermionic chemical potential is determined according to equation~(\ref{eq:thomas_fermi_coupled_N_trans2}). The scheme continues with updates to bosonic, fermionic, $\ldots$ density distributions until sufficient convergence is obtained.

Figure~\ref{fig:densitydist} shows solutions to equations (\ref{eq:thomas_fermi_coupled}) for various scattering length combinations. We will limit the discussion to repulsive interactions between bosons ($a_{BB}>0$) first. Let us start the discussion by looking at the non-interacting case depicted in fig.~\ref{fig:non_int}. In this case, we have the fairly broad distribution of the fermionic cloud which is due to the Pauli exclusion principle preventing two atoms to come too close to each other. The BEC, on the other hand, shows the characteristic inverted parabolic distribution. As we turn on a weak attractive heteronuclear interaction (see fig.~\ref{fig:att_int}), we observe that the shape of the BEC is almost unchanged, whereas the Fermi sea shows a peak in the center in the overlap region with the BEC. For relatively weak interaction, we can almost stop the iterative solution process to equations (\ref{eq:thomas_fermi_coupled}) after applying (\ref{eq:thomas_fermi_coupled_trans2}) once, considering only the effect of the BEC on the fermions. The high density of the BEC creates a mean field `dimple' in which the fermions are trapped. This leads to the characteristic increase in fermionic density in the overlap region. 

A similar picture arises for a relatively weak repulsive interaction between bosons and fermions (fig.~\ref{fig:rep_int}). In that case, the fermions get pushed out of the center of the harmonic trap by the BEC, and we observe a lowered fermionic density in the overlap region. 

As the Fermi-Bose interaction becomes stronger and repulsive, the fermions can become completely expelled from the overlap region with the condensate, and we observe the onset of phase separation as seen in fig.~\ref{fig:phase_sep}. For approximately equal particle numbers of bosons and fermions, phase separation occurs as a dense core of bosons surrounded by a dilute shell of fermions, although there can be more exotic scenarios. 

\begin{figure}
  \centering
  \includegraphics[width=1.0\columnwidth]{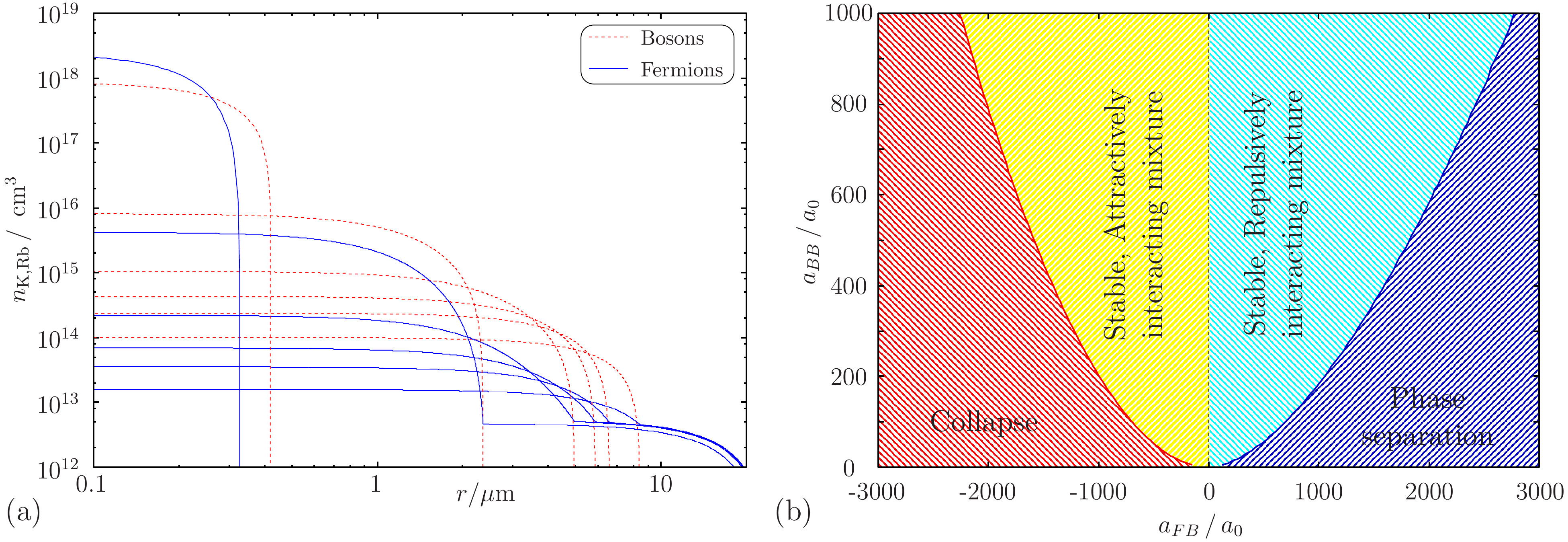}
  \caption{(a) Iterations of the mean field model in the case of instability (collapse).
           (b) Phase diagram of a harmonically trapped Fermi-Bose mixture for $N_B=10^6$ and $N_F=5\cdot10^5$.
          }
  \label{fig:collapse_phase_diag_combine}
\end{figure}

For very strong and attractive interaction, fig.~\ref{fig:collapse_phase_diag_combine}(a) shows subsequent steps in the iterative solution scheme to the coupled mean field problem. In this case, the presence of the condensate pulls a substantial amount of fermions into the center of the trap. In the limit of strong interactions, there is a sizeable back action onto the condensate, which in turn contracts even more, leading again to an increased fermionic distribution at the center of the trap. Beyond certain critical conditions, there will not be any self-consistent solution to equations (\ref{eq:thomas_fermi_coupled}). This is the regime of the mean field collapse of the mixture, in analogy to the mean field collapse observed for pure BECs with attractive interactions. If a stable mixture is brought into the regime of the mean field collapse by either increasing the number of particles or by increasing the interaction, the sample will collapse in a 3-body implosion, where the ever and ever increasing density at the center of the trap is accompanied by strong 3-body losses. We will later see how these losses can be observed in the experiment.

To summarize the behavior of trapped mixtures, let us have a look at fig.~\ref{fig:collapse_phase_diag_combine}(b), which shows a simple phase diagram of a trapped mixture. The phase diagram shows areas of stable mixtures for small values of the heteronuclear interaction; for large and attractive interaction, we observe the onset of collapse; for large and repulsive interaction, we observe the onset of phase separation. Note that the phase diagram is for an equilibrium situation at $T=0$ only; in particular, the presence of excitations can lead to collapse earlier than expected from equilibrium mean field theory (see below). Also, the presence of a dense thermal cloud at $T\ne 0$ and the corresponding mean field confinement will lead to collapse earlier than predicted by our cartoon picture.

So far, we have only talked about the case of repulsive interaction between bosons. In the case of attractive interaction ($a_{BB}<0$), the condensate has a very strong tendency to collapse all by itself already for very small particle numbers.

\section{Harmonically trapped Fermi-Bose mixtures with large particle numbers}

\begin{figure}
  \centering
  \includegraphics[width=0.5\columnwidth]{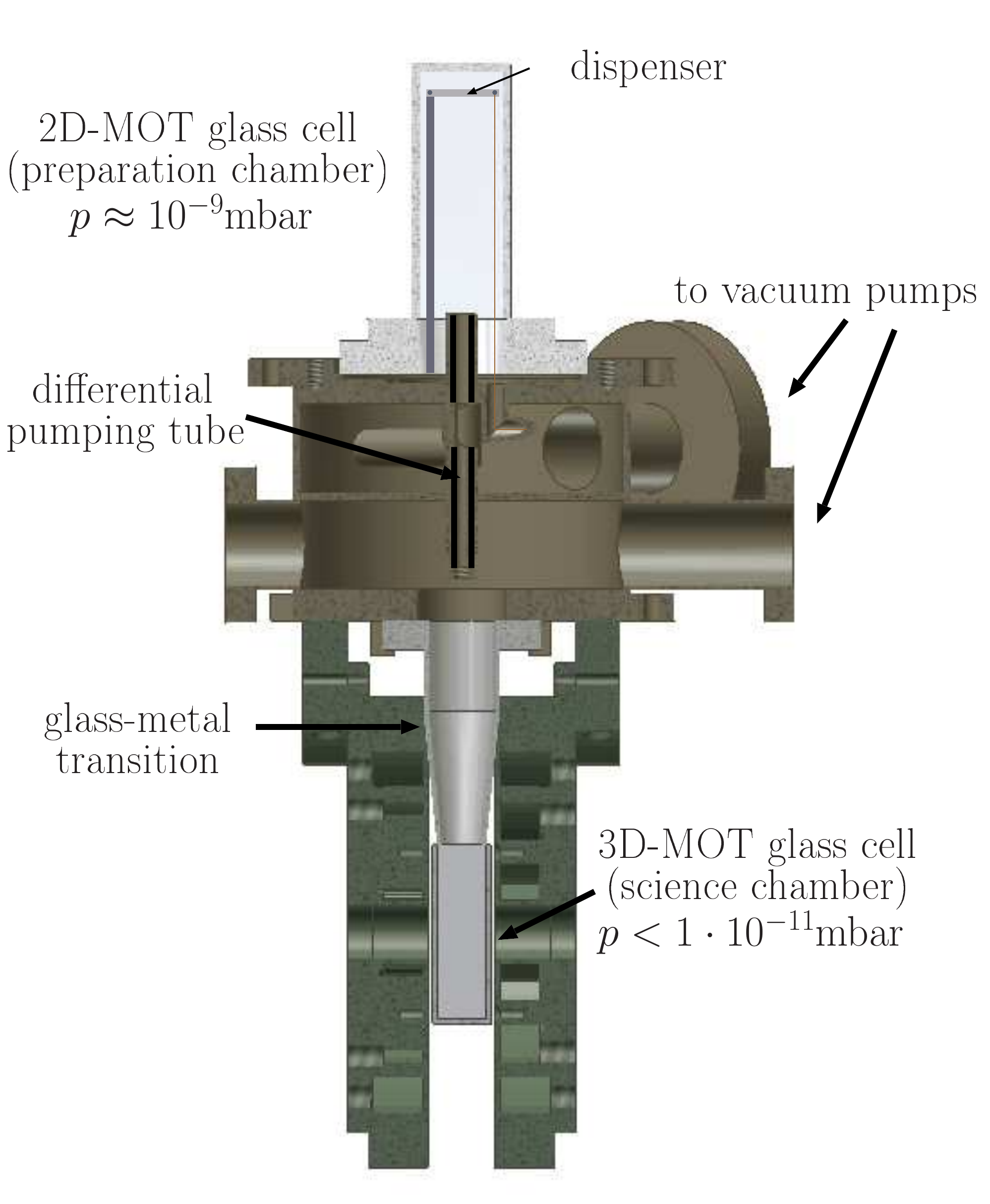}
  \caption{Sketch of the basic experimental setup, excluding all laser systems, optics and electronics. In the 2D magneto-optical trap, a cold atomic beam is produced, sent through a differential pumping stage and recaptured in a 3D-MOT. Atoms are transferred into a magnetic trap, sympathetically cooled into the degenerate regime and then transferred into optical traps and optical lattices for further experiments.}
  \label{fig:vacuum}
\end{figure}

After this discussion of basic properties of trapped Fermi-Bose mixtures, let us describe the basic experimental setup and  route to producing trapped mixtures with large particle numbers. Figure~\ref{fig:vacuum} illustrates the basic concept of the experiment. A vapor of \pot\ and \rub\ is produced within a rectangular glass cell (2D-MOT cell) at a pressure of between $10^{-8}$ and $10^{-9}$~mbar using alkali metal dispensers~\cite{Wieman1995a}; in the case of \pot, we use enriched sources~\cite{DeMarco1999a} because of the low natural abundance of \pot. From this background vapor, a cold atomic beam is produced using two pairs of horizontal, orthogonal laser beams forming a two-dimensional magneto-optical trap~\cite{Dieckmann2D,JuergenThesis} for both species. The beam is collimated in the horizontal direction and directed along the vertical axis of the apparatus.
The flux towards the lower part of the apparatus is enhanced by using a resonant pushing beam which is optimized to match the velocity distribution with the capture range of a 3D two-species magneto-optical trap located in the lower part of the apparatus in a second glass cell. This region is separated from the 2D-MOT region by a differential pumping stage, thereby achieving UHV conditions in the 3D-MOT region. From the cold atomic beam, atoms are accumulated in the 3D-MOT for typically ten seconds with final particle numbers on the order of $1\cdot 10^{10}$ for \rub\ and $2\cdot10^8$ for \pot\ ($5\cdot10^7$ in the presence of \rub). In the case of \pot, we use a dark SPOT scheme~\cite{dark_spot,MarlonThesis,jmo} to overcome losses due to light-assisted collisions between atoms. This scheme has been crucial for achieving large particle numbers of \pot\ in the presence of \rub\ in the combined MOT.

After accumulation of atoms in the MOT and subsequent polarization gradient cooling, the atoms are prepared in a spin state appropriate for magnetic trapping. In our experiment, we use a mixture of \pot\ in the $\left|F=9/2,m_F=9/2\right>$ state and \rub\ in the $\left|F=2,m_F=2\right>$ state. Atoms in these states are prepared using $\sigma^+$ light on the corresponding cycling transitions starting from ensembles in the $F=9/2$ and $F=2$ manifolds, respectively. We then switch on the decompressed magnetic trap with trapping frequencies of $\omega^{\mathrm{Rb}}_x=\omega^{\mathrm{Rb}}_y=\omega^{\mathrm{Rb}}_z=2\pi\cdot11$~Hz to allow for almost mode-matched and efficient loading of the magnetic trap~\cite{DinterThesis}. The magnetic trap is then compressed adiabatically in the radial direction during $1.5$\,s to final frequencies of $\omega^{\mathrm{Rb}}_x=\omega^{\mathrm{Rb}}_y=2\pi\cdot257$\,Hz, while $\omega^{\mathrm{Rb}}_z$ remains constant. We typically prepare $3\cdot10^9$ \rub\ atoms and $2\cdot10^7$ \pot\ atoms in the magnetic trap.

Inside the magnetic trap, we perform forced evaporative cooling of \rub. At a given magnetic field, the energy difference between neighboring Zeeman states is $g_{F_{\mathrm{Rb}}}/g_{F_{\mathrm{K}}}=9/4$ higher for \rub\ than for \pot; the evaporative cooling therefore predominantly removes \rub, while \pot\ is cooled in the thermal bath of \rub. Any spin-polarized fermionic gas alone would suffer from Pauli-blocking of $s$-wave collisions~\cite{DeMarco2001a} at the temperatures in the magnetic trap, making evaporative cooling of such a system very inefficient.

The \pot--\rub\ system is characterized by a relatively large and attractive background interaction~\cite{florence_fermions_on_bec}, although there was some initial controversy on the exact interaction strength. The large attractive interaction favors efficient evaporative cooling but, on the other hand, make the system susceptible to the mean field collapse of the mixture discussed above.

Let us now look at various stages of the evaporative cooling process as the \rub\ component condenses and the interaction between \pot\ and \rub\ affects both density distributions. When the \rub\ atoms are not completely evaporated, various regimes of mixtures are accessible, ranging from dense thermal $^{87}$Rb clouds of $10^7$ $^{87}$Rb atoms right at the phase transition point interacting with a moderately degenerate Fermi gas of $2\cdot 10^6$ $^{40}$K atoms to deeply degenerate mixtures with almost pure condensates. We achieve $>1\cdot 10^6$ atoms in the condensate coexisting with $7\cdot 10^5$ $^{40}$K atoms, limited by the onset of the collapse discussed below.

Throughout this discussion, the \pot\ images were recorded with a time of flight $T_E$ of 3 to 5\,ms. With an axial \pot\ trap frequency of $\omega^{\mathrm{K}}_{z}=2\pi\cdot 16.6$~Hz, this means that $\omega^{\mathrm{K}}_{z}\cdot T_E$ is on the order of 0.3 to 0.5, and with the relatively large size of the Fermi gas in the trap, we can still deduce information on the axial {\it spatial} distribution in the magnetic trap from the time of flight image. In the experiment, this has allowed the extraction of information about the outer regions of the Fermi gas where it is noninteracting and about the inner regions where it overlaps and interacts with the dense Bose-Einstein condensate. Note that the expansion of the samples is also affected by the attractive interaction between \pot\ and \rub; however, in a cigar-shaped trap, the main effect is on the radial expansion where the confinement is strong~\cite{Ferlaino2004a,Hu2003a}.

The \rub\ atoms, on the other hand, are imaged after a total time of flight of 20\,ms. All of the experiments described in this chapter are based on imaging using the two-species detection technique (see~\ref{sec:imaging}) which has allowed us to extract information about both species on the same CCD chip in one run.

\begin{figure}
  \centering
  \includegraphics[width=0.5\columnwidth]{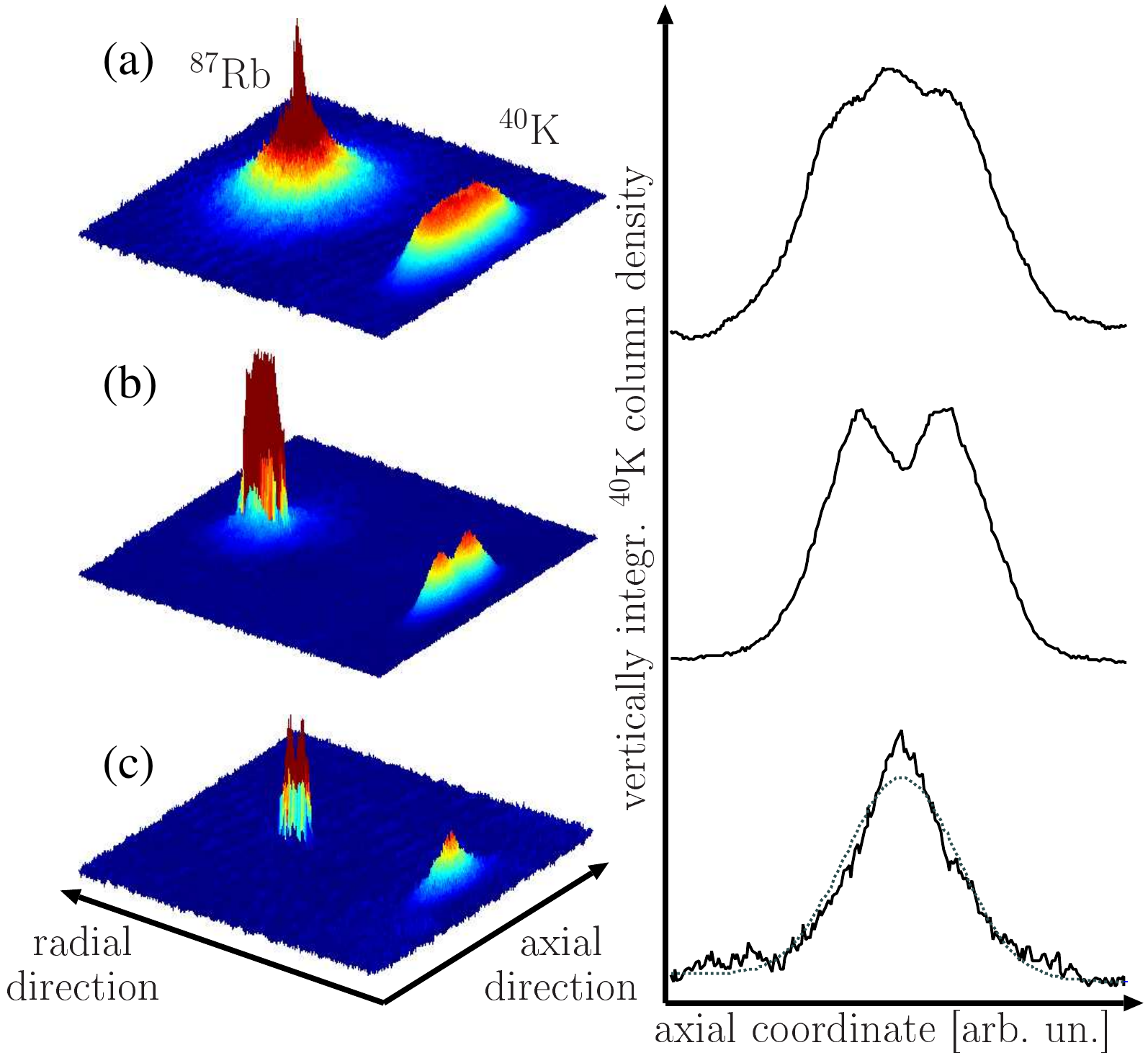}
  \caption{Density profiles of interacting \pot--\rub\ mixtures in various stages of the evaporative cooling process~\cite{Ospelkaus2006b}. {\it Copyright (2006) by the American Physical Society.}}
  \label{fig:collapse_profile_total}
\end{figure}

During the final stage of evaporative cooling, various regimes of interacting mixtures are observed. These different regimes become most evident when analyzing the axial 1D profile of the interacting clouds (see fig.~\ref{fig:collapse_profile_total}). Here, we have analyzed in more detail three different stages. In (a), the Fermi cloud exhibits a small peak in the center on top of a flat profile. This peak is due to the Fermi-Bose attraction increasing the fermionic density in the center of the trap where the BEC is formed. Once critical conditions for the onset of the collapse have been met as in (b), the collapse localized in the overlap region removes the overlapping fraction of the Fermi gas. The hole which is left behind can be refilled from the outer regions of the trap, and, as we shall see, this refilling process can even lead to a second collapse of the mixture. The peaked structure which is shown in (c) is stable over relatively long timescales of several 100\,ms once the total \pot\ number has been reduced to undercritical values; for comparison, we have included the result of a Fermi-Dirac fit to the experimental data which clearly fails to account for the interactions with the Bose-Einstein condensate in the center. In contrast to the situation depicted in (a), which results from overcritical initial conditions and is on the edge to collapse, the condition of subfigure (c) is a signature of Fermi-Bose interactions in a stable mixture and corresponds to fig.~\ref{fig:att_int} in our introductory discussion.

\subsection*{Observation of a mean field induced collapse}

\begin{figure}
  \centering
  \includegraphics[width=0.6\columnwidth]{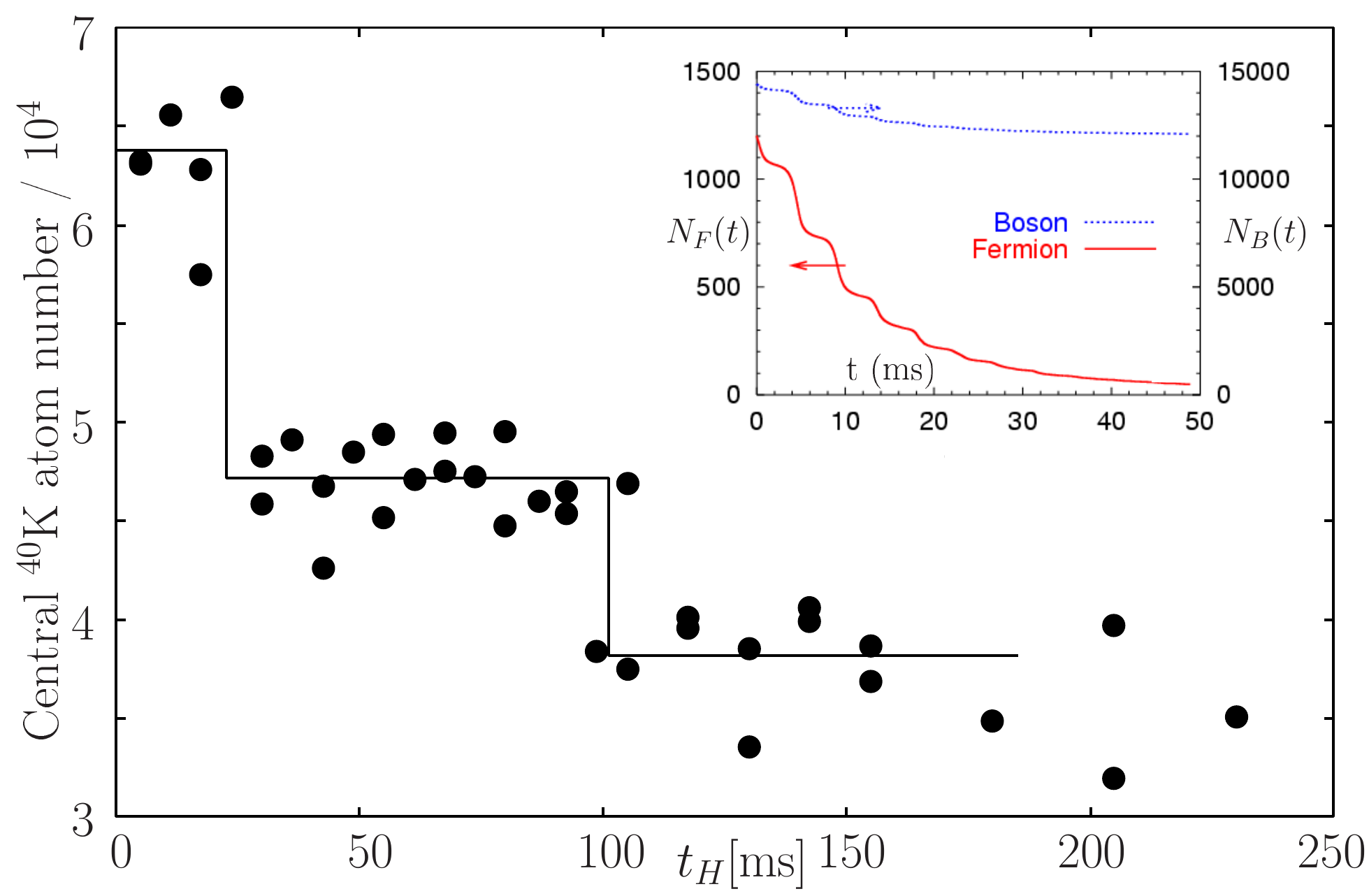}
  \caption{Sudden drop in atom number as a result of the initial collapse, followed by one revival of the collapse out of a non-equilibrium situation~\cite{Ospelkaus2006b}. Inset taken from~\cite{Adhikari2004a}. {\it Copyright (2006) by the American Physical Society.}}
  \label{fig:revivals}
\end{figure}

In the discussion of the density profiles, we have claimed that the appearance of the hole in the center of the trap (see fig.~\ref{fig:collapse_profile_total}(b)) is due to the mean field collapse of the mixture. The distinction from localized loss processes in the presence of high densities is most apparent when looking at the associated time scales. In order to study the time scale of atom loss, we will discuss a situation where an only slightly overcritical mixture has been prepared. During the first few milliseconds of the hold time, where the rf knife is held at a fixed frequency, the condensate will still grow out of the thermal cloud and reach critical conditions after some initial delay. The result of the measurement is shown in fig.~\ref{fig:revivals}. The figure shows the particle number contained in the central part of the \pot\ absorption image which helps enhance the visibility of the effect. Between 10 and 20\,ms after the start of the measurement, a sudden drop reduces the integrated particle number in this area to about two thirds of its original value. The atom number then remains constant for some time and at 100\,ms again undergoes a second sudden drop, which is due to a refilling of the cloud in the center from the outer regions and a subsequent second collapse. Looking at fig.~\ref{fig:revivals}, one might argue that this refilling should become apparent in the center particle number going up again after the initial collapse. There is no such evidence from the data. One must however take into account that the central atom number from the time of flight image only approximately reflects the particle number in the overlap region. The occurrence of excitations and strong unequilibrium conditions following the initial collapse may modify the expansion behavior and in particular lower the collapse conditions for the second occurrence of this mean-field instability. The message of fig.~\ref{fig:revivals} is the step-like feature which shows a very sudden drop of the particle number  characteristic for this rapidly contracting mean-field implosion. From fig.~\ref{fig:revivals}, we can constrain this timescale to below 10~ms, which is incompatible with pure 3-body loss in the absence of the collapse. The inset in fig.~\ref{fig:revivals} shows results from numerical modelling of the mixture in the regime of mean field instability based on hydrodynamic equations~\cite{Adhikari2004a}. The numerical simulations qualitatively predict the same behavior for the particle number as a function of time. Note that the simulations assume different trap parameters, resulting in the different time scales. Also, the sharpness of the steps is a function of the 3-body loss rate assumed in the simulation.

One of the experimental challenges in measuring data as in fig.~\ref{fig:revivals} is that over many runs, the initial conditions must be stable enough so that the time evolution can be well controlled. This is most important with respect to initial particle numbers and transfer from the MOT to the magnetic trap. When large overcritical mixtures are initially prepared, the steps shown in fig.~\ref{fig:revivals} can no longer be resolved individually due to initial fluctuations and the very complex dynamics. Instead, in this limit the collapse is observed as a very rapid overall decay when measurements with independently prepared samples are visualized together in a way similar to fig.~\ref{fig:revivals}. It would be an interesting perspective to analyze the behavior of the cloud in the regime of instability using phase-contrast imaging which would allow several subsequent nondestructive measurements of the same sample. 

\subsection*{Stability analysis of Fermi-Bose mixtures}

\begin{figure}
  \centering
  \includegraphics[width=0.6\columnwidth]{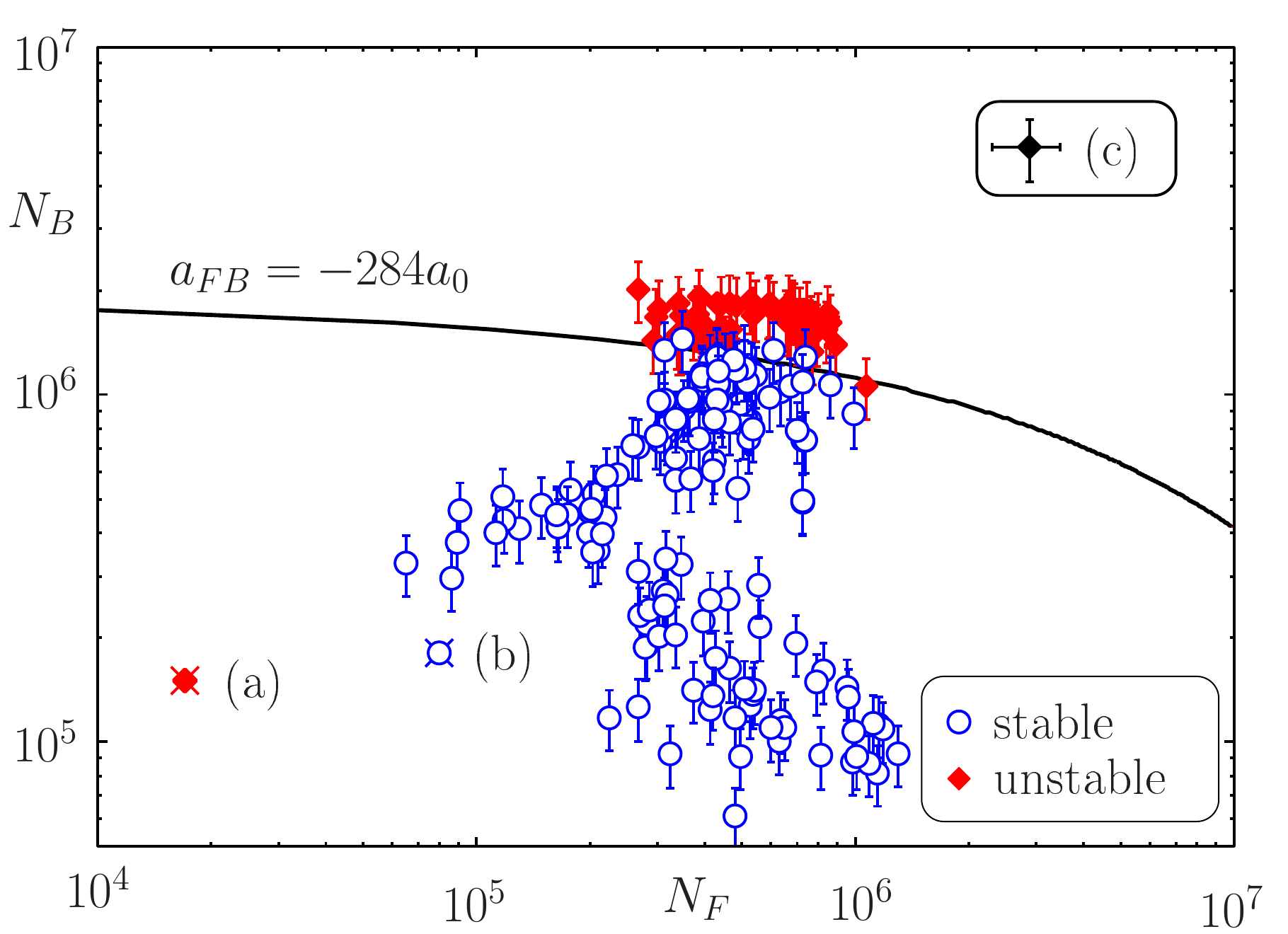}
  \caption{Stability of \pot--\rub\ mixtures~\cite{Ospelkaus2006b} as a function of bosonic and fermionic particle numbers $N_B$ and $N_F$ with error bars (c). For comparison, we include stable (b) and unstable (a) particle number combinations observed in other experiments (see text). {\it Copyright (2006) by the American Physical Society.}}
  \label{fig:stability}
\end{figure}

During the above discussion, we have mentioned conditions of criticality of mixtures and critical particle number of the mean field collapse several times. Here, we summarize our experimental findings on stable and unstable mixtures and discuss the relation to the value of the $s$-wave scattering length for collisions between \pot\ and \rub. Two criteria for observation of mean-field instabilities have been used here:
\begin{itemize}
\item Particle number combinations where the overall decay of the mixture is much too fast for normal 3-body decay in connection with the appearance of the pronounced hole are considered as unstable.
\item The observation of the step-like drop as discussed in context with fig.~\ref{fig:revivals} is a sign of instability.
\end{itemize}
Other situations are identified as stable. The resulting atom number combinations are plotted in fig.~\ref{fig:stability}. Stable atom number combinations are plotted in blue, and combinations found to be unstable in the experiment in red. Atom numbers observed as stable are e.\ g.\ 
\begin{equation}
N_F=7\cdot10^5;\quad N_B=1.2\cdot10^6;\quad\bar\omega=2\pi\cdot91\,\mathrm{Hz}
\end{equation}
The error bars (c) result from an estimate of the \rub\ and \pot\ atom number uncertainty of 20\% and 30\%, respectively. The uncertainty in the \pot\ atom number is less important than for \rub, since in the regime considered here the dependence of critical conditions on $N_F$ is much smaller than on $N_B$. Also shown in fig.~\ref{fig:stability} are particle number combinations from two other experiments, namely the one at LENS (a) and the JILA \pot--\rub\ setup (b). The particle number combination (a) of
\begin{equation}
N_F=2\cdot10^4;\quad N_B=1.5\cdot10^5;\quad\bar\omega=2\pi\cdot91\,\mathrm{Hz}
\end{equation}
has been reported as critical for the onset of collapse in ref.~\cite{florence_mean_field_analysis}, but is an order of magnitude smaller than the stable particle number combination observed in our experiment. Data point (b) is the stable particle number reported at JILA in ref.~\cite{jila_interaction_strength} without any sign of instability:
\begin{equation}
N_F=8\cdot10^4;\quad N_B=1.8\cdot10^5;\quad\bar\omega=2\pi\cdot86\,\mathrm{Hz}
\end{equation}
which raises the question why mixtures which are observed as stable in this experiment and by the JILA experiment have been reported as unstable at LENS.

When comparing critical particle numbers of different experiments, a few words on the role of the trap parameters are appropriate. Conditions for instability are of course influenced by the trapping potential. In general, the stronger the external confinement, the lower the critical particle numbers or the critical interaction are. The scale for the onset of instability is set by the geometric mean of the trap frequencies $\bar\omega$ which is given above for all three experiments. The mean harmonic trap frequency used in this experiment and at LENS are equal, whereas the JILA trap is just slightly shallower.

Within the Thomas-Fermi approximation for bosons and fermions, we have shown in our introductory discussion that for fixed harmonic mean trap frequency, the aspect ratio does not influence the point of onset of collapse. This trap and the LENS trap are therefore directly comparable within this approximation, and the JILA trap is only slightly shallower and should therefore be slightly less subject to collapse than the two other traps for the same particle number.

The only difference between the LENS trap and the trap used here is the trap aspect ratio (our trap is slightly more elongated). This  may play a role in the following situations: 
\begin{itemize}
\item {\bf Breakdown of the Thomas-Fermi approximation.} This may occur when going to very elongated quasi-1D geometries (which is not the case in any of the experiments discussed here). The breakdown of the Thomas-Fermi approximation in the radial direction would result from radial trap frequencies on the order of kHz. The general effect of such a geometry is a stabilization with respect to collapse. This stabilization is responsible e.\ g.\ for the stabilization of bright solitons in a narrow window before the onset of collapse \cite{bf_bright_solitons}. Another reason may be that particle numbers become below, say, $10^4$. Neither of these scenarios applies to the experiments discussed here.
\item {\bf Gravitational sag.} A trap which is radially more tight features a reduced {\it differential gravitational sag} between atoms of different mass compared to the isotropic situation. This differential sag is given by:
\begin{equation}
-g\left(\frac1{\omega_\mathrm{Rb}^2}-\frac1{\omega_\mathrm{K}^2}\right)
\end{equation}
and vanishes for equal trap frequencies of both components. Since the differential gravitational sag reduces the spatial overlap of the components, some of the mean field energy is already ``used up'' in compensating for the gravitational sag. The larger gravitational sag in the Florence experiment thus results in the LENS trap becoming ``less subject to collapse'' for the same critical particle numbers, while the opposite is observed in the experiment. The role of the gravitational sag has also been discussed in~\cite{fb_sag}.
\end{itemize}
Therefore, for the purpose of comparing the above-mentioned experimental results, the difference in trap aspect ratio does not play a role and cannot account for the order of magnitude difference in critical particle numbers.

It has been pointed out by the LENS group~\cite{florence_mean_field_analysis} that with the very sensitive relationship between critical particle numbers and the Fermi-Bose interaction parameter $a_{FB}$, the observation of the collapse instability can be a very sensitive way to determine $a_{FB}$. This picture is true with respect to equilibrium conditions; however, the observation of instabilities during evaporative cooling, where excitations can never be completely avoided, complicates the picture, as we shall see. In this sense, the observation of a given particle number combination as stable imposes an upper limit on the interaction parameter.

\subsection*{Excitations}

\begin{figure}
  \centering
  \includegraphics[width=0.6\columnwidth]{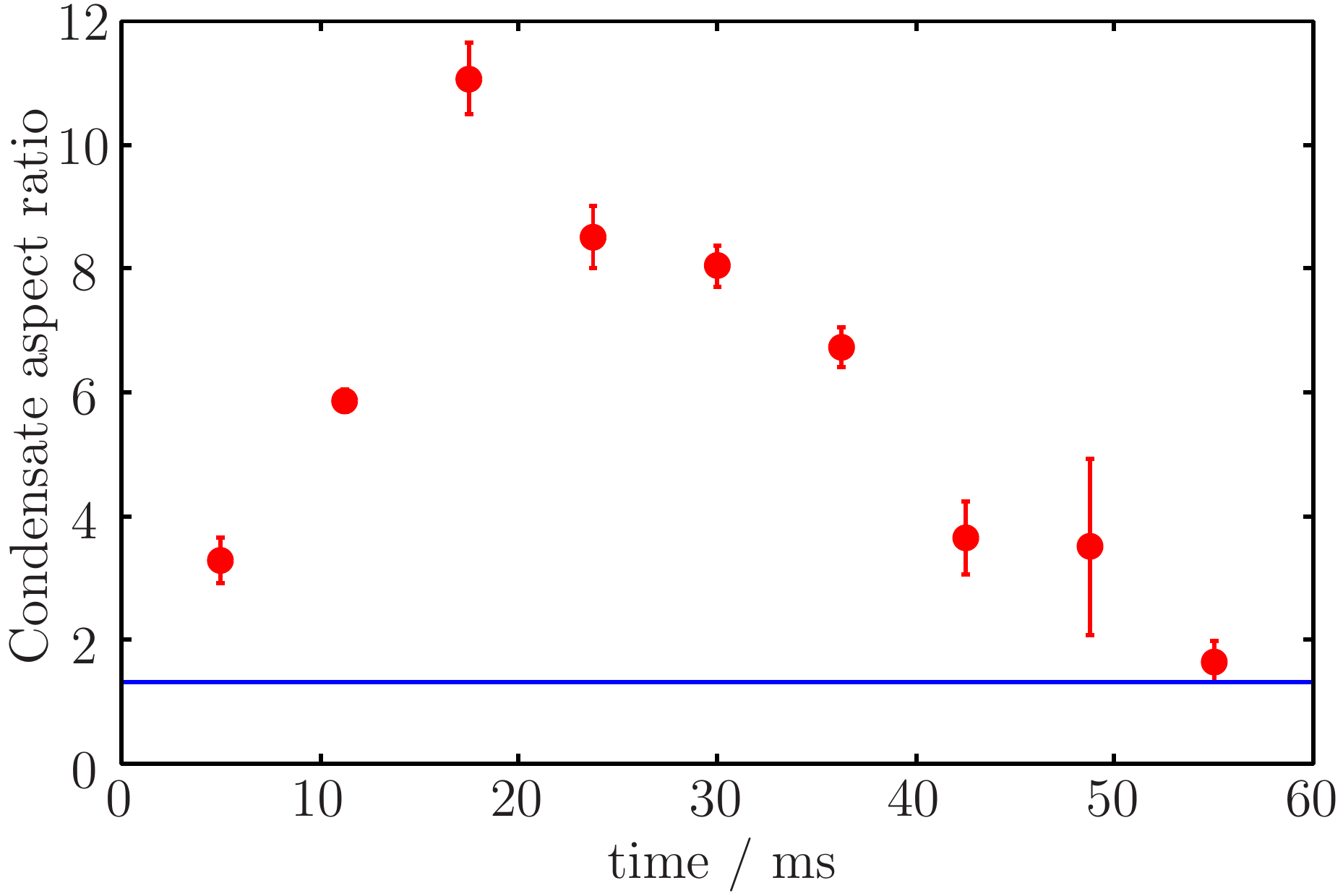}
  \caption{Aspect ratio of a condensate forming in a \pot--\rub\ mixture, compared to the pure BEC expected aspect ratio of~1.3 (blue line)~\cite{jmo}. {\it Reprinted with permission of the publisher, Taylor and Francis Ltd.}}
  \label{fig:aspect_ratio}
\end{figure}

In our discussion of the implications of observed stability limits, we have mentioned several times that the presence of excitations in the sample during evaporative cooling means that an observed stability limit is always an upper limit on the interaction parameter. To illustrate this, fig.~\ref{fig:aspect_ratio} shows the aspect ratio of the condensate before the onset of collapse in some of our experiments. The strong deviations (factor of $\approx 10$) from the equilibrium aspect ratio of $1.3$ may be due to fast evaporation compared to the axial trap frequency, thereby creating strong axial excitations.

These excitations can in turn locally increase the density and lead to critical conditions even for particle number combinations which are stable in an equilibrium situation. For example, a twofold increase of the Thomas-Fermi radius of the condensate due to excitations would lead to a fivefold increase in the central density of the condensate, leading to the same densities required for a an equilibrium collapse at $a_{FB}=-200\,a_0$. The same is true for any experiment observing the collapse during the evaporation ramp: In the absence of more sophisticated theory taking into account the excitations, observation of certain particle number combinations as stable provides an upper limit on scattering parameters.

When putting the observed experimental stability limit from this experiment (which produced the largest degenerate mixtures so far) into mean field theory, we obtain $-a_{FB}<284~a_0$ as an upper limit on the interaction parameter. This upper limit coincides with the value deduced from the first \pot--\rub\ spectroscopy experiment at JILA and the corresponding resonance assignment. The identification of more Feshbach resonances and an upgraded collisional model performed at LENS as well as an improved magnetic field calibration has resulted in a scattering length of $-215(10)~a_0$~\cite{Ferlaino2006b}, and the upper limit reported here is compatible with that value. Since the same argument is valid for any experiment, there is no contradiction between the observed stable particle number combinations and $a_{FB}=-215(10)~a_0$. Rather, the above comparison demonstrates limits of a comparison between equilibrium mean field theory and the experimental observation during evaporation, where strong dynamics and excitations can occur.

It is an important result from this experiment that large stable \pot--\rub\ particle numbers can be produced, which was not clear {\it a priori}, and to have identified experimental signatures of the mean field collapse in Fermi-Bose mixtures, which have later been seen in experiments with tunable interactions, as we shall see now.

\section{Tuning of heteronuclear interactions}

So far, we have investigated the behavior of the mixture in the harmonic trap at a fixed background scattering length. In the early days of ultracold quantum gases, a lot of interest in a particular atomic species was motivated by its static scattering properties; in the case of \rub\ e.\ g., the moderately large background scattering length, combined with favorable collisional losses (and technical considerations such as cooling laser sources), is the main reason for the fact that the overwhelming majority of BEC experiments relies on \rub\ these days. This picture changed dramatically with the advent of Feshbach resonances.

\begin{figure}
  \centering
  \includegraphics[width=0.6\columnwidth]{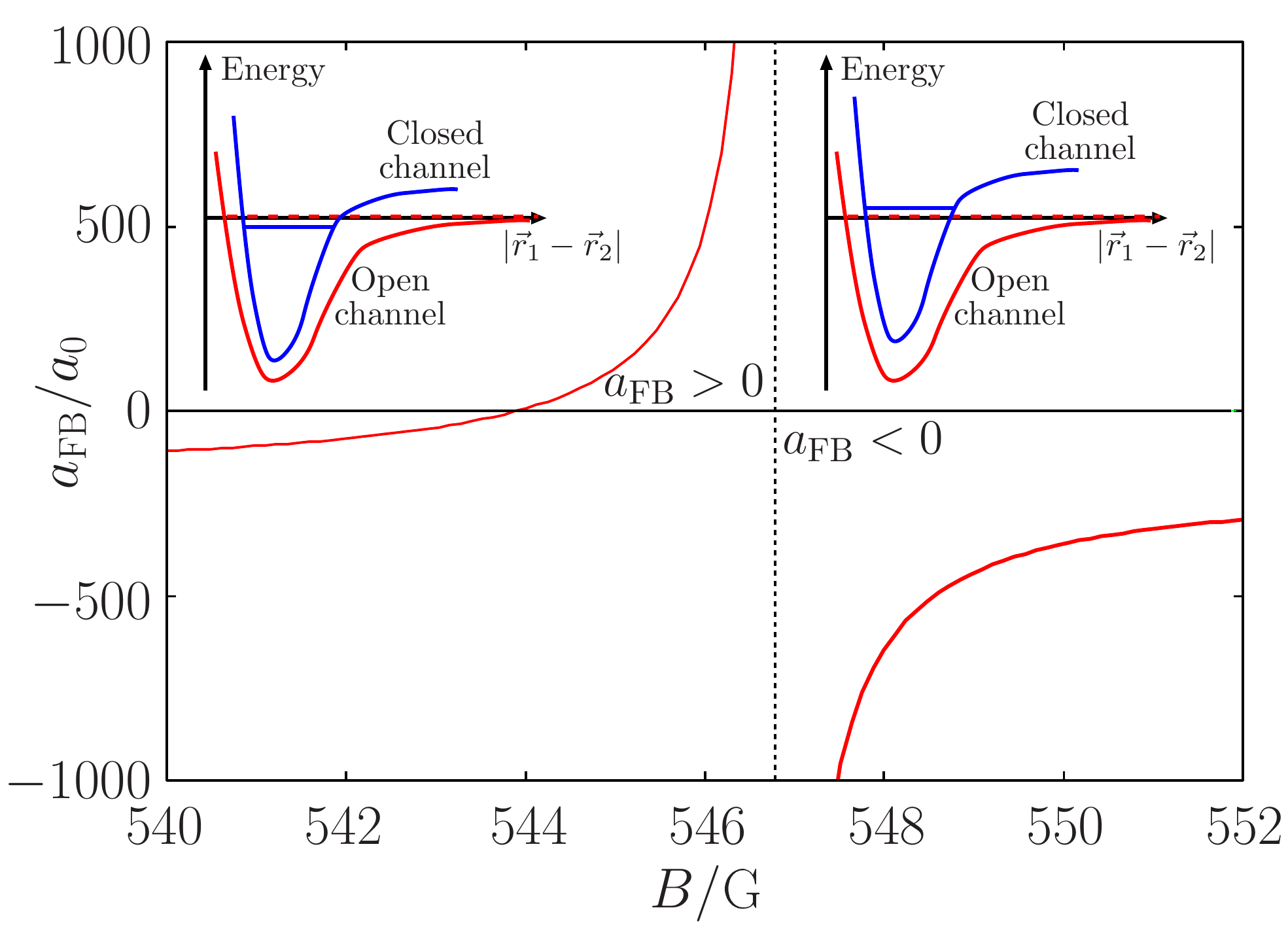}
  \caption{Cartoon picture of a Feshbach resonance. Such resonances occur when the energy of two atoms scattering from a molecular potential (`open channel') is close to resonance with a  quasibound state in a different molecular potential curve (`closed channel'). In general, both channels have different magnetic moments, allowing the bound state in the closed channel to be tuned relative to the open channel disscociation limit. This allows the s-wave scattering length (plotted here) between atoms in the open channel to be tuned from large and attractive to large and repulsive. In particular, when the bound molecular state is slightly below the dissociation threshold (immediately below the magnetic field reconance center position $B_0$, indicated by the vertical dashed line), s-wave interactions between atoms are repulsive ($a_{\mathrm{FB}}>0$), and weakly bound molecules can exist in the closed channel. When the molecular bound state is slightly above the resonance (right of the dashed line), no stable Feshbach molecules are formed, and the interaction between atoms is attractive ($a_{\mathrm{FB}}<0$).}
  \label{fig:feshbach_resonance}
\end{figure}

A Feshbach resonance is a phenomenon originally introduced in the context of nuclear physics ~\cite{Feshbach1958a}. In cold atomic gases, these resonance phenomena have been discussed since 1992, especially for Cs \cite{Tiesinga1992a, Tiesinga1993a} which at the time seemed to be the most promising candidate for the observation of BEC in a dilute atomic gas. A Feshbach resonance  occurs when the energy of the colliding atoms in an open channel  resonantly couples  to a quasibound molecular state  in a closed channel. A Feshbach resonance allows tuning of the interaction properties of an atomic gas from large and repulsive to large and attractive attractive and vice versa. The enhanced interaction properties near the Feshbach resonances are often accompanied by enhanced inelastic collisions. These inelastic collisions provide an easy tool for the identification of resonances, but can at the same time affect the timescales for experiments in the vicinity of the resonances. Feshbach resonances have been suggested to be inducible via  static electric fields~\cite{Marinescu1998a}, rf radiation~\cite{Moerdijk1996a}, via far-resonant or near-resonant laser fields~\cite{Fedichev1996a,Bohn1997a} or via homogeneous magnetic fields~\cite{Tiesinga1992a, Tiesinga1993a,Moerdijk1995a, Boesten1996a}. Nowadays, magnetically induced Feshbach resonances are a very versatile and common tool in cold atom laboratories. They occur when the magnetic moment of the  bound state in the closed channel is different from the magnetic moment of the open channel and the energy of the bound state can thus be continuously tuned through the open channel dissociation limit (see fig.~\ref{fig:feshbach_resonance}).

The first observation of Feshbach resonances in cold atom scattering has been reported almost simultaneously by P. Courteille and coworkers in $^{85}$Rb~\cite{Courteille1998a} and by S. Inouye and coworkers in a BEC of $^{23}$Na~\cite{Inouye1998a}. Tuning of interactions was demonstrated by Inouye and coworkers in the vicinity of the resonance by  a  static homogeneous magnetic field. Magnetically induced Feshbach resonances have later  been reported in almost any alkali atom:  $^{6}$Li~\cite{Dieckmann2002a} and $^{7}$Li~\cite{Khaykovich2002a, Strecker2002a}, $^{23}$Na~\cite{Inouye1998a}, $^{40}$K~\cite{Loftus2002a}, $^{85}$Rb~\cite{Courteille1998a,Roberts1998a}, $^{87}$Rb~\cite{Marte2002a} and $^{133}$Cs~\cite{Vuletic1999a}. Recently, optically induced Feshbach resonances have been reported in~\cite{Fatemi2000a, Theis2004a}. 

The availability of an additional knob to tune the scattering length has had important consequences for the development of the field. On the one hand, the fact that interaction parameters no longer have to be taken for granted has enabled new species to be condensed which would otherwise have been very difficult to nearly impossible, such as $^{85}$Rb~\cite{Cornish2000a} or $^{133}$Cs~\cite{Weber2003a}. Tuning of interactions is needed in order to achieve efficient rethermalization in this case.
On the other hand, a large class of experiments exploring degenerate gases had been studying the properties of the system in terms of atom numbers or trap frequencies, whereas in many situations the parameter which is physically interesting is the interaction, and Feshbach resonances allow this knob to be turned just by changing the magnetic field. $^{7}$Li, for example, is a system with a strong attractive background interaction, allowing the production of BECs of only a few thousand atoms before the onset of collapse~\cite{Bradley1995a}. Phenomena like the collapse have first been observed by varying the atom number in the sample~\cite{Gerton2000a}, thereby only indirectly affecting interactions through the density. Feshbach resonances allow these phenomena to be directly observed as a function of interactions~\cite{Roberts2001a}.

Feshbach resonances have allowed to enter the regime of strongly correlated systems and have made a whole class of experiments possible which have no analogon in neighboring fields such as solid state physics. One prominent example is the BCS-BEC crossover in two-component Fermi gases interacting at a Feshbach resonance~\cite{Regal2004a, Bartenstein2004a,Zwierlein2004a,Bourdel2004a,Kinast2004a}. Apart from the BCS-BEC crossover, Feshbach resonances have been the key to success in ground breaking experiments such as solitons~\cite{Strecker2002a, Khaykovich2002a}, collapse~\cite{Roberts2001a}, molecule formation~\cite{Donley2002a,Herbig2003a,Xu2003a,Duerr2004a,Regal2003a,Strecker2003a,Cubizolles2003a,Zwierlein2003a,Jochim2003a,Thalhammer2006a}.

In any case, the impressively large variety of experiments relying on Feshbach resonances have exclusively made use of \textit{homonuclear} Feshbach resonances, that is Feshbach resonances arising between spin states of a \textit{single} atomic species. As of early 2006, the study of \textit{heteronuclear} Feshbach resonances in collisions between certain spin states of two \textit{different} atomic species (either bosonic or fermionic) was still in its infancy. Heteronuclear Feshbach resonances had been identified for the first time in Fermi-Bose mixtures of $^6$Li~-~$^{23}$Na at MIT~\cite{Stan2004a} and in \pot-\rub\ at JILA~\cite{Inouye2004a}. The identification has been reported through increased inelastic collisions and atom loss at the resonance, but no tuning of interactions has been reported.

The control of heteronuclear interactions offers fascinating perspectives in Fermi-Bose mixed systems and Fermi-Fermi systems of different mass: In Fermi-Fermi systems, heteronuclear Feshbach resonances would open the way to the exploration of the BCS-BEC crossover in a two-species fermion mixture with unequal mass~\cite{Iskin2006a,Wu2006a}. In Fermi-Bose systems, a different and complementary approach to fermionic superfluidity has been suggested in which the interaction between fermionic atoms is provided by bosonic atoms taking over the role of phonons in the solid state superconductor \cite{Bijlsma2000a, Viverit2002a, Efremov2002a, Albus2004a}. When loaded into three-dimensional optical lattices, a wealth of different phases with no equivalent in condensed matter systems have been predicted to exist~\cite{Albus2004a, Lewenstein2004a, Cramer2004a, Lewenstein2006a}. In addition,  heteronuclear Feshbach molecules~\cite{Ospelkaus2006d,Papp2006a,Zirbel2008b} which have been produced for the first time within this work (see text) could be used as a starting point for the creation of polar molecules with novel anisotropic interactions.

Here we demonstrate the tunability of heteronuclear interactions in the vicinity of a Feshbach resonance arising between \pot\ and \rub~\cite{Ospelkaus2006c} (see also~\cite{Zaccanti2006a}). To provide evidence for tuning of interactions and thus a  variation of the scattering length around these heteronuclear resonances, the interaction energy of the Bose-Einstein condensate coexisting with a quantum degenerate Fermi gas is measured in the vicinity of an $s$-wave heteronuclear Feshbach resonance. Additional and complementary information is gained from a measurement of interaction effects of the Fermi gas on an expanding Bose-Einstein condensate and vice versa. As a first application of heteronuclear Feshbach resonances, the phase diagram for harmonically trapped Fermi-Bose mixtures ranging from collapse for large attractive interactions to phase separation in the case of strong repulsive interactions~\cite{Molmer1998a} has been studied. Tuning of interactions has been the prerequisite for the first formation of ultracold heteronuclear molecules as reported later in this article.

\subsection*{Tuning of heteronuclear interactions in stable mixtures}

Evidence for tuning of interactions in the heteronuclear \pot--\rub\ mixture is provided in  two different and complementary ways:
\begin{itemize}
\item[1.] In a first experiment, we  study tuning of interactions by a measurement of the mean-field energy of the Bose-Einstein condensate as a function of  magnetic field in the vicinity of a Feshbach resonance.  Due to the heteronuclear interaction, the Bose-Einstein condensate is confined in the combined potential of the external dipole trap and the heteronuclear mean-field potential (see our introductory discussion). The latter becomes stronger as the heteronuclear interaction increases. Hence, the effective confinement of the Bose-Einstein condensate which determines its mean-field energy is changed. A measurement of this interaction-dependent mean-field energy is performed by a sudden switch off of all confining potentials including the additional mean-field potential and an observation of the time-of-flight expansion of the condensate. A sudden switch off of the mean-field potential is realized in a good approximation by suddenly switching off the Feshbach field, reducing the heteronuclear scattering length to its background value. A related study has been done in the very first demonstration of tuning of homonuclear interactions in a  Bose-Einstein condensate of $^{23}$Na~\cite{Inouye1998a} in the vicinity of a Feshbach resonance.
\item[2.] In a second experiment, we study the  influence of the heteronuclear interaction on the time-of-flight expansion of the Bose-Einstein condensate and the Fermi gas. When the heteronuclear interaction is left on during time-of-flight, the expansion of the two clouds is either slowed down due to attractive interaction or influenced by repulsive interactions.  The study is performed by a sudden switch off of the external dipole trapping potential while the Feshbach field is left on during time of flight.
\end{itemize}
For our studies, we have used the broadest of the available $s$-wave resonances which occurs between the \pot\ $\left|9/2,-9/2\right>$ state and the \rub\ $\left|1,1\right>$ state. The resonance position is at $B_0$=546.8(1)~G as determined by the transition between strong attractive and repulsive interactions (see below). At the resonance, the scattering length varies dispersively as a function of the external magnetic field~\cite{Moerdijk1995a}
\begin{equation}
\label{eq:scattering_length_at_Feshbach}
a_{\mathrm{FB}}=a_{\mathrm{bg}}\cdot\left(1-\frac{\Delta B}{B-B_0}\right).
\end{equation}
In equation~(\ref{eq:scattering_length_at_Feshbach}), $B_0$ is the resonance center, $a_{\mathrm{bg}}=-185\,a_0$~\cite{Zaccanti2006a} is the background scattering length between bosons and fermions in the respective spin state and $\Delta B$ is the width of the resonance which in our case has been predicted to be -2.9\,G~\cite{Ferlaino2006a}. Note that the width of the Feshbach resonance can be negative. In our case, this means that the scattering length will diverge to $a \rightarrow +\infty$ below the resonance $(B<B_0)$ and $a\rightarrow -\infty$ above the resonance $(B>B_0)$.  A sketch of the expected variation of the scattering length as a function of magnetic field for the \pot--\rub\ $s$-wave resonance at $B=546.8(1)$\,G is given in fig.~\ref{fig:feshbach_resonance}.

\paragraph{Experimental Procedure} For the preparation of ``Feshbach-resonant'' mixtures, we start with a magnetically trapped thermal ensemble of \pot\ and \rub\ in the low-field seeking but non-resonant $\left|9/2,9/2\right>$ and $\left|2,2\right>$ states.  Slightly before reaching the critical temperature for condensation  $T_c$ of the bosonic \rub\ ensemble, we ramp up the optical dipole potential of a crossed dipole trap within 500\,ms. This novel ``magic''  trap (see~\ref{sec:magic_trap}) is designed to eliminate the differential gravitational sag between \rub\ and \pot\ by operating at the magic wavelength of 807.9\,nm. In the combined optical and magnetic potential, a last step of rf-induced evaporation is performed within 1\,s and a quantum degenerate mixture is prepared. While ramping down the magnetic trapping potential, care has to be taken not to induce spin flips in either \pot\ or \rub\ affecting the polarization of the atomic ensemble. To this end, an offset field of a few ten Gauss is applied throughout the ramp-down process. Residual heating of the cloud during the ramp down is subsequently compensated for by a last step of purely optical evaporation in the crossed dipole trap. We typically end up with a quantum degenerate mixture of $5\cdot10^4$ $^{40}$K and up to $2\cdot10^5$ $^{87}$Rb atoms and no discernible bosonic thermal fraction.

In a final step, the \pot\ $\left|9/2,9/2\right>$ and \rub\ $\left|2,2\right>$ mixture is converted into a $\left|9/2,-7/2\right>$ and $\left|1,1\right>$ mixture using rf and microwave Landau-Zener transfer schemes (see~\ref{sec:rf_setup}). The final transfer to the resonant $\left|9/2,-9/2\right>$ state in \pot\ is performed at two fixed magnetic field values by an rf Landau-Zener transfer after the field has settled. For studies above the resonance $(B>B_0)$, where the heteronuclear interaction is expected to be attractive ($a_{\mathrm{FB}}<0$), the final transfer is performed at a magnetic field of 550.5\,G. For  studies below the resonance center $((B_0-\Delta B)<B<B_0)$, where the heteronuclear interaction is expected to be repulsive ($a_{\mathrm{FB}}>0$), the transfer is done at $B=543.8$\,G. At these magnetic field values, the heteronuclear scattering length is  still moderately attractive in both cases and not too different from the background scattering length between \pot\ and \rub\ in the  $\left|9/2,-7/2\right>$ and $\left|1,1\right>$ mixture. This is important in order  not to induce oscillations in both clouds due to the sudden change of the scattering length. After the final transfer at these two magnetic field values, the field is ramped to varying values in the immediate vicinity of the resonance either from above or below. The ramp is performed within 50\,ms, thereby changing $a_{\mathrm{FB}}$ adiabatically  with respect to any other time scale in the experiment: $\dot a_{FB}/ a_{FB} \ll \omega_i$ so that the clouds remain in equilibrium within the field ramp\footnote{In case of a thermal ensemble, one would  also have to require $\dot a_{FB} / a_{FB} \ll \gamma_{el}$, where $\gamma_{el}$ is the elastic collision rate. In our experiment, we obtain $\omega_{i}<\gamma_{el}$ and $\dot a_{FB} / a_{FB} \ll \omega_i$ is the stronger requirement. Note that in the immediate vicinity of the resonance, the adiabaticity condition may break down as $\dot a/ a$ becomes comparable to $\omega_i$}~\cite{Inouye1998a, Kagan1997a}. We then probe the expansion of the cloud at different values of the heteronuclear scattering length and under different conditions.

\subsubsection*{Evidence for tuning I: Mean-field  energy of the BEC}
While the kinetic energy of a Bose-Einstein condensate in the trap is negligible (Thomas-Fermi limit), the energy of the BEC is dominated by its interaction energy given by
\begin{equation}
\label{eq:interaction_energy}
E_I/N=\frac{2\pi \hbar^2}{m}a_{\mathrm{BB}}\left< n_\mathrm{B}\right>.
\end{equation}
In this equation, $a_{\mathrm{BB}}$ denotes the Bose-Bose scattering length, $\left< n_\mathrm{B}\right>$ is the average density of the condensate, $N$ the number of atoms of mass $m$.  This mean-field energy of the condensate can be measured by time-of-flight expansion. During time-of-flight, the stored interaction energy is converted into the kinetic energy of a freely expanding condensate.

At first glance, the interaction energy of the condensate, proportional to the mean density and the bosonic scattering length, seems to be independent of the value of the heteronuclear scattering length. However, the effect of heteronuclear interaction enters through the density of the Bose-Einstein condensate which is affected by the presence of the fermionic cloud. The trapped fermions, along with the heteronuclear interaction, act as an additional heteronuclear mean-field potential $U_{F\rightarrow B}(\vec r)$ (equation~(\ref{eq:hetnucpotFonB})) forming an effective potential  composed of the dipole trapping potential $V_B$ and the mean-field potential $U_{F\rightarrow B}(\vec r)$ and determines the density distribution of the bosonic cloud $n_B(\vec r)$ via the Thomas Fermi equation~(see equation~(\ref{eq:thomas_fermi_coupled})).

In the case of attractive heteronuclear interactions ($a_{FB}<0$), the density distribution of the fermionic cloud interacting with the BEC is increased compared to the non-interacting case. The effective confinement of the BEC is thus increased with increasing attractive heteronuclear interaction in the vicinity of the resonance, enhancing the density and the mean-field energy according to equation~(\ref{eq:interaction_energy}). Along with the intrinsic repulsion of the Bose-Einstein condensate, characterized by the positive bosonic scattering length $a_{\mathrm{BB}}=100.4(1)a_0$~\cite{Kempen2002a}, the effect is observable through the increasing kinetic energy of the bosonic cloud after time-of-flight.

In the case of repulsive interactions, the condensate acts as a repulsive bump in the center of the fermionic cloud, pushing the fermionic cloud out of the center of the trap. The fermionic density overlapping with the center of the BEC will be reduced, whereas the fermionic density in the border areas of the potential increases. In this case, the additional fermionic curvature acting on the condensate will push the bosons towards the center of the trap, thereby again increasing density and mean field energy of the condensate. However, the effect is smaller than the enhancement of density in the attractive case.

\subsubsection*{Some numerical results}

\begin{figure}
  \includegraphics[width=1.0\columnwidth]{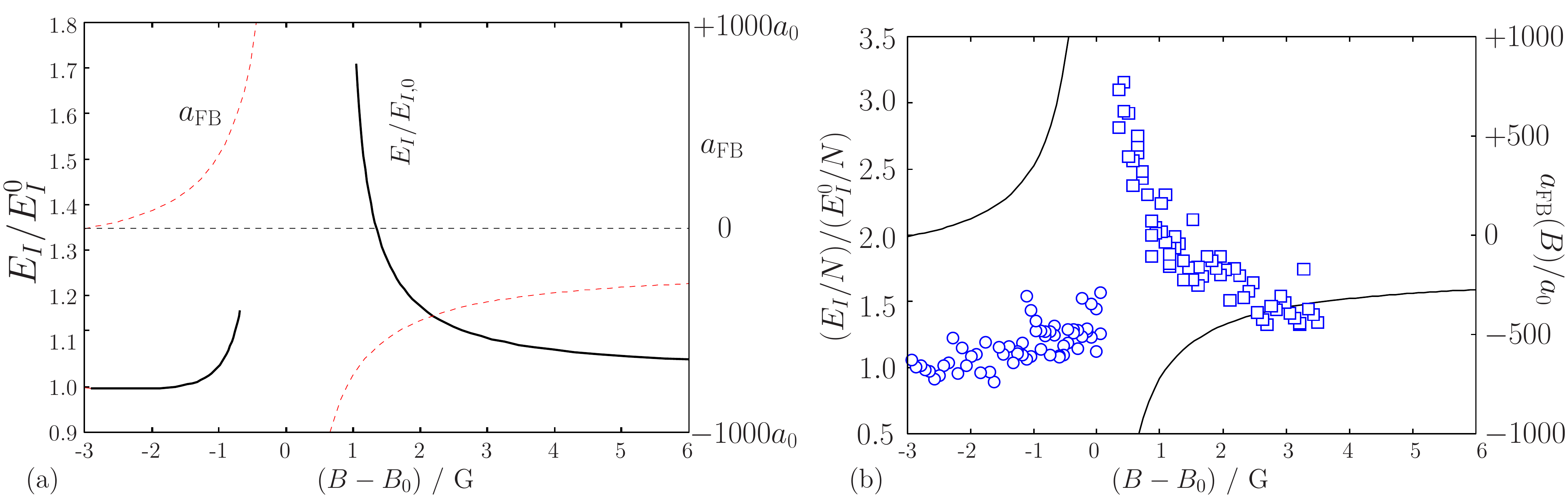}
  \caption{(a) Theoretically predicted variation of the mean-field energy of the condensate as a function of $B-B_0$ in the vicinity of the $546.8(1)$\,G Feshbach resonance. The mean-field energy is normalized to the interaction energy of a pure bosonic condensate $E_I^0$. (b) Interaction energy of a BEC per particle in the vicinity of a heteronuclear resonance normalized to the interaction energy of a pure condensate. On both sides of the resonance, the interaction energy increases although the effect is much more pronounced on the attractive side of the resonance where both the fermionic and the bosonic density distribution is enhanced. On the repulsive side of the resonance, the interaction energy increases only slightly. In the same diagram, we have plotted the expected variation of the heteronuclear scattering length across the resonance according to equation (\ref{eq:scattering_length_at_Feshbach}).\label{fig:mf_energy}}
\end{figure}

The above  intuitive picture can be supported by numerical simulations. The simulations rely on the mean-field model summarized at the beginning of this article (see~\cite{Molmer1998a}) and provide a self-consistent calculation of the densities of the harmonically trapped fermionic and bosonic cloud, considering the additional heteronuclear mean-field potentials $U_{F\rightarrow B}$ and $U_{B\rightarrow F}$. Figure~\ref{fig:mf_energy}(a) shows the theoretically expected variation of the mean-field energy $E_I$ of a condensate of $10^5$ bosons in the vicinity of the Feshbach resonance at 546.8(1)\,G, interacting with  a fermionic cloud of $5\cdot10^4$ atoms via a varying heteronuclear scattering length $a_{\mathrm{FB}}(B)$. In the figure, we have plotted the interaction energy of the condensate as a function of $B-B_0$. The mean field energy has been calculated according to equation~(\ref{eq:interaction_energy}) and is normalized to the mean field energy $E_I^0$ of a pure condensate, i.\ e.\ equation~(\ref{eq:interaction_energy}), with $n_\mathrm{B}$ given by the Thomas-Fermi profile of a pure condensate. The behavior of the scattering length in the vicinity of the resonance is determined by equation~(\ref{eq:scattering_length_at_Feshbach}) with an assumed  width of the resonance of $\Delta B=-2.9$\,G. We have plotted the interaction energy only for stable interacting mixtures. In the same diagram, a sketch of the expected heteronuclear scattering length is plotted.

On the repulsive side of the resonance (low-field side), the interaction energy of the condensate increases by 20\% compared to the non-interacting case. At the same time, the heteronuclear scattering length varies between 0 at $\approx-3$\,G and $\approx400\,a_0$ before the onset of phase separation. On the high field side of the resonance, where heteronuclear interactions are attractive, the effect of heteronuclear interactions is much more pronounced: the mean-field energy of the condensate increases by almost a factor of two while the scattering length is increased by approximately the same factor. The effect of heteronuclear interactions on the mean-field energy of the condensate is thus strongest on the attractive side of the resonance whereas only a slight effect is observed on the repulsive side.

\subsubsection*{Experimental data}
In the above discussion, we have seen that the mean field energy of a Bose-Einstein condensate in a Fermi-Bose mixture does depend on the value of the heteronuclear scattering length, although the dependence can only be calculated numerically. The interaction energy can be measured in a time-of-flight expansion experiment where all confining potentials are suddenly switched off, thus converting the stored mean-field energy into kinetic energy. The kinetic energy per particle is just given by $E_K/N=\frac{1}{2}mv_{\mathrm{rms}}^2$ where $v_{\mathrm{rms}}$ is the mean-root-squared velocity. Both the atom number $N$ and the velocity $v_{\mathrm{rms}}$ can be extracted from time-of flight images, more precisely $v_{\mathrm{rms}}$ from the size of the cloud after time of flight.

Figure~\ref{fig:mf_energy}(b) shows a measure of the condensates mean-field energy per particle as a function of the magnetic field in the vicinity of the 546.8(1)\,G heteronuclear Feshbach resonance. The experimental procedure for the preparation of resonant mixtures in the vicinity of the resonance has been described above. After the final magnetic field ramp  from either  above or below the resonance center to final field values near the resonance and a 10\,ms hold time, both the Feshbach field and the external optical potential are suddenly switched off and \pot\ and \rub\ are detected after time-of-flight.

On the low-field side of the resonance, the interaction energy increases only slightly with increasing repulsive heteronuclear interaction. Within the investigated magnetic field range below the resonance, the scattering length is expected to vary from 0 to $+1000\,a_0$. In the same magnetic field interval, the interaction energy is observed to increase only by 25\%. A comparable effect has been predicted by the numerical simulations discussed above. On the high-field side of the resonance, the effect of heteronuclear interactions is much more pronounced. The heteronuclear scattering length is expected to vary between $-300\,a_0$ and $-1000\,a_0$ within the investigated magnetic field range. As can be seen from the data, the  mean-field energy per particle stored in the condensate increases in the same interval by a factor of 3. This is in qualitative agreement with theory which predicts a pronounced increase in interaction energy of the condensate by a factor of 2. The remaining quantitative disagreement between theory and experiment might be due to a finite magnetic field settling time after the adiabatic ramps.

For magnetic field values above $B>551$\,G, the heteronuclear scattering length is expected to recover slowly to its background value of $-185\,a_0$. Consequently, the interaction energy per particle is expected to saturate somewhat above the interaction energy of a pure condensate. On the low magnetic field side, however, the scattering length changes sign from repulsive interaction to attractive at $B\approx543.9$\,G. Below this zero-crossing, the scattering length will become negative and eventually recover to the negative background interaction. Hence, the interaction energy per particle is expected to increase again below the zero-crossing converging towards the same interaction energy as far above the resonance.

\subsubsection*{Evidence for tuning II: Interaction effects during expansion}
In a second experiment, we have studied  the influence of the heteronuclear interaction on the time-of-flight expansion of the Bose-Einstein condensate and the Fermi gas. When the heteronuclear interaction is left on during time-of-flight, the expansion of the two clouds is either slowed down due to attractive interactions or influenced by repulsive interactions. The external dipole potential is suddenly switched off, whereas the Feshbach field is left on during time-of-flight expansion.

The effect of attractive interactions on the time-of-flight expansion of the bosonic and fermionic cloud has already been studied in~\cite{Ferlaino2004a} in a Fermi-Bose mixture of \pot\ and \rub. However, in these experiments, the heteronuclear interaction was determined by the background interaction strength and could not be tuned. The observations in~\cite{Ferlaino2004a} are therefore restricted to the observation of an affected time-of-flight expansion of the bosonic and fermionic cloud, respectively, in comparison to the expansion of a pure bosonic or fermionic cloud.

\begin{figure}[tbp]
  \centering
  \includegraphics[width=0.6\columnwidth]{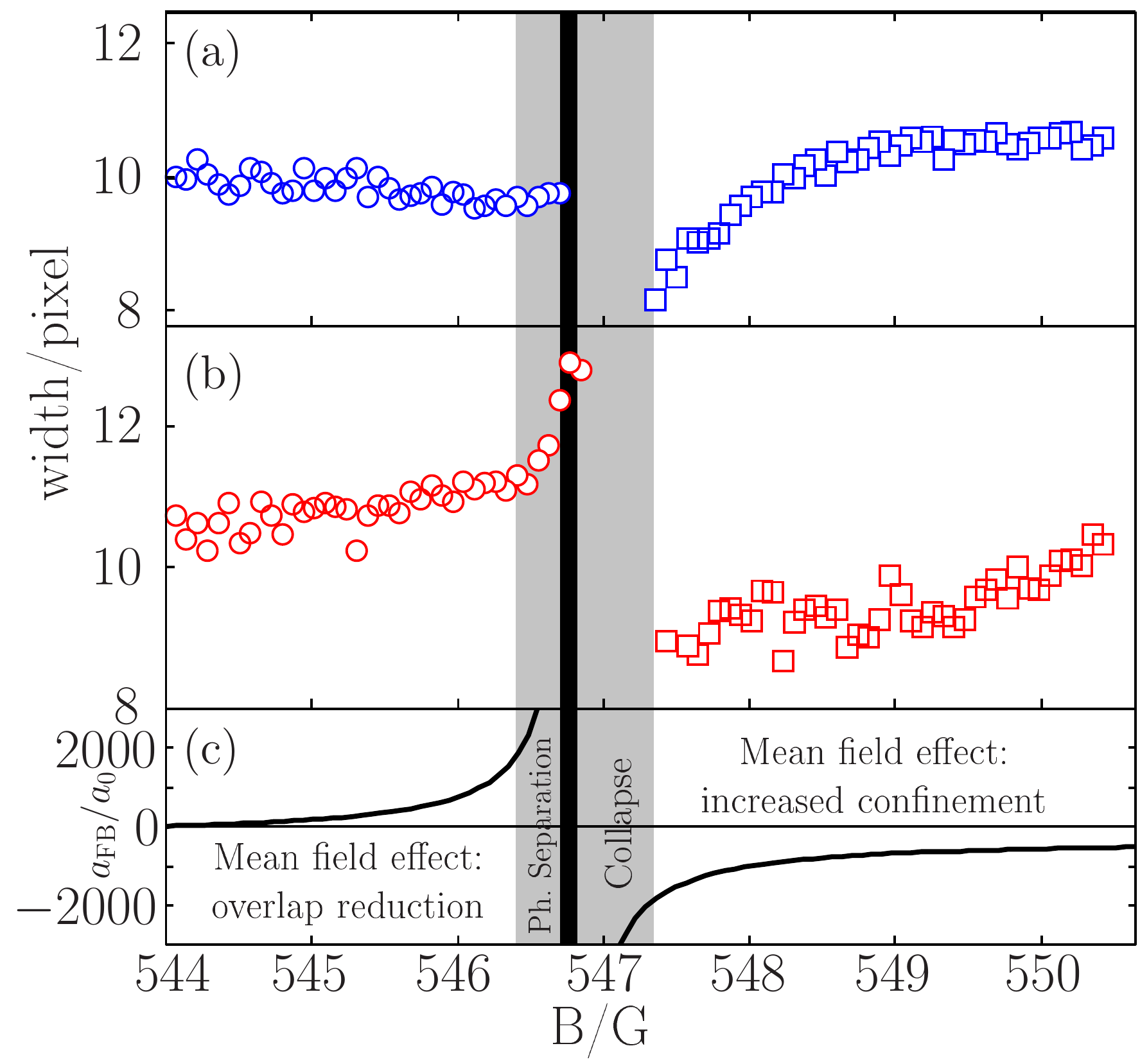}
  \caption{Observed width of the components after time-of-flight expansion as a function of magnetic field when the resonant interaction is left on during expansion~\cite{Ospelkaus2006c}. (a) Width of the bosonic cloud. (b) Width of the fermionic cloud. (c) Heteronuclear scattering length. The region shaded in grey indicates the magnetic field regions where instability with respect to collapse and phase separation have been observed (see below). The black vertical line marks the observed transition from attractive to repulsive interactions (resonance center). (1 pixel $\hat=$ 3.2\,$\mu$m; time of flight 25.2\,ms for \rub\ and 7.5\,ms for \pot). {\it Copyright (2006) by the American Physical Society.}}
  \label{fig:expansion_with_field}
\end{figure}

Figure~\ref{fig:expansion_with_field} shows the width of the bosonic cloud (a) and fermionic cloud (b) after 7.5\,ms and 25.2\,ms time-of-flight, respectively. To account for enhanced particle loss in the vicinity of the resonance, the observed width data should be normalized by $N^{1/5}$ and $N^{1/6}$ for the bosonic and fermionic cloud, respectively.\footnote{The equilibrium width of a non-interacting Fermi gas grows with $N^{1/6}$ whereas the width of a Bose-Einstein condensate increases with $N^{1/5}$} However, the observed particle number loss is on the order of 20\% in the studied magnetic field range and the correction due to particle number variations is therefore below 5\%. On the repulsive side of the resonance, where the effect of interaction on the time-of-flight expansion is very weak compared to the attractive case,  particle loss is even smaller. In the same diagram, we have plotted the expected variation of the heteronuclear scattering length across the resonance.

On the high-field side of the resonance, where heteronuclear interactions are attractive, we observe the width of both the fermionic and the bosonic cloud to decrease when approaching the resonance. This is a clear signature for increasing heteronuclear attraction. The simultaneous expansion of both clouds is slowed down due to the attractive potential that is present throughout the expansion process. In principle, the effect of the heteronuclear attractive interaction on the expansion of the BEC and the Fermi gas is twofold. On the one hand, the densities of both components are significantly increased inside the trap. This enhancement of density which corresponds to a tighter confinement of either of the clouds, will lead to a faster expansion (as discussed for the bosonic cloud above). On the other hand, the condensate and degenerate Fermi gas will interact during expansion and most importantly in the early stages of  time of flight, thereby reducing significantly the expansion rate. This slow-down is determined by the relative dynamics between the condensate and the degenerate Fermi gas and dominates the experiments.

On the repulsive side of the resonance, however, the effects of heteronuclear interaction are less pronounced. The width of the condensate after time-of-flight stays almost constant, whereas the width of the Fermi gas after time of-flight increases slightly. In the immediate vicinity of the resonance, this increase exhibits a change of slope. When the fermionic density at the trap center vanishes at even higher repulsion, the potential felt by the Bose cloud will rather be that of the pure external trapping potential with quite a sharp transition to a very steep higher order potential created by the fermionic density in the outer shell, at the edges of the condensate. We identify this region with the regime from 546.4\,G to the center of the resonance at 546.8\,G where, as seen in fig.~\ref{fig:mf_energy}(b), the width of the condensate saturates. This may indicate that at complete phase separation, the repulsive interaction leads to a rapid expansion of the Fermi gas suddenly accelerated outside when the external potential is switched off and the repulsive bump of the BEC in the center maintained.

\subsection*{Observing collapse and phase separation}

As we have seen so far, interactions between fermions and bosons can fundamentally affect the properties of the system such as the density of the components and expansion properties. However, when increasing the scattering length above a certain critical interaction strength, two instabilities can occur: collapse for the case of attractive interactions and phase separation for the case of repulsive interactions. In our experiments, we have  been able to access the complete phase diagram of the mixture and observe clear signatures of both collapse and phase separation. The phase diagram has been accessed through tuning of interactions in the vicinity of the Feshbach resonance. Although the mean-field collapse of a harmonically trapped Fermi-Bose mixture has been studied before, in those experiments critical conditions have been achieved by increasing particle numbers in the system, holding the value of the heteronuclear interaction fixed (see~\cite{Modugno2002a,Ospelkaus2006b,OspelkausC2006a} and the discussion above in this article), whereas in the experiments reported here, the collapse is induced by tuning of the heteronuclear scattering length.

\begin{figure}[tbp]
  \centering
  \includegraphics[width=0.8\columnwidth]{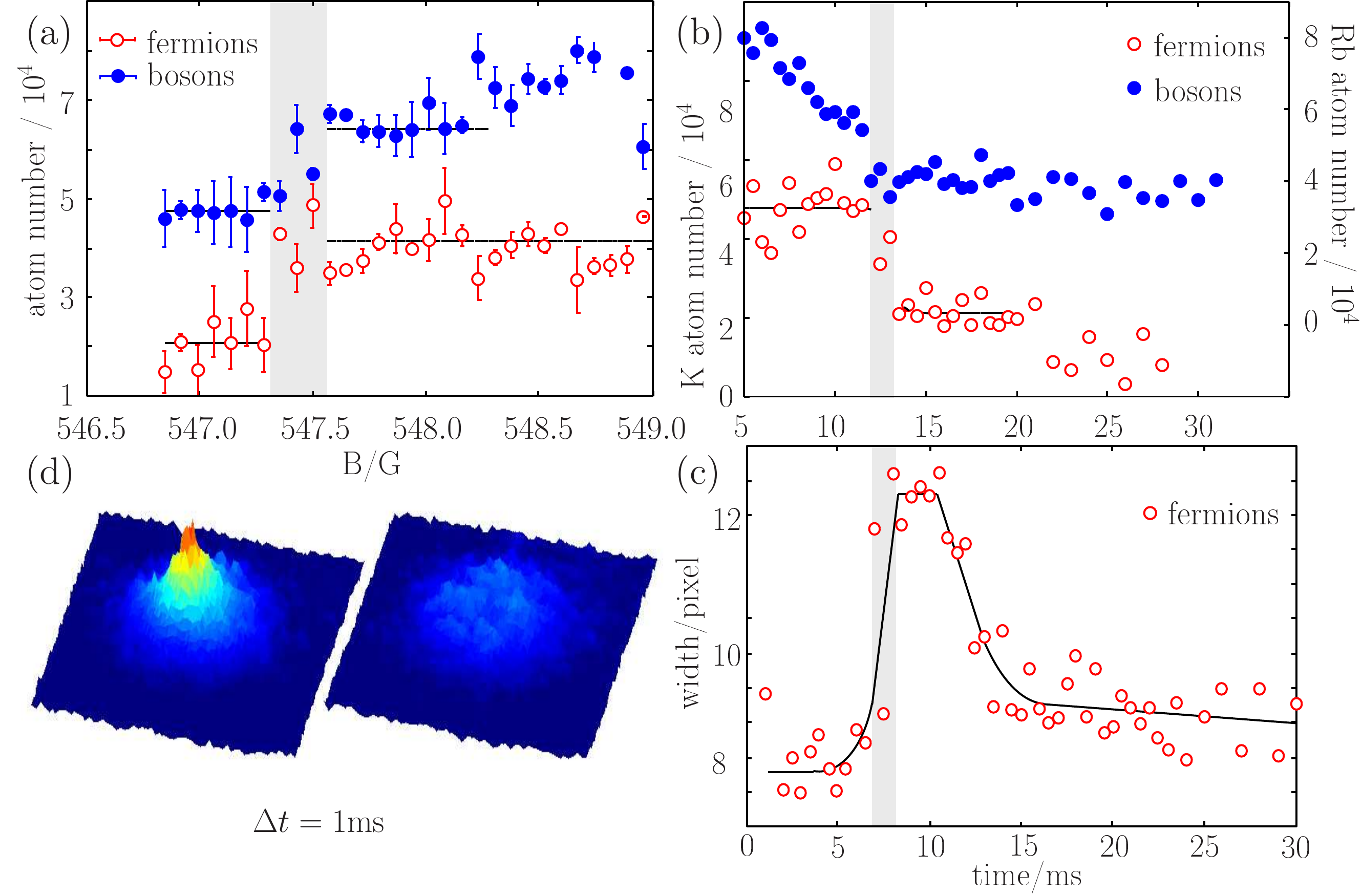}
  \caption{Induced mean-field collapse of the mixture~\cite{Ospelkaus2006c}.
           (a) Sudden drop of atom numbers for critical heteronuclear interactions.
           (b) The onset of collapse is retarded by a timescale given by the trap frequency.
           (c) Sample time of flight images showing the bimodal distribution in the fermionic component and the sudden collapse of the system in the overlap region with the BEC.
           (d) As the collapse happens, the sample is excited and heated, visible in the width of the fragments. After the collapse, the mixture finds a new equilibrium.
           {\it Copyright (2006) by the American Physical Society.}
           \label{fig:induced_collapse}
          }
\end{figure}

\subsubsection*{Collapse}

For attractive interactions above a certain critical value, the Fermi-Bose mixture is expected to become unstable with respect to collapse. In the vicinity of the resonance occurring at 546.8\,G, attractive interactions are strongest immediately above the resonance center. In fig.~\ref{fig:expansion_with_field}, the corresponding region  has been shaded in grey. This region immediately above the resonance center has been studied in detail in fig.~\ref{fig:induced_collapse}. Figure~\ref{fig:induced_collapse}(a) shows the atom numbers of both the fermionic and the bosonic cloud as a function of magnetic field when adiabatically approaching the resonance from above. In the vicinity of 547.4\,G (at a detuning of about 0.6\,G above the resonance), we observe a sudden drop in both the fermionic and the bosonic atom number. We loose almost 50\% of the fermionic atoms and a significant fraction of the bosonic cloud. In contrast to previous work, reporting the onset of instability as a function of atom number (see preceding discussion and~\cite{Modugno2002a}), the collapse is now due to tuning of interactions above a certain critical interaction strength in an otherwise undercritical mixture. The precise magnetic field value at which the collapse happens is of course dependent on the initially prepared atom number. In the experiments, we observe the onset of the collapse at a magnetic field value corresponding to a theoretically predicted scattering length of $-1000\,a_0$.

To gain some information about the time scale on which the collapse happens, we have repeated the experiment in a slightly varied version: In this case, we have stopped the magnetic field ramp at the observed critical magnetic field value and then varied the hold time of the mixture, thereby observing the collapse happen as a function of time. This is illustrated in fig.~\ref{fig:induced_collapse}(b). We cannot fully exclude some residual variation of the scattering length due to settling of the magnetic field after the nominal stop of the ramp. The onset of the collapse, again visible as a sudden drop in atom number, is retarded by a timescale given roughly by the trap frequency and happens on a timescale $<$1~ms. The dynamics associated with the collapse can be seen in fig.~\ref{fig:induced_collapse}(c) and (d). Figure~\ref{fig:induced_collapse}(d) shows two time of flight images of the fermionic cloud prior to the collapse and after the collapse has happened (when leaving the Feshbach field on during expansion). Prior to the collapse, the density of the fermionic cloud is strongly enhanced. The fermionic density profile after time of flight shows a bimodal distribution with a broad and dilute pedestal and a very dense center. The latter part of the fermionic cloud is trapped inside the mean-field potential of the BEC. When critical conditions for collapse are achieved, this central core suddenly disappears. During the collapse, the overlap region of the fermionic cloud with the BEC is destroyed by a three-body implosion, leaving behind only the broad dilute pedestal. The three-body implosion causes significant heating and excitation in the remaining sample, reflected in the evolution of the width of the fermionic cloud as the collapse happens --- see fig.~\ref{fig:induced_collapse}(d). After the collapse, the mixture assumes a new equilibrium through evaporation and particle loss which is reflected in the decreasing width of the fermionic cloud after the collapse has happened.

\subsubsection*{Phase separation}

\begin{figure}[tbp]
  \centering
  \includegraphics[width=0.7\columnwidth]{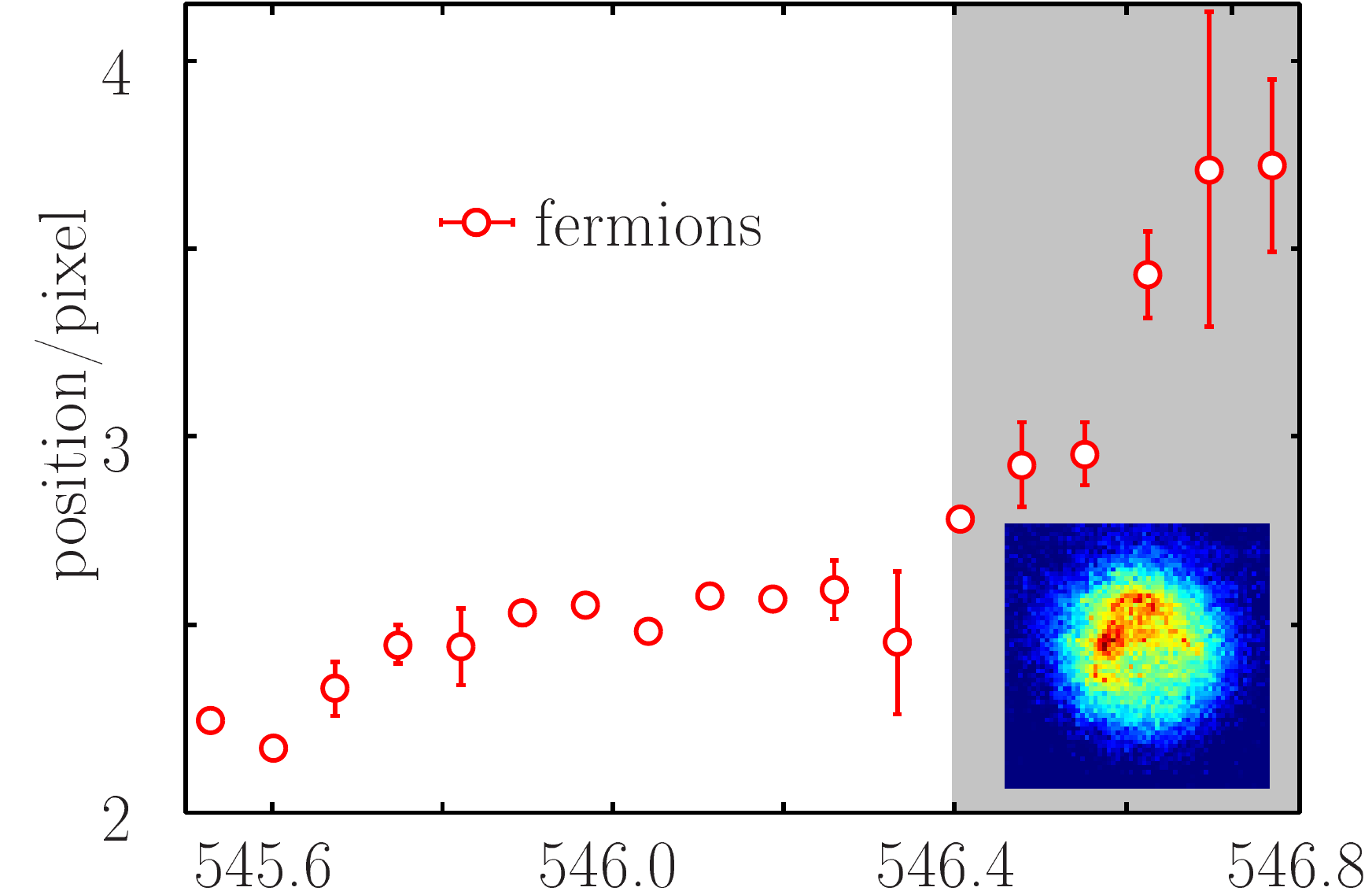}
  \caption{Vertical position of the Fermi gas as a function of magnetic field~\cite{Ospelkaus2006c}. The grey shaded area indicates the assumed region of full phase separation, where the fermionic density is expected to vanish at the BEC core. Due to gravitational symmetry breaking, the fermions are pushed above the BEC, an effect amplified by time of flight expansion. The inset shows the corresponding fermionic density distribution where most of the density is concentrated in the upper part of the image. {\it Copyright (2006) by the American Physical Society.}}
  \label{fig:phase_separation}
\end{figure}

Phase separation due to repulsive interactions in a composite system of harmonically trapped fermions and bosons has been discussed in theory \cite{Molmer1998a,Roth2002a}, but never been explored in experiment. Tuning of heteronuclear interactions has enabled us to enter the regime of repulsive heteronuclear interactions, where phase separation occurs. In the limit of vanishing differential shift due to gravity and for our experimental parameters, phase separation will occur as a shell of Fermions surrounding a dense BEC core (as we have seen in the introductory discussion).

Inside a harmonic trap and in the presence of gravity, atoms experience a gravitational sag given by $-g/\omega^2$. For systems with different masses, such as the \pot--\rub\ system, this will in general lead to a differential gravitational sag between the components, as the trap frequencies may be different. Although our optical dipole trap has been designed to eliminate the differential gravitational sag, a slightly different gravitational sag has still been present in these experiments and breaks the symmetry of the system which therefore favors phase separation to occur in the vertical direction. As a consequence, the position of the fermionic component in the time of flight image is shifted upwards as a function of detuning from resonance, with an even stronger slope in the region of complete phase separation (see fig.~\ref{fig:phase_separation}). An important aspect is that the shift in position between fermions and bosons in the trap is amplified by the repulsive interaction during expansion if we leave on the interaction. Thus, the small initial symmetry-breaking in the direction of gravity is strongly enhanced and clearly visible in absorption images such as in the inset of fig.~\ref{fig:phase_separation}, where the Fermionic density is concentrated in the upper part of the image.

\section{Fermi-Bose mixtures in optical lattices}

Quantum degenerate gases in optical lattices provide unique opportunities for engineering of many-body Hamiltonians discussed in the context of condensed matter systems and even beyond in a very pure and controllable environment: The periodic potentials realized by optical lattices are almost defect-free. By tuning the depth of the lattice, the strength
of the crystal can be controlled, thereby controlling e. g. tunnelling with respect to onsite interaction. Playing with lattices of different geometries and dimensionalities, one, two- and three-dimensional geometries can be realized and continuously transformed from one into another. In addition, the use of Feshbach resonances allows the precise control
of interactions between the atoms even allowing for continuous tuning from repulsive interactions to attractive interactions and vice versa~\cite{Tiesinga1992a,Tiesinga1993a,Moerdijk1995a,Boesten1996a,Courteille1998a,Inouye1998a}. The possibility of realizing strongly correlated systems with ultracold atoms loaded into 3D optical lattices has been pointed out in a seminal paper by D. Jaksch and coworkers in 1998~\cite{Jaksch1998a} who discussed the realization of the Bose Hubbard Hamiltonian~\cite{Fisher1989a} with ultracold bosons loaded into 3D optical lattices. The experimental observation of the quantum phase transition from a superfluid to a Mott-Insulator state with repulsively interacting bosons followed in a pioneering experiment by M. Greiner and coworkers in 2002~\cite{Greiner2002a}. Following experiments with either bosonic or fermionic atoms in optical lattices demonstrated the enormous degree of control inherent in these intriguing physical systems (see review~\cite{Lewenstein2006a} and references therein).

In the introduction, we have mentioned the bright possibilities for heteronuclear quantum gases in 3D optical lattices as a novel quantum many-body system. In a heteronuclear Fermi-Bose mixture, the different quantum statistical behavior of the two components are predicted to give rise to fundamentally novel quantum many-body phases (see introduction). Here, we report on the first realization of a Fermi-Bose quantum many-body system confined in a 3D optical lattice~\cite{Ospelkaus2006e}. In these studies, we investigate the coherence properties of bosonic atoms when interacting with a varying fraction of fermionic impurities. The interaction between fermions and bosons is held fixed at the attractive background interaction strength. When ramping up the optical lattice, we observe the phase coherence properties of the bosonic cloud to be strongly affected  by a very small fraction ($N_F/N_B\approx 3\%$) of fermionic atoms and find a fermion concentration dependent shift of the coherence properties with respect to the properties of a pure bosonic ensemble. A very small admixture of fermionic impurities induces a significant loss of coherence at much lower lattice depths as compared to the pure bosonic case. While  the coherence properties of the pure bosonic system (loss of coherence with increasing lattice depth) can be explained in terms of the superfluid to Mott-insulator transition and the associated many-body wavefunctions in the superfluid and the Mott-insulating regime, the nature of the observed shift of the coherence properties towards lower optical lattice depth in the mixture is still a point of intense discussions in the community.  Possible scenarios include thermodynamic effects like adiabatic heating when ramping up the optical lattice, disorder-induced localization scenarios or a shift of the quantum critical point of  the bosonic superfluid to Mott-insulator transition due to attractive interactions with the fermionic atoms ($a_\mathrm{FB}=-215(10)\,a_0$~\cite{Ferlaino2006b}). In parallel to this work, similar studies and results have been reported at ETH Z\"urich~\cite{Gunter2006a}.

\subsection{Influence of fermions on bosonic coherence}
The coherence properties of a bosonic cloud  can be revealed by a study of the bosonic interference pattern after a sudden switch off of the lattice and time-of-flight expansion. Whereas the observation of sharp interference peaks is a striking signature of long-range phase coherence of the bosonic cloud, decreasing interference peaks and the appearance of an incoherent background indicates the reduction of the coherence length. In a pure bosonic component, the loss of phase coherence with increasing lattice depth accompanies the transition from a superfluid to a Mott-insulating state.

These experiments have been performed using a 3D optical lattice setup based on an Yb:YAG disc laser operating at a wavelength of 1030\,nm. The lattice setup consists of three orthogonal, retro-reflected laser beams with orthogonal polarizations and a finite detuning of several MHz with respect to each other to minimize crosstalk between the different axes. Details of the lattice setup, including alignment, focussing, frequency and intensity stabilization can be found in~\ref{sec:lattice_setup} and~\cite{WilleThesis,SuccoThesis,OspelkausC2006a,SilkeThesis}.

In a first experiment, we have qualitatively studied the influence of a small admixture of fermionic atoms on the bosonic cloud. In these experiments, we have prepared $2\cdot10^4$ fermionic \pot\ atoms, coexisting with a pure Bose-Einstein condensate of $10^5$ \rub\ atoms  with no discernible thermal cloud in a combined magnetic and optical potential with trapping frequencies of $ \omega=2\pi\cdot(50, 150 , 150)\,$Hz. Finally, the lattice is ramped up to various lattice depths between $2.5\,E_r^\mathrm{Rb}$ and $25\,E_r^\mathrm{Rb}$.

\begin{figure}[tbp]
  \centering
  \includegraphics[width=0.9\columnwidth,clip]{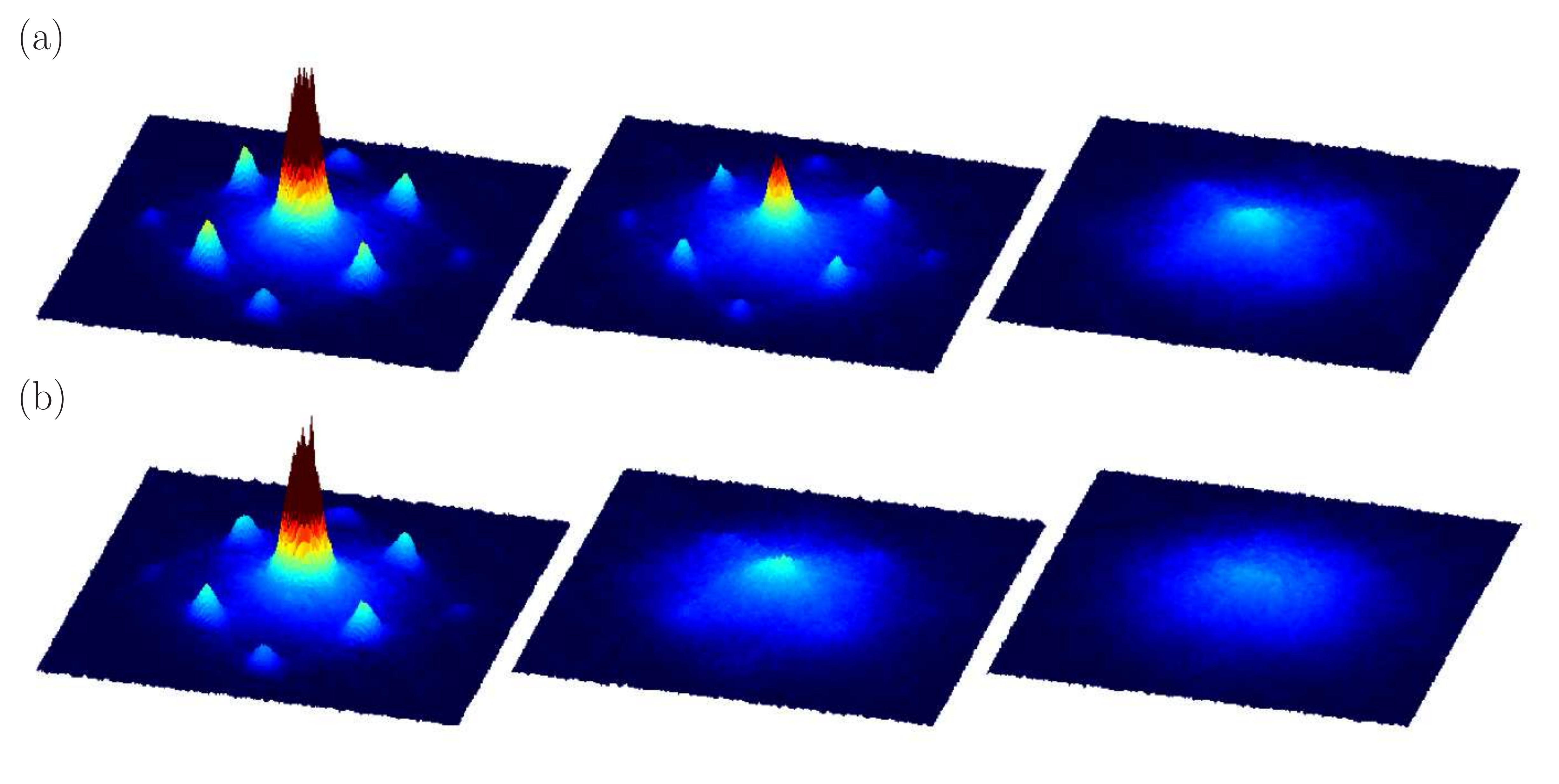}
  \caption{Time of flight absorption images of the bosonic component 15 ms after switching off the lattice and trap potentials~\cite{Ospelkaus2006e} (The colors and pseudo-3d representation encode the particle column density integrated along the imaging direction). (a) Pure bosonic ensemble of $10^5$ atoms at three different lattice depths. (b) Bosonic ensemble interacting with a fermionic impurity fraction of $20\%$ for the same lattice depths as in (a). We observe a striking loss of interference contrast in the images of the bosonic cloud interacting with fermions  compared to the pure bosonic cloud. {\it Copyright (2006) by the American Physical Society.}}
  \label{fig:lattice_images}
\end{figure}

Figure~\ref{fig:lattice_images} shows three sample images of the interference pattern of the pure bosonic cloud (top row) in comparison to the interference pattern of the bosonic cloud interacting with $2\cdot10^4$ fermionic impurities (bottom row). The images have been taken after $15~\mathrm{ms}$ time-of-flight. To ensure that experimental conditions are really comparable in the two cases, we have prepared the pure bosonic component with the same experimental sequence as the mixture prior to ramping up the optical lattice: In both cases, we have prepared a deeply degenerate mixture of fermionic and bosonic atoms in the combined magnetic and optical potential. The pure bosonic cloud is then realized immediately before ramping up the optical lattice by a removal of the  fermionic atoms from the trap with a short resonant light pulse, leaving behind a pure Bose Einstein condensate.
The three sample images of fig.~\ref{fig:lattice_images} have been taken at lattice depths of $12.5~E_r^\mathrm{Rb}$, $20~E_r^\mathrm{Rb}$ and $25~E_r^\mathrm{Rb}$, respectively. The lattice depths are given in units of the recoil energy for the $^{87}$Rb component $E_r^{\mathrm{Rb}}=(\hbar^2k^2)/(2m_{\rm Rb})\approx h\cdot2.14\,\textrm{kHz}$ where $k=2\pi/\lambda$.

In both cases, starting with a pure condensate or with a mixture of bosonic and fermionic atoms, we clearly observe a loss of interference contrast with increasing lattice depth marking the breakdown of long range phase coherence. As already mentioned, in case of the pure
bosonic gas, the loss of coherence accompanies the well-known superfluid to Mott-insulator phase transition \cite{Jaksch1998a,Greiner2002a,Greiner2003a} which occurs as a result of competition between the minimization of kinetic energy, parameterized by the tunnelling matrix element $J$ which tends to delocalize the atomic wavefunction over the crystal and the minimization of interaction energy $U$ (fig.~\ref{fig:lattice_images}(a)). As can be clearly seen from fig.~\ref{fig:lattice_images}(b), the presence of fermionic impurities induces a loss of coherence
at much lower lattice depths than for a pure BEC. At $12.5~E_r^\mathrm{Rb}$, we   observe sharp interference peaks in the pure bosonic cloud and the mixture. At $20~E_r^\mathrm{Rb}$, however, the interference pattern vanishes almost completely in case of the mixed system, whereas  significant interference is still visible in the case of pure \rub.

\subsection{Characterizing the phase coherence properties of the bosonic cloud}
\label{sec:characterizing_coherence}

For a quantitative analysis of the observed phenomena, we  adopt two different approaches  appropriate for the  characterization of  the coherence properties of the bosonic cloud. These approaches have both been  developed in the context of studies on the superfluid to Mott-insulator transition in an ensemble of repulsively interacting bosons:
\begin{itemize}
  \item[1.] First, we  characterize the interference pattern using the visibility of the interference fringes~\cite{Gerbier2005a, Gerbier2005b}. The definition of visibility, in this context, is motivated by the definition of the interference contrast in optics. The visibility has been related to the correlation properties of the  many-body wavefunction of the atomic cloud~\cite{Gerbier2005a,Gerbier2005b}.
  \item[2.] Second, we analyze the width of the central $p=0$ peak which is directly related to the inverse coherence length of the ensemble apart from some deviations due to the repulsive interactions between the bosonic atoms~\cite{Kollath2004a}.
\end{itemize}

\subsubsection*{Visibility}

\begin{figure}[tbp]
  \centering
  \includegraphics[width=0.5\columnwidth]{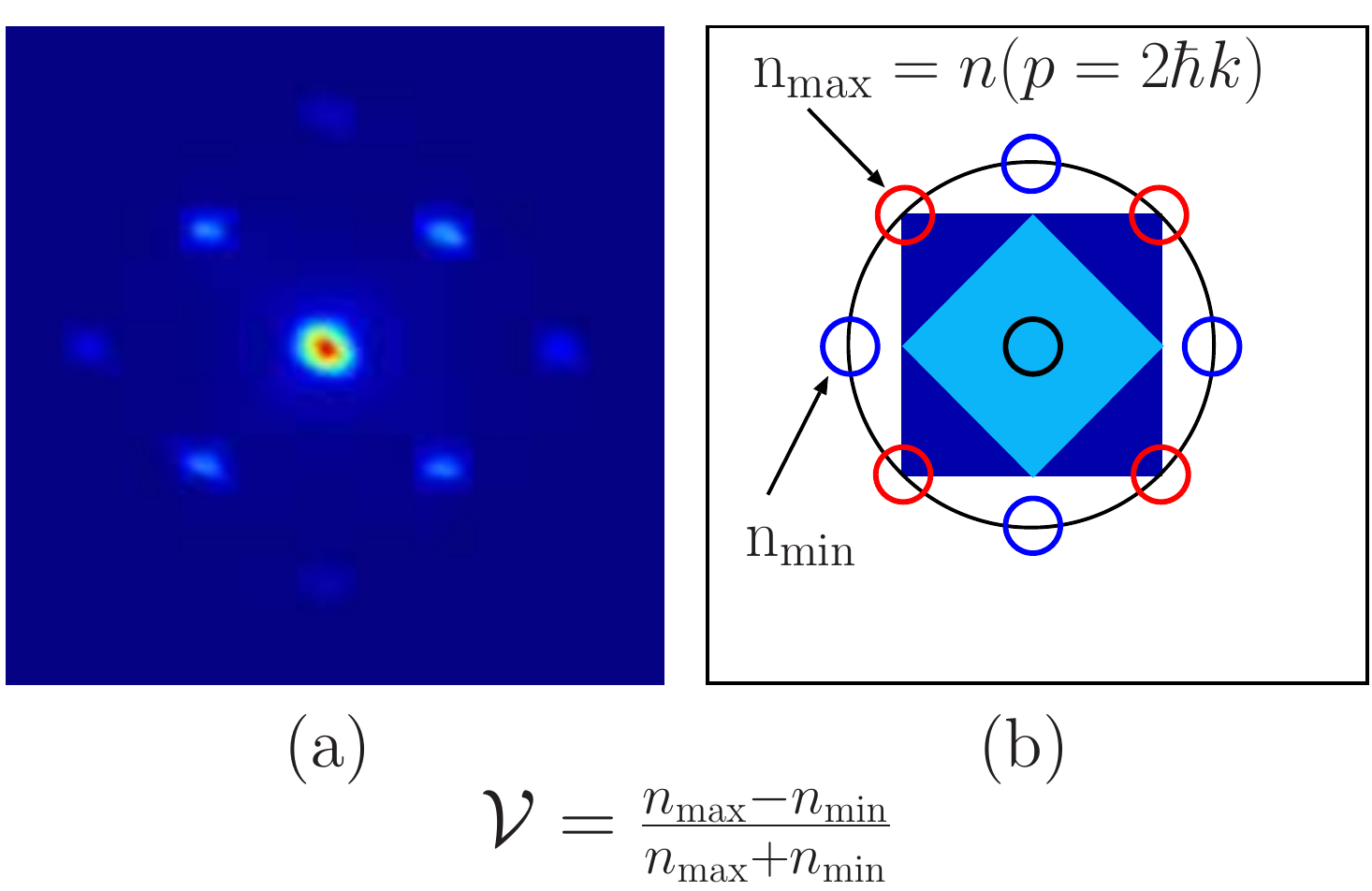}
  \caption[Definition of the visibility]{Definition of the visibility of the bosonic interference pattern $\mathcal{V}$~\cite{Gerbier2005a,Gerbier2005b}. The visibility is calculated from the atom numbers in the first order interference fringes in comparison to the atom numbers in equivalent areas at intermediate positions between these maxima. The definition is illustrated by a typical interference pattern for shallow optical lattices (a) and with a sketch (b). Apart from the interference fringes, the sketch illustrates the first (light blue) and second Brillouin zone (dark blue).}
  \label{fig:visibility}
\end{figure}

Figure~\ref{fig:visibility}(a) shows a typical interference pattern of a bosonic ensemble after time of flight released from a shallow optical lattice where a macroscopic fraction of the system  occupies the lowest Bloch state and phase coherence is present across the lattice. An intuitive definition of the contrast of the interference pattern is  given by~\cite{Gerbier2005a, Gerbier2005b} 
\begin{equation}
\label{eq:visibility}
\mathcal{V}=\frac{n_{\mathrm{max}}-n_{\mathrm{min}}}{n_{\mathrm{max}}+n_{\mathrm{min}}}
\end{equation}
where $n_{\mathrm{max}}$ is the total atom number in the first order interference fringes reflecting the $p=2\hbar k$ momentum component and  $n_{\mathrm{min}}$  is the sum of number of atoms in equivalent areas at intermediate positions between the maxima.

At first sight, the relation of the visibility to the bosonic many-body wavefunction is  not clear. However, it can be derived from the density distribution of the atoms after time of flight which is  given by~\cite{Pedri2001a, Kashurnikov2002a,Zwerger2003a, Gerbier2005b} 
\begin{equation}
n(\vec r)\propto\left|\hat w\left(\vec k =\frac{m\vec r}{\hbar t}\right)\right|^2\mathcal{S}\left(\vec k = \frac{m\vec r}{\hbar t}\right)
\end{equation}
where $\hat w$ is the Fourier transformation of the Wannier function $w(\vec r)$ and 
\begin{equation}
\mathcal{S}(\vec k)=\sum_{i,j}\exp(i\vec k \cdot (\vec r_i - \vec r_j))\left< b_i^\dagger b_j\right>
\end{equation}
is the quasimomentum distribution when we restrict ourselves to the lowest Bloch band.
As $\mathcal{S}$ is the Fourier transformation of the correlation function $\left<b^\dagger_ib_j\right>$, sharp interference fringes are only possible when the correlation function varies slowly across the lattice. As outlined in~\cite{Gerbier2005a}, this is the case of  long-range phase coherence. $\left<b^\dagger_ib_j\right>$ extends  only over a few lattice sites in the case of short range coherence, resulting in a reduction of contrast and visibility. Note that based on the definition of equation~(\ref{eq:visibility}), the Wannier envelope $|\hat w|^2$ cancels out and the visibility $\mathcal{V}$ is directly related to the Fourier transform of the correlation function.

\begin{figure}[tbp]
  \includegraphics[width=1.0\columnwidth]{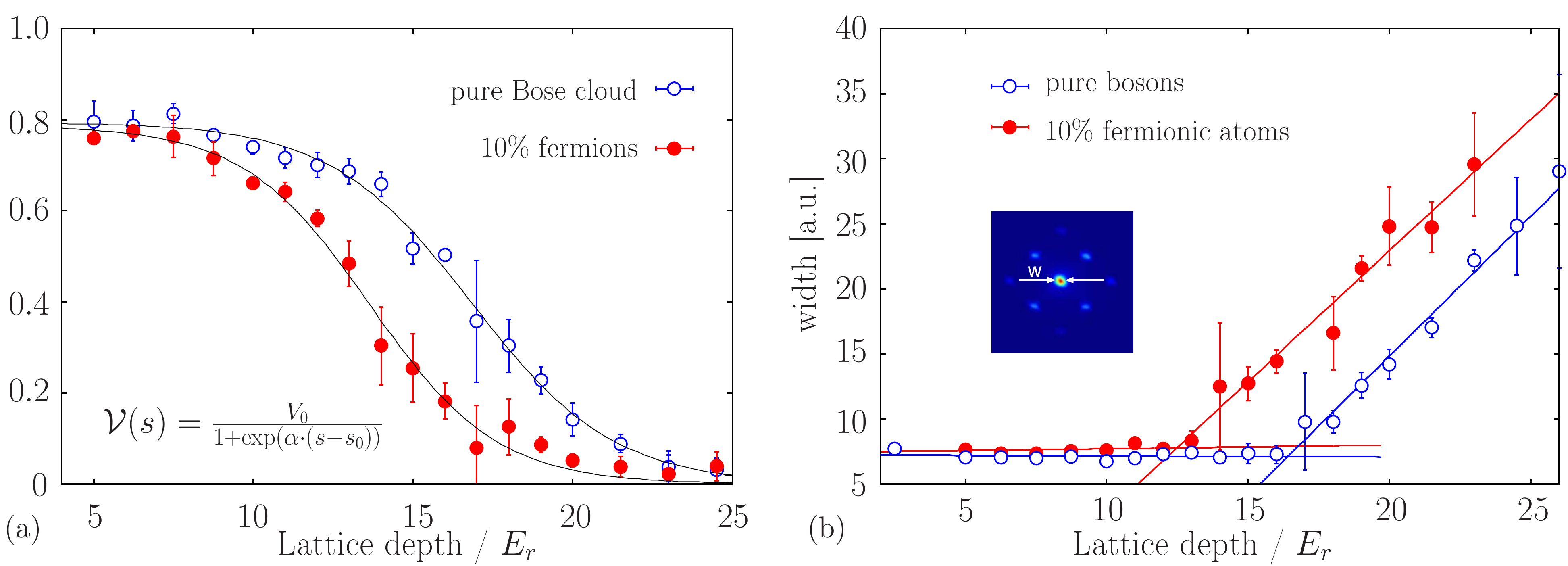}
  \caption{(a) Visibility data for both a pure bosonic and a mixed system with an impurity fraction of $10\%$. We have fitted a phenomenological fit function to the data given by $\mathcal{V}(s)=\mathcal{V}_0/(1+\exp(\alpha\cdot(s-s_0))$. The initial visibility depends on the particular choice of the circles used to extract $n_{\mathrm{max}}$ and $n_{\mathrm{min}}$.
           (b) Width of the central interference fringe of the bosonic cloud both for a pure bosonic cloud and a mixed system with $10\%$ impurity fraction. The width data shows a sudden increase above a certain lattice depth. The associated characteristic depth can be extracted from the data as the crossing point of two straight lines fitted to the data at low lattice depth and at high lattice depth.
  \label{fig:sample_visibility_width}}
\end{figure}

Figure~\ref{fig:sample_visibility_width}(a) shows the visibility of a bosonic cloud of $10^5$ atoms loaded into an optical lattice as a function of lattice depth. The data has been obtained starting from a pure condensate in the magic trap with trapping frequencies of $\omega=2\pi\cdot 50$~Hz in the axial and radial direction. The optical lattice intensity is ramped up with a linear ramp of $0.5~E_r^\mathrm{Rb}/$ms to various final lattice depths. $n_{\mathrm{max}}$ and $n_{\mathrm{min}}$ as defined in fig.~\ref{fig:visibility} are extracted  from the time-of-flight images using a circle with a radius of 4 pixels around the center of the first order interference peaks and the displaced intermediate region, respectively. In the same figure, we have plotted the visibility of the bosonic cloud in a mixed system with $10\%$ impurity fraction.

Let us first study the visibility of the pure bosonic cloud. Starting from a lattice depth of $10~E_r^\mathrm{Rb}$, we observe a continuous decrease in the visibility of the bosonic interference pattern which becomes particularly steep above $15~E_r^\mathrm{Rb}$. Note that in our system, the quantum critical points for the onset of the superfluid to Mott-insulator transition in a pure bosonic system are  lattice depths of $13.5~E_r^\mathrm{Rb}$, $15.5~E_r^\mathrm{Rb}$, $16.9~E_r^\mathrm{Rb}$,  $18.0~E_r^\mathrm{Rb}$ and $18.8~E_r^\mathrm{Rb}$  for $n=1,\ldots,n=5$, respectively. The observed loss of interference contrast is ascribed to the increasing fraction of atoms which enter the  Mott-insulating phase. The quasi-continuous decrease in visibility, instead of a sharp one above just one lattice depth corresponding to the quantum critical point of the transition, can be explained by the inhomogeneity of the system where atoms in different shells enter the Mott-insulating state with $n$ atoms per site at different lattice depths.

\paragraph*{Phenomenological fit function}
For the analysis of the visibility data, we use a phenomenological fit function defined by 
\begin{equation}
\label{eq:fit_function}
\mathcal{V}(s)=\frac{\mathcal{V}_0}{1+\exp\left(\alpha\cdot(s-s_0)\right)}
\end{equation}
where $\mathcal{V}_0$ is the initial visibility in the limit $s\rightarrow 0$, $s_0$ is a measure for the  ``critical'' lattice depth and $\alpha$ is an additional fit parameter accounting for  the steepness of the loss of contrast.\footnote{For typical fit results with $\alpha\approx 0.4$ and $s_0\approx 15$, we obtain $\exp(-\alpha s_0)\approx 0.002$ and therefore $\mathcal{V}(s\rightarrow 0)\approx \mathcal{V}_0$} As can be seen in fig.~\ref{fig:sample_visibility_width}(a), the phenomenological function describes the experimental data quite well.
In particular, the fit function accounts for the limiting cases of the superfluid state with a theoretical visibility of $\mathcal{V}=1$ and the pure Mott-state with a visibility of $\mathcal{V}=0$.

\paragraph*{Comparison of pure bosonic to mixed systems}
Based on the phenomenological fit function of equation~(\ref{eq:fit_function}), we can now compare the decrease in visibility for a pure bosonic cloud to the decrease in case of a mixed system. The corresponding data is plotted in fig.~\ref{fig:sample_visibility_width}(a). Each data point corresponds to $5$ to $10$ averaged measurements. The errorbars denote the standard deviation of the 5-10 measurements. To both the visibility data of the bosonic and the visibility data of the mixed system with $10\%$ impurity fraction, we have fitted the phenomenological fit function of equation~(\ref{eq:fit_function}). The obtained fit parameters are summarized in table~\ref{tab:fit_results}.
\begin{table}[htbp]
  \centering
  \begin{tabular}{|c||c|c||c|}\hline
    System            &  $s_0/E_r^\mathrm{Rb}$  & $\alpha$      & $\mathcal{V}_0$ \\\hline\hline
    pure Bose cloud   &  $16.8\pm0.2$           & $0.45\pm0.03$ & $0.8\pm0.02$    \\\hline
    $10\%$ impurities &  $13.7\pm0.3$           & $0.5\pm0.05$  & $0.8\pm0.02$    \\\hline
  \end{tabular}
  \caption[Comparison of fit results to the visibility data of a pure bosonic and a mixed system ]{Comparison of the parameters of the phenomenological function for a fit to two data sets, one for a pure bosonic system and one for a system with $10\%$ impurity fraction. The errors correspond to the fit errors only. Note that e.g. the lattice depth calibration is estimated to be accurate within 5$\%$, resulting in an error bar on $s_0$ of $\approx1~E_r^\mathrm{Rb}$.}
  \label{tab:fit_results}
\end{table}
Whereas $\mathcal{V}_0$ and the fit parameter $\alpha$ are equal in the two cases within the fit uncertainty, we obtain a significantly lower ``critical'' lattice depth in the case of the mixed system than in the pure bosonic case. We observe a shift in the characteristic parameter $s_0$ of $\Delta s=s_0^\mathrm{pure}-s_0^\mathrm{mixed}\approx- 3 E_r^\mathrm{Rb}$ towards lower lattice depth when comparing the mixed to the pure system. We have checked that we do not observe a significant variation in $s_0^\mathrm{mixed}$ for the mixed system when fixing $\alpha$ and $\mathcal{V}_0$ to the values of the pure bosonic system. 

To estimate the reliability of the analysis based on the phenomenological fit function, we have analyzed the visibility data a second time, this time based on an analytically known dependence of the visibility in the limit of deep optical lattices.
Based on perturbation theory in the limit of very weak tunnelling, the ground state of the system can be approximated by
\begin{equation}
\left|\Psi\right>\approx\left|\Psi\right>_{\mathrm{MI}}+\frac{J}{U}\sum_{<i,j>}b_i^\dagger b_j\left|\Psi\right>_{\mathrm{MI}}
\end{equation}
where $\left|\Psi\right>_{\mathrm{MI}}$ is the Mott-insulating state. The analytical expression for the  ``column'' integrated visibility is then given by~\cite{Gerbier2005b}
\begin{equation}
\mathcal{V}\propto \frac{zJ}{U}
\end{equation}
where $z$ is the number of nearest neighbors ($z=6$ in case of a simple cubic 3D lattice). The above approximation is valid in the limit of deep optical lattices where tunnelling is strongly suppressed.

To compare  our analysis of the visibility data based on the phenomenological fit function to an analysis based on the known  dependence of the visibility in the limit of deep optical lattices, we have analyzed the visibility both for  the pure bosonic cloud and the mixture in the limit of deep optical lattices.

\begin{figure}[tbc!]
  \centering
  \includegraphics[width=0.6\columnwidth]{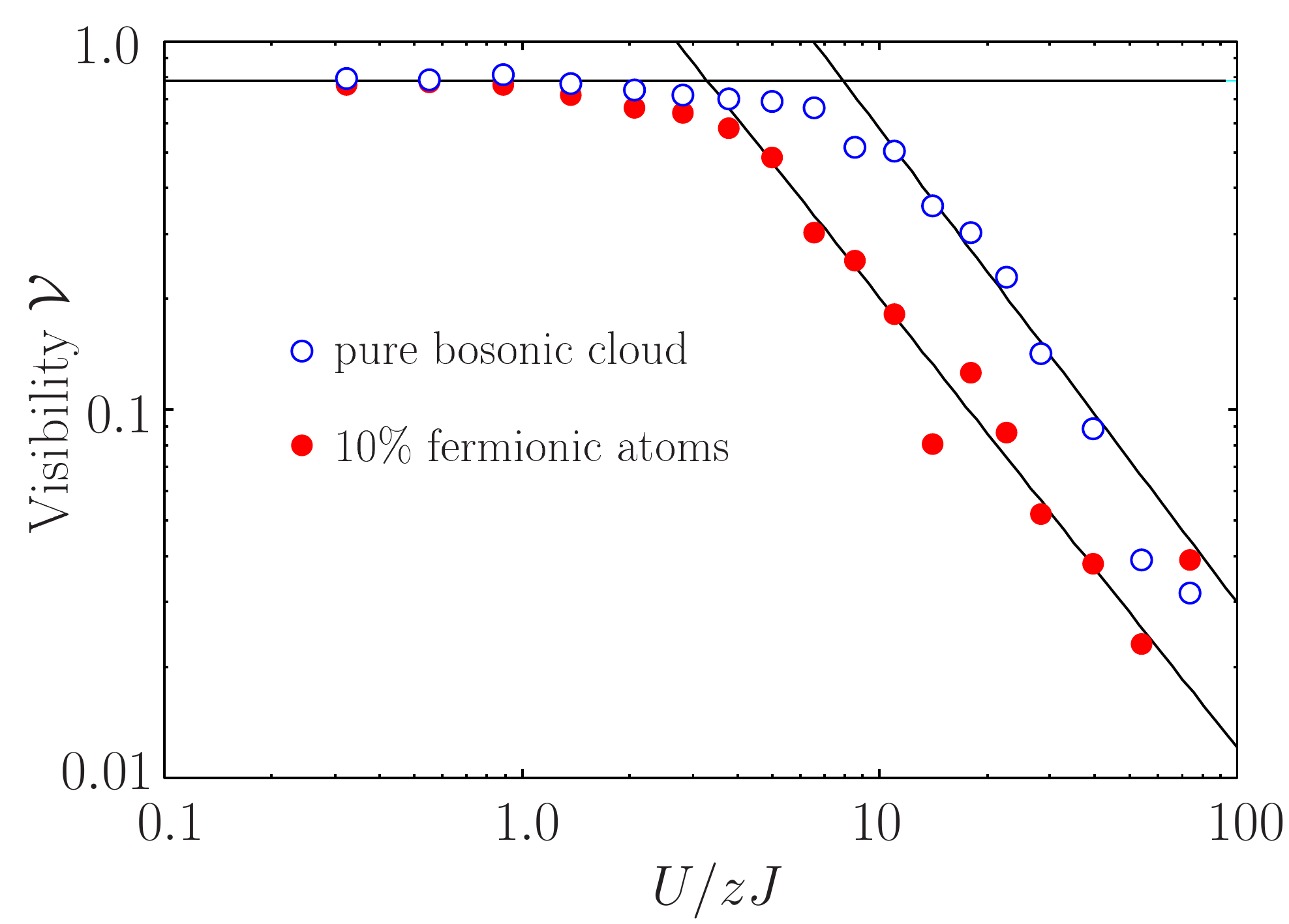}
  \caption[Analysis of visibility data based on~\cite{Gerbier2005a}]{ Visibility data  both for a pure bosonic and a mixed system with an impurity fraction of $10\%$. This time, we have plotted the visibility against $U/zJ$ where $U/J=\frac{ka_\mathrm{BB}}{\sqrt{2}}\exp(2\sqrt{s})$ with $s=V_0/E_r^\mathrm{Rb}$.}
  \label{fig:sample_visibility_gerbier}
\end{figure}

Figure~\ref{fig:sample_visibility_gerbier} shows the visibility data discussed above, this time plotted as a function of $(U/zJ)$ which is directly related to the lattice depth in the tight-binding limit via
\begin{equation}
\frac{U}{zJ}=\frac{k a_{\mathrm{BB}}}{\sqrt{2}z}\exp(2\sqrt{s}).
\end{equation} To extract a shift from the data, we have fitted a function $C \left(\frac{U}{zJ}\right)^{\nu}$ in the limit of deep optical lattices ($U/zJ>10$) and a constant in the limit of shallow optical lattices. From these fits, we can again extract critical lattice depths for both the pure bosonic cloud and the mixture. Here, the critical depth is defined as the intersection point between the two fits. 
We observe a critical $(U/zJ)_c\approx 9.9$ in the case of the pure bosonic cloud which corresponds to a lattice depth of $15.6~E_r^\mathrm{Rb}$ and a critical depth of $(U/zJ)\approx 3.3$ corresponding to $V_0\approx 11.6~E_r^\mathrm{Rb}$ in the case of the mixture. The exponent $\nu$ is comparable to each other in both cases. We observe $\nu=-1.3(1)$ for the pure bosonic cloud and $\nu=-1.2(2)$ in case of the mixture. From the analysis, we extract a shift of $\approx-4~E_r^\mathrm{Rb}$ which is comparable to the shift extracted from the phenomenological fit function. However, the extracted shift depends critically on the chosen fit interval for the deep lattice limit and varies easily by $\pm 1~E_r^\mathrm{Rb}$, whereas the analysis based on the phenomenological fit function is independent of the fit interval.

\subsubsection*{Width of the central interference fringe}

The phase coherence properties of the bosonic cloud loaded into the optical lattice can also be characterized by the width of the central interference fringe $w$ which is directly related to the correlation length $\zeta$ of the system via $\zeta\propto w^{-1}$. The analysis is again motivated by fundamental properties of the superfluid to Mott-insulator transition of a pure bosonic system. Whereas, in the superfluid phase, the correlation length diverges, $\zeta$  becomes finite in the Mott-insulating state and is on the order of the inverse energy gap of the system $\zeta\propto U^{-1}$~\cite{Kollath2004a}. 

Figure~\ref{fig:sample_visibility_width}(b) shows the half-width of the central interference fringe as a function of lattice depth for the pure bosonic and the mixed system for the same set of data as in the visibility analysis. In both cases, the width stays constant for some time or even slightly decreases for shallow optical lattices. Above a certain lattice depth, the width increases rapidly. The characteristic lattice depth for the sudden increase is extracted from the data by fitting two linear fits to the descending and ascending branches of the data. The characteristic lattice depth is then defined by the intersection point of the two linear curves.

\begin{table}[htbp]
  \centering
  \begin{tabular}{|c||c|c||c|}\hline
    System            & $s_0/E_r^\mathrm{Rb}$ \\\hline\hline
    pure Bose cloud   & $16.5\pm1.6$          \\\hline
    $10\%$ impurities & $12.5\pm1.5$          \\\hline
  \end{tabular}
  \caption[Comparison of width data of a pure bosonic and a mixed system ]{Comparison of the ``critical depth'' of a pure bosonic cloud to a mixed system with $10\%$ impurity fraction as extracted from the width data.}
  \label{tab:fit_results_width}
\end{table}

Table~\ref{tab:fit_results_width} summarizes the characteristic lattice depths for the two cases. The error bars are again based on the fit uncertainties. Whereas the critical lattice depth of the pure bosonic cloud is given by 16.5(1.6)~$E_r^\mathrm{Rb}$, in the mixed system, we observe a critical depth of $12.5(1.5)~E_r^\mathrm{Rb}$. The critical depth is thus shifted by -4.0(2.0)~$E_r^\mathrm{Rb}$ towards shallower optical lattices. The result is comparable to the respective results extracted from the visibility data.

\subsection{Particle number dependence}

Based on the techniques of section~\ref{sec:characterizing_coherence}, we have studied the change of coherence properties in the bosonic cloud as a function of particle number of the fermionic impurities. To this end, we have prepared mixtures of $\approx 10^5$ bosons with a variable fraction of admixed fermionic atoms in the optical lattice. In the experiments, we have prepared mixtures with $3\%$, $7\%$, $10\%$ and $20\%$ fermionic atoms corresponding to $3\cdot 10^3$, $7\cdot 10^3$, $1\cdot10^4$ and $2\cdot10^4$ fermions, respectively, in the combined magnetic and optical potential with trapping frequencies of $\vec \omega=2\pi\cdot(50,150,150)$\,Hz (for the experiments reported in~\cite{Ospelkaus2006e}) or in the ``magic trap'' with trapping frequencies of $\omega=2\pi\cdot50$\,Hz. When choosing the above fractions of fermionic impurities, we have taken care to realize mixtures where the extension of the fermionic atoms is mostly still within the extension of the bosonic cloud in the optical lattice (details in~\cite{SilkeThesis}).

\begin{figure}[tbp]
  \centering
  \includegraphics[width=0.6\columnwidth]{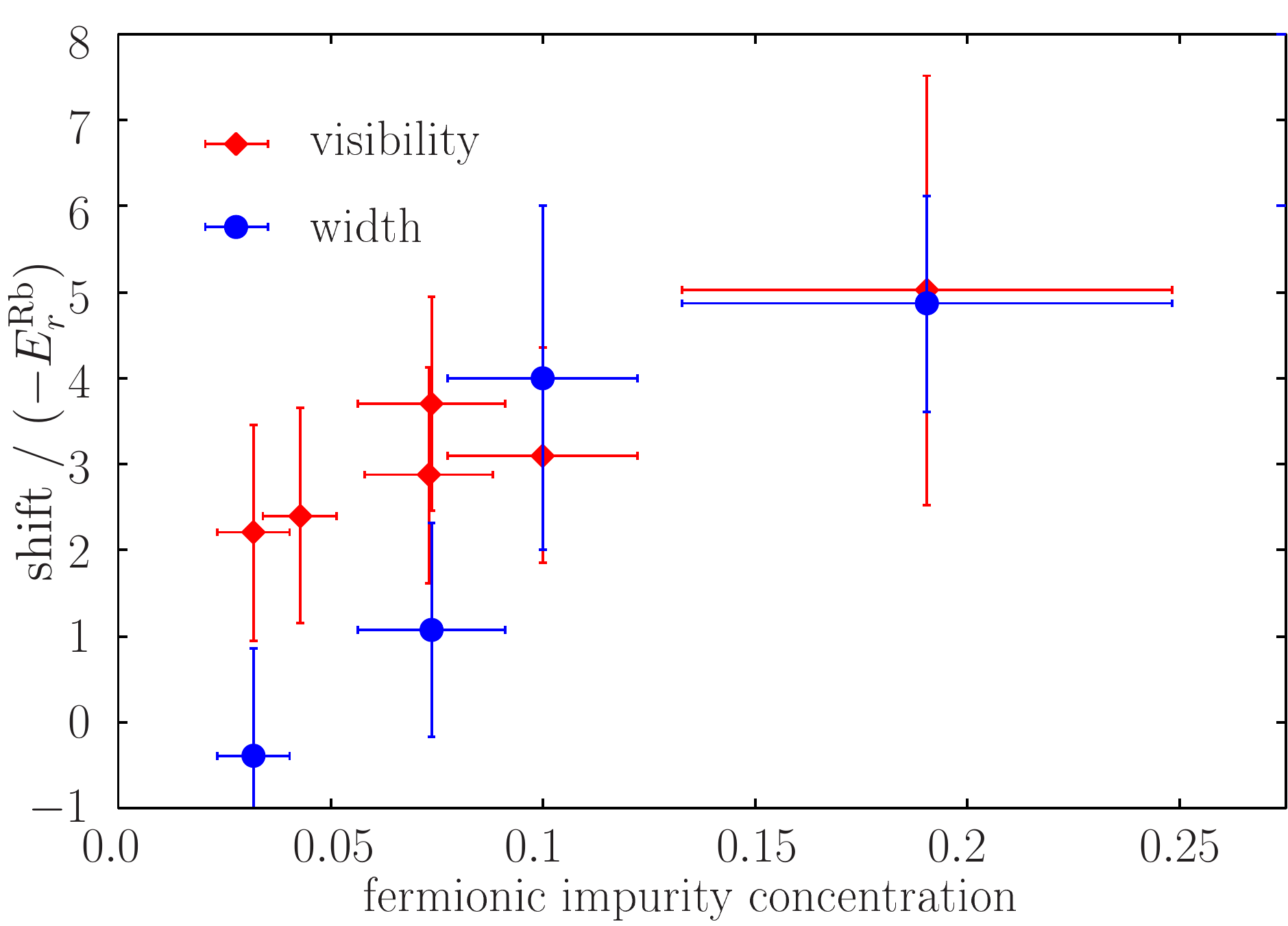}
  \caption[Summary of observed shift]{Observed shift of the ``critical'' lattice depth characterizing the loss of coherence in the bosonic cloud as a function of fermionic impurity concentration.}
  \label{fig:concentration}
\end{figure}

Figure~\ref{fig:concentration} summarizes our findings on the particle number dependence. We have analyzed samples with $10^5$ bosons with varying impurity concentration between $3\%$ and $20\%$. In case of the visibility data, the error bars in the diagram correspond to the ``sample'' rate of the underlying data. A sample rate of $1-1.25~E_r^\mathrm{Rb}$ has been translated into an assumed maximum error of the same size; this error is larger than the error based on the fit errors. In case of the width data, we have calculated the error bars based on the underlying fit errors of the involved parameters. As can be seen from fig.~\ref{fig:concentration}, an increasing impurity concentration leads to a considerable shift of the bosonic coherence properties towards shallower optical lattices. The shift increases with impurity concentration and reaches $-5~E_r^\mathrm{Rb}$ for an impurity concentration of  20\%. Note that the different data points correspond to ensembles that have been prepared on different days and with two different trap geometries. Whereas most of the measurements have been done with an external harmonic confinement given by a combined magnetic and optical potential with an aspect ratio of $1:3$, some of the data points have been recorded with an almost spherically symmetric confinement of $\omega=2\pi\cdot50~$Hz of the `magic trap' discussed in our experiments on tuning of interactions. In all these cases, we observe qualitatively and quantitatively the same characteristic behavior.

\subsection{Studies on systematic effects}

To exclude trivial irreversible heating effects as a source of the observed systematic shift in the coherence properties of the bosonic cloud, we have studied the  reversibility of the observed coherence loss. To this end, we have ramped up the optical lattice to final lattice depths between $15~E_r^\mathrm{Rb}$ and $20~E_r^\mathrm{Rb}$, where the coherence of the bosonic cloud is either completely lost or at least significantly distorted,  and have subsequently  ramped down the lattice with the same ramp speed to varying lattice depths between $5~E_r^\mathrm{Rb}$ and $7.5~E_r^\mathrm{Rb}$. Apart from some signs of non-adiabaticity visible in a slight asymmetry of the obtained visibility data with respect to the reversal point at deep optical lattices, we have been able to restore the initial visibility of the bosonic cloud almost completely. The maximum observed irreversible loss of visibility is on the order of  $10\%$ for both the Fermi-Bose and the pure bosonic case. In case that the overall ramp sequence has not been fully reversible, allowing for an additional equilibration time of $5-10$~ms after ramp-down has resulted in a near complete recovery of visibility.

Note that the recovery of the visibility is not necessarily an indicator of adiabaticity, whereas an incomplete recovery would be a sure indicator. Also, there could be technical noises which have a more drastic effect on the mixture than on the pure bosonic system.

\subsection{Discussion}
There is currently an intense discussion on the nature of the observed shift of the coherence properties of the bosonic cloud due to a small admixture of fermionic atoms. Possible scenarios include a shift of the superfluid to Mott insulator transition of the bosonic cloud due to interactions with the fermionic impurities, thermodynamic effects such as adiabatic heating and effects of disorder induced localization scenarios. In the following section, we will briefly review the current status of the discussion and provide the interested reader with some simple estimates. 

\subsection*{Shift of the superfluid to Mott-insulator transition}
One of the possible scenarios is a shift of the superfluid to Mott insulator transition of the bosonic cloud due to the admixture of attractively interacting fermions. However, there is even a controversy about the sign of the expected shift depending on the underlying theoretical model: the Fermi-Bose Hubbard approximation or an effective potential approach.

\paragraph*{Fermi-Bose Hubbard approximation}
The change of the quantum critical point of the bosonic superfluid to Mott-insulator transition in a homogeneous system due to the presence of  fermionic atoms  has  been discussed in a recent work by Pollet and coworkers~\cite{Pollet2008a}. Based on the Fermi-Bose-Hubbard Hamiltonian~\cite{Cramer2004a}, Pollet \textit{et al.} have discussed in zero and first order perturbation theory the change in the quantum critical point for the superfluid to Mott insulator transition point $(U/zJ)_c$ due to interactions with the fermionic cloud. Here, we summarize the main results: Let us consider the Bose-Fermi Hubbard Hamiltonian of a homogeneous system 
\begin{eqnarray}
H_{BF}& = & -J_B\sum_{<i,j>}b_i^{\dagger}b_j+\frac{1}{2} U_{BB}\sum_i n_i^B(n_i^B-1)\\
      & - & J_F\sum_{<i,j>}f_i^{\dagger}f_j\\\nonumber
      & + &  U_{BF}\sum_i n^B_i n_i^F\nonumber
\end{eqnarray}
(notation see~\cite{Cramer2004a}). Under the influence of the fermions as a perturbation of the pure bosonic Hubbard Hamiltonian,  the lowest order effect that the fermions have on the bosons can be derived when replacing the fermionic particle number operators $n_i^F$ by their expectation values. In this case, the perturbation to the pure bosonic Hubbard Hamiltonian is: $U_\mathrm{FB}\sum n^B_i \left<n_i^F\right>$. To lowest order perturbation theory, the fermionic expectation values  $\left<n_i^F\right>$ are calculated by solving the pure fermionic Hubbard Hamiltonian (neglecting the effects of the bosons on the fermionic cloud). In the homogeneous system considered here,  $\left<n_i^F\right>$ is independent of the lattice site $\left<n_i^F\right>=\left<n_j^F\right>=\left<m\right>$. To lowest order perturbation theory, the fermions thus induce an overall shift of the system's energy and do not change the phase diagram of the bosonic cloud at all.

The next order effect can be derived when adopting the results of linear response theory (see \cite{Pollet2008a}). In this case, one assumes the unknown bosonic density $n_B(\vec q)$ to induce a fermionic density  $\left<n_F(\vec q)\right>=U_{FB}\Xi(T,\vec q)n_B(\vec q)$ where $\Xi(T, \vec q)$ is the Lindhard response function known from the theory of screening when calculating electronic charge distributions in condensed matter physics~\cite{AshcroftMermin,Buchler2004a,Pollet2008a}. The induced fermionic density distribution results in a back-action on the bosons  and thus in an effective interaction for the pure bosonic cloud given by $\hat U^\mathrm{eff}_\mathrm{BB}=U_\mathrm{BB}+U_{\mathrm{FB}}^2\Xi(T,\vec q)$. As the Lindhard function is always negative, the interaction matrix element between two initially repulsively interacting bosons is effectively reduced by an amount $U_{\mathrm{FB}}^2\Xi(T,\vec q)$ independent of the sign of $U_\mathrm{FB}$. This induced attractive interaction is in analogy  to phonon induced attraction between electrons in conventional superconductors and is at the heart of recent proposals on boson-induced Cooper pairing of fermions in optical lattices~\cite{Albus2004a}. Based on these considerations, the superfluid to Mott-Insulator transition is shifted towards deeper optical lattices independent of the sign of Bose-Fermi interaction $U_{FB}$.

The situation becomes more complicated when considering an inhomogeneous situation where the atoms experience both an optical lattice and an external harmonic trapping potential. In the case of attractive interactions,  the equilibrium densities of both components will be enhanced  in the center of the confining potential.  However, enhanced density results in an increased mean  bosonic occupation number on a single lattice site, thereby shifting the transition from a  superfluid to a Mott insulating state  towards deeper optical lattices.

\paragraph*{Mean-field approach --- Induced effective potential.}
Let us now come back to the scenario of a homogeneous optical lattice without additional overlapped harmonic potential. It has been suggested by Marcus Cramer~\cite{Cramer2006priv} to consider the fermionic density distribution in the lattice  as an additional mean-field potential $U^{F}_{MF}(\vec r)$ which enters into the bosonic many-particle Hamiltonian via the confining external potential $V_{\mathrm{effective}}(\vec r)=V_{\mathrm{periodic}}(\vec r)+U^{F}_{MF}(\vec r)$. To lowest order perturbation theory, the additional fermionic mean-field potential is determined by the unperturbed fermionic density distribution given by $n_F(\vec r-\vec r_i)=\left<m\right>\sum_{i}|w^\mathrm{F}(\vec r -\vec r_i)|$, resulting in 
\begin{equation}
U^F_{MF}=n_F(\vec r)U_{BF}=\left<m\right>U_{BF} \sum_{i}|w^F(\vec r -\vec r_i)|^2,
\end{equation}
where $w^F$ is the fermionic Wannier function for the case of a pure sinusoidal external lattice potential of depth $V_\mathrm{Lattice}$. Taking into account this additional  potential, the bosonic many-particle Hamiltonian reads:
\begin{eqnarray}
\label{eq:eff_ham}
H_{B} & = &\int d^3r \Psi_B^\dagger(\vec r)\left(\frac{p^2}{2m}+V_{\mathrm{periodic}}(\vec r)+n_F(\vec r)U_{\mathrm{BF}}\right)\Psi_\mathrm{B}(\vec r) \\\nonumber
     & + & \frac{1}{2}U_\mathrm{BB}\int d^3r\Psi_\mathrm{B}^\dagger(\vec r)\Psi_\mathrm{B}^\dagger(\vec r)\Psi_{\mathrm{B}}(\vec r)\Psi_{\mathrm{B}}(\vec r)
\end{eqnarray}
The fermions thus alter the properties of the pure external periodic potential and change both the depth and geometry of a single lattice well. From the effective many-body Hamiltonian of equation~(\ref{eq:eff_ham}), we can derive an effective Bose-Hubbard Hamiltonian. The  presence of the fermionic mean-field potential then changes both onsite interaction and tunnelling matrix elements of the bosonic atoms.

In a recent numerical simulation, D. L\"uhmann and coworkers~\cite{Luehmann2008a} have shown that the attractive interaction between \pot\ and \rub\ of $-215\,a_0$ will lead to a mutual trapping of particles. In particular, the bosonic effective potential will be strongly deformed and enhanced by the presence of the fermions and will thus result in an effective shift of the superfluid to Mott phase transition for the bosonic cloud. The authors derive a shift of the transition between $2.2\,E_r$ to $5.2\,E_r$ towards shallower lattices, depending on the filling factor of the bosons in the optical lattice and also observe a strong dependence on the interaction properties between bosons and fermions characterized by $a_{\mathrm{FB}}$. Note that the main discussion in this paper assumes a homogeneous optical lattice. The case of an inhomogeneous optical lattice, i.\ e.\ an optical lattice with additional harmonic confinement  in three dimensions (as realized in the experiment) is not covered.

\subsection*{Adiabatic heating or cooling}
When adiabatically loading either bosonic or fermionic quantum gases into an optical lattice, the density of states for the components changes markedly,  thereby e.~g. changing the characteristic temperatures for degeneracy in bosonic and fermionic quantum gases, $T_c$ and $T_F$. Adiabatic loading of atoms occurs at constant  entropy $S$, not at constant temperature, and the absolute temperature may change.  The change in   degeneracy characterized by $T/T_c$ and $T/T_F$ depends on the starting conditions such as temperature and atom number as well as the trap geometry and the interactions present in the system.  Note that the temperature change is reversible as it occurs at constant entropy and is a purely thermodynamic effect. However, increasing  $T/T_c$ with increasing lattice depth may lead to a decreasing condensate fraction which might result in loss of coherence of the bosonic cloud.

Adiabatic temperature change at constant entropy due to adiabatic loading of atoms into an optical lattice has been discussed in the literature for a homogeneous non-interacting Fermi  gas~\cite{Blakie2005a} and for homogeneous and inhomogeneous interacting and non-interacting bosonic clouds~\cite{Blakie2004a, Rey2006a}. 

Recently, we have~\cite{Cramer2008a} analyzed thermodynamic properties of the adiabatic loading of an attractively interacting harmonically trapped Fermi-Bose system into an optical lattice. The analysis in the limit of vanishing optical lattice and deep optical lattice reveals that a significant adiabatic heating of the mixture is expected during the adiabatic loading process. While regions of adiabatic heating and cooling can be identified in M. Cramer's numerical simulations~\cite{Cramer2008a} depending on the starting conditions of the loading process, the presence of fermions leads either to a more distinct heating of the mixture or a less distinct cooling. In any case, the Fermi-Bose mixture will have a significantly higher temperature at the end of the loading process than the pure bosonic system. Although this temperature increase is a thermodynamic effect due to the adiabaticity of the loading process and can thus be reversed, it will definitely result in a decrease of the condensate fraction of the bosonic cloud in the optical lattice and thereby affect the phase coherence properties of the bosonic cloud in the lattice as characterized in the visibility and the width of the bosonic interference pattern.

\subsection*{Disorder-enhanced localization scenarios}

\begin{figure}[tbp]
  \centering
  \includegraphics[width=0.5\columnwidth]{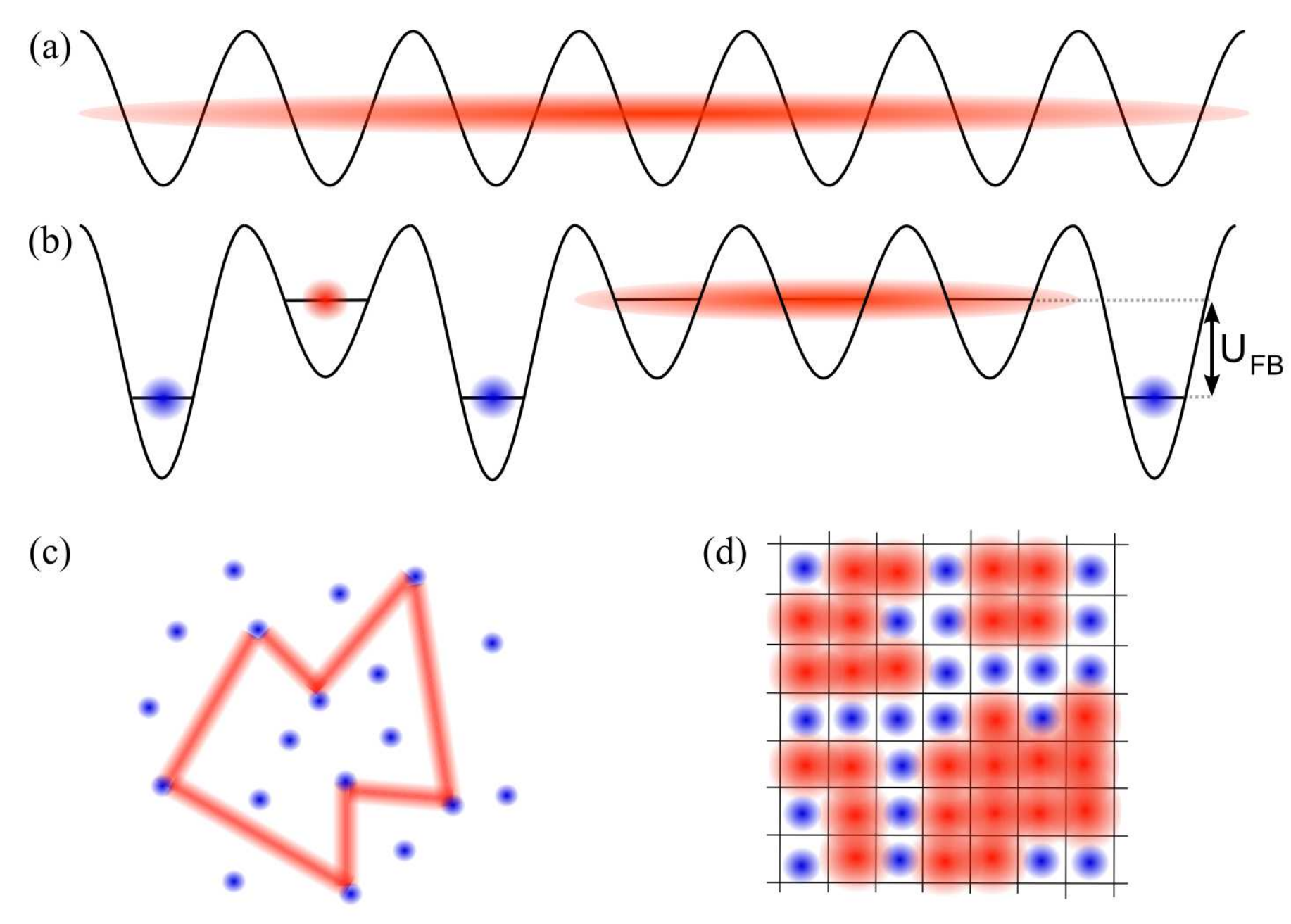}
  \caption{Schematic localization scenarios~\cite{Ospelkaus2006e}. (a) Pure bosonic superfluid in an optical lattice. (b) Shift of the effective potential depth due to fermionic impurities. (c) Localization by interfering paths of the bosonic wavefunction scattered by randomly distributed fermionic impurities. (d) Localization due to percolation. A random fermion distribution hampers the establishment of a coherent connection and causes the localization of bosonic ensembles in superfluid ``islands''. {\it Copyright (2006) by the American Physical Society.}}
  \label{fig:localization_scenarios}
\end{figure}

Let us finally review intuitively  possible disorder-related localization scenarios  which might be observable in Fermi-Bose mixtures and would possibly also lead to reduced coherence in the bosonic cloud.  For illustration purposes, let us start  with a pure bosonic superfluid (fig.~\ref{fig:localization_scenarios}(a)). Adding fermionic impurities and considering the attractive \pot--\rub\ Fermi-Bose interaction energy as an additional potential for the bosons, the ``defects'' caused by the fermionic impurities can be described by a local change of the effective optical lattice depth for the bosons due to the interparticle interaction (fig.~\ref{fig:localization_scenarios}(b)). If the energy level shift caused by the interaction energy is large enough, the superfluid bosonic wavefunction will not extend into this defect region, but will be scattered by the impurity. If scattering becomes frequent, interference effects along a closed scattering path are predicted to suppress transport and lead to a localization scenario similar to Anderson localization  (fig.~\ref{fig:localization_scenarios}(c)). A further increase in the impurity density will lead to the formation of ``forbidden walls''. Once the walls in this quantum percolation scenario lead to a sufficiently complicated labyrinth like structure for the bosonic wavefunction, a single coherent superfluid phase can no longer be sustained and several separated domains will be formed (fig.~\ref{fig:localization_scenarios}(d)).

Definitely, further experiments are required to shine more light on the observed phenomena. A possible shift of the Mott-insulating transition can be probed by a spectroscopic investigation of the excitation spectrum of the mixture similar to probing of the excitation gap of the pure bosonic Mott insulator in~\cite{Greiner2002a,Stoferle2004a}. Lattice site occupation can be probed by rf spectroscopic techniques~\cite{Campbell2006a}. The correlation properties of the fermionic and bosonic components can be revealed by noise correlation techniques~\cite{Greiner2005a,Folling2005a,Rom2006a}. With the availability of heteronuclear Feshbach resonances, it is straightforward to extend our studies to almost arbitrary interaction strength between the fermionic and bosonic constituents, and scattering-length dependent measurements have recently been performed by Best {\it et al.}~\cite{Best2008a}. Also, the coherence properties of Bose-Bose mixtures with repulsive (interspecies-) interactions have been investigated by Catani \textit{et al.}~\cite{Catani2008a}. In this case, the discussion is complicated by the limited overlap of the clouds due to repulsive interactions; in fact, a significant shift of coherence reduction to lower lattice depths is observed even for close to vanishing overlap between the clouds. In any case, further studies will have to concentrate on isolation and identification of underlying phases and associated phase transitions in this rich system.
\section{Ultracold heteronuclear molecules in a 3D optical lattice}

Being able to perform experiments with ultracold dense polar molecular gases is one of the outstanding challenges in atomic, molecular and optical physics~\cite{cold_molecule_conference_itamp_cua}. The increasing interest in this problem is motivated by a wide range of of research directions which would directly benefit from the availability of such molecular samples. 

\subsection*{The quest for ultracold polar molecules}

A diatomic polar molecule is characterized by the presence of a permanent internal dipole moment occurring as a result of the electronic charge distribution not having the same ``center of mass'' as the nuclear charge distribution. The dipole moment associated with this electronic charge distribution gives rise to a long-range dipolar interaction between molecules. This interaction is novel in two senses: it is long range, and the interaction depends on the relative alignment of the constituents. In a sample of aligned dipoles, it is energetically more favorable for two dipoles to sit in front of each other than side by side. To date, all quantum gas experiments have studied effects of interactions which are short-range and isotropic, with the notable exception of experiments performed with ultracold Cr~\cite{ChromiumCondensate} which has a noticeable {\it magnetic} dipole moment. Studies of dipole-dipole interacting quantum gases are expected to reveal novel fascinating physics. For example, in the presence of dipolar interactions, the elementary excitations and the stability of the gas in various dimensions are expected to strongly depend on the dimensionality, the external confinement and the ratio of long range to short range interactions~\cite{Yi2000a,Goral2000a,Santos2000a,Goral2002a,Giovanazzi2002a, Koch2008a}. To name only a few of the predicted effects, quantum hall states of dipolar gases have been considered, both for fermionic and bosonic particles \cite{QuantumHallDipolarFermions,QuantumHallDipolarBosons}, phases of gases with dipolar interactions in optical lattices \cite{Goral2002a} and novel solitonic behavior \cite{SolitonsDipolarCondensates}.

It has been suggested by D.\ DeMille in 2001~\cite{DeMille2002a} to use polar molecules for quantum computation. The idea is to encode quantum information in the orientation of the molecular dipole moment relative to an external field. Using the dipole alignment of a single molecule, a single qubit can be realized. A qubit register, i.\ e.\ multiple qubits can be realized by loading dipolar molecules into a 1D optical lattice. In order to make this quantum information storage register a quantum computer, some degree of coupling needs to be introduced between the qubits. This is where the long range dipolar interaction becomes important: it couples molecules at different lattice sites and is used as the quantum bus. Following the initialization of the quantum register and the proper register operation, the readout of individual qubit states can be achieved by adding an electric field gradient along the direction of the optical lattice to separately address lattice sites through the Stark effect. Whereas trapped atomic ions are currently the most advanced system from an experimental point of view, the suggested quantum computation scheme is being evaluated as a promising candidate for future quantum computers and simulators.

Paramagnetic polar molecules are currently being considered for precision experiments testing upper limits on the permanent electric dipole moment of the electron (EDM). While the standard model sets an upper limit on such an EDM which is vanishingly small (less than $10^{-40}\,e\mathrm{cm}$), supersymmetric theories predict values of the EDM which are getting within reach of current experimental efforts. Improving the upper limit of the EDM would thus eventually enable certain new theories to be ruled out, while finding a non-zero EDM would be an exciting signature of T-violating physics beyond the standard model. It has initially been pointed out by Sandars~\cite{sandars_edm_of_an_atom} that the electronic EDM signature (a linear Stark effect) could be enhanced in heavy paramagnetic atoms, which is expressed by the so-called ``enhancement factor''. Several experiments are under way testing limits on the EDM using e.\ g.\ atomic fountains, atoms confined in solid helium matrices and beam experiments. In almost all atomic experiments, the achievable sensitivity is affected by the electric fields that can be applied without initiating a discharge. In paramagnetic heavy polar molecules, an enormous degree of polarization can be achieved due to the internal electric field with only moderate external fields. Experiments on polar molecules and the EDM are currently under way using e.\ g.\ PbO, TlF, and also molecular ions. Measurements of the EDM with polar molecules would greatly benefit from the availability of ultracold samples with long interrogation times, and the counting rate would be substantially increased by having all molecules in a well defined quantum state.

Actively pursued routes to molecular degeneracy fall into two different categories, depending on whether the starting point is a deeply bound molecular sample (direct, or `molecular' approach) or whether molecules are assembled from precooled and possibly degenerate atoms (atomic or `indirect' approach).

The key obstacle in obtaining ultracold or even degenerate molecular samples compared to atomic Bose-Einstein Condensates or degenerate Fermi gases is the lack of laser cooling techniques for molecular gases. The main issue is the wealth of molecular energy levels, enhanced by rotational and vibrational degrees of freedom and the corresponding lack of clear cycling transitions for laser cooling of molecules. Currently followed routes to molecular degeneracy starting with deeply bound molecules (direct approach) therefore either rely on cryogenic techniques bringing the sample into contact with a thermal reservoir or on interaction with external electric fields slowing down the external degrees of freedom or filtering out low velocity classes. All of these can be followed by subsequent evaporative cooling of a trapped sample, but the temperature gap into the ultracold and even degenerate regime has yet to be bridged. Among the techniques being actively pursued are buffer gas cooling~\cite{BufferGasMolecules},
Stark deceleration~\cite{StarkDeceleration} and velocity filtering~\cite{VelocityFiltering}.

The technical difficulties associated with slowing down the external degrees of freedom of molecules have made another route to molecular degeneracy attractive. This approach is based on ultracold atomic samples which are routinely produced in laboratories today. The goal is to assemble ultracold molecules in both their external and internal (rovibrational) ground state from ultracold atoms by means of a coherent photoassociation step~\cite{Peer2007a,Sage2005a}.

The bottleneck in these photoassociation schemes will be the small Franck-Condon overlap between free atoms and deeply bound molecular states~\cite{Sage2005a, Wang2004a}. At this points, heteronuclear Feshbach molecule formation as demonstrated for the first time in this work~\cite{Ospelkaus2006d} is a key step in obtaining a suitable initial state for coherent Raman photoassociation schemes. The Feshbach association provides us with a well-defined initial molecular state with known molecular distance and a much better Franck-Condon overlap with deeply bound molecular states.

Recent years have seen a lot of progress in molecule formation based on Feshbach resonances. Molecule creation at Feshbach resonances is based on the fact that the occurrence of the resonance itself is due to the coupling between a bound molecular state energetically close to the incoming energy of two free atoms.

When the energy of the bound molecular state is above the free-atom threshold, the coupling between the two close to degenerate states gives rise to strong attractive interactions between atoms. In free space, no two-body bound state exists near threshold (see fig.~\ref{fig:feshbach_resonance} right).

When the energy of the bound molecular state is below the open channel, the interaction between atoms is large and repulsive, and pairs of atoms can occupy the two-body bound state and form a weakly bound molecule with typical binding energies between a few 10~kHz and a few MHz. Starting from an atomic sample, such molecules can e.\ g.\ be created by sweeping a magnetic field from attractive interactions through the resonance center position over to repulsive interactions (see fig.~\ref{fig:feshbach_resonance} left).

In a groundbreaking experiment performed at JILA~\cite{Donley2002a}, this magnetic field sweep technique was first used to create a coherent atom-molecule superposition. In this and following experiments with bosonic atoms and Feshbach resonances, it was observed that the resulting molecular lifetime was rather low and on the order of about 1\,ms, limited by collisional losses. A big step forward in Feshbach chemistry was achieved when it was realized that bosonic molecules created from two fermionic atoms using the magnetic field sweep technique would be collisionally stable due to the Pauli exclusion principle~\cite{petrov_dimer_stability}. The experimental demonstration of long-lived bosonic molecules created from fermionic atoms was soon to follow \cite{Regal2003a,Cubizolles2003a,jila_atom_molecule_lifetime,Strecker2003a,Jochim2003a}. 

Parallel to this development, it was realized that the lifetime limitation of molecules created from bosonic atoms \cite{Donley2002a,stanford_cesium_molecules,Xu2003a,Herbig2003a,Duerr2004a} can be overcome in 3-dimensional lattices by creating molecules from atom pairs in isolated wells of the lattice and thus inhibiting inelastic collisions \cite{Thalhammer2006a,mpq_lattice_molecules}. Creation of molecules in optical lattices highlighted the role of the external trapping potential in molecule formation. A model system of two particles in a harmonic trap was already considered in 1997~\cite{TwoColdAtoms}, and in 2005 experiments with optical lattices demonstrated that the existence of the external trapping potential shifts the free-atom threshold, resulting in the existence of two-body bound states on the attractive side of the Feshbach resonance where no stable molecules exist in free space~\cite{eth_confinement_molecules}, also seen later in the behavior of repulsively interacting pairs~\cite{innsbruck_repulsive_pairs}.

Today, many people consider molecule creation through magnetic field sweeps as {\it the} way to create molecules at Feshbach resonances. In recent years, however, a number of techniques has been used to create Feshbach molecules. For example, it seems to be a unique property of $^{6}$Li that molecule formation can be achieved by evaporative cooling of thermal lithium atoms on the repulsive side of the Feshbach resonance without any magnetic field sweeps \cite{Zwierlein2003a}. Another technique which has been demonstrated is magnetic field modulation at a Feshbach resonance~\cite{jila_bcs_bec_excitation_spectrum}. When the modulation frequency corresponds to the binding energy of the bound molecular state at the time average value of the magnetic field, free atoms can be converted into molecules on the repulsive side of a Feshbach resonance. Another possibility of creating molecules is rf association of atoms into molecules as developed within this work.

\subsection*{Ultracold heteronuclear Feshbach molecules}

While {\it heteronuclear} Feshbach resonances had already been predicted in 2003 \cite{krb_feshbach_prediction} (for \pot---\rub) and experimentally identified through collisional losses in the middle of 2004 \cite{Inouye2004a,Stan2004a}, all of the fascinating physics which has become accessible through Feshbach resonances has been limited to homonuclear systems. As a crucial step towards the exploration of dipole-dipole interacting molecular systems, novel quantum computation schemes and fundamental measurements, this work~\cite{Ospelkaus2006d} presents the first experimental demonstration of ultracold long-lived heteronuclear molecules. Molecules are created within the single well of an optical lattice at a heteronuclear Feshbach resonance by means of a novel technique developed for this purpose and based on rf spectroscopy~\cite{jila_rf_spectroscopy}. Lifetime and molecule creation efficiency have been measured and are consistent with a physical picture of lattice occupation. Both molecules stable in the absence of any external potential, attractively interacting atom pairs and repulsively interacting pairs with a positive ``binding energy'' have been identified. A detailed understanding of this model system has been developed based on a pseudopotential approach for the atomic interaction in cooperation with F.\ Deuretzbacher, K.\ Plassmeier, F.\ Werner, D.\ Pfannkuche and F.\ Werner~\cite{two_particles_hh}. The model consistently treats both the anharmonicity of the lattice potential and the general case of particles with unequal trapping frequencies; excellent agreement with numerical simulations has been obtained.

As discussed above, various techniques have been employed for molecule production at Feshbach resonances in the literature. Within this work, rf association in an optical lattice as a novel technique for molecule production has been used. The idea of this association scheme is to start with a non Feshbach-resonant mixture in one spin state. In another spin state, a Feshbach resonance occurs at the same magnetic field, and the difference between this state's energy and the non-interacting limit in the same spin state is precisely the binding energy. By shining in an rf photon which provides the undisturbed energy corresponding to the spin flip plus the additional interaction or binding energy, we can populate the two-body bound molecular state (see inset in fig.~\ref{fig:sample_rf_spectrum}). In the presence of an optical lattice, molecule production through rf association provides us with a precise measurement of the binding energy at the same time, thus extracting a maximum of information about the energy spectrum. Rf association allows molecule production without rapid magnetic field sweeps and uncertainties associated with magnetic field settling, thereby removing technical complications which can make a measurement of the binding energy quite cumbersome.

\begin{figure}
  \centering
  \includegraphics[width=0.6\columnwidth]{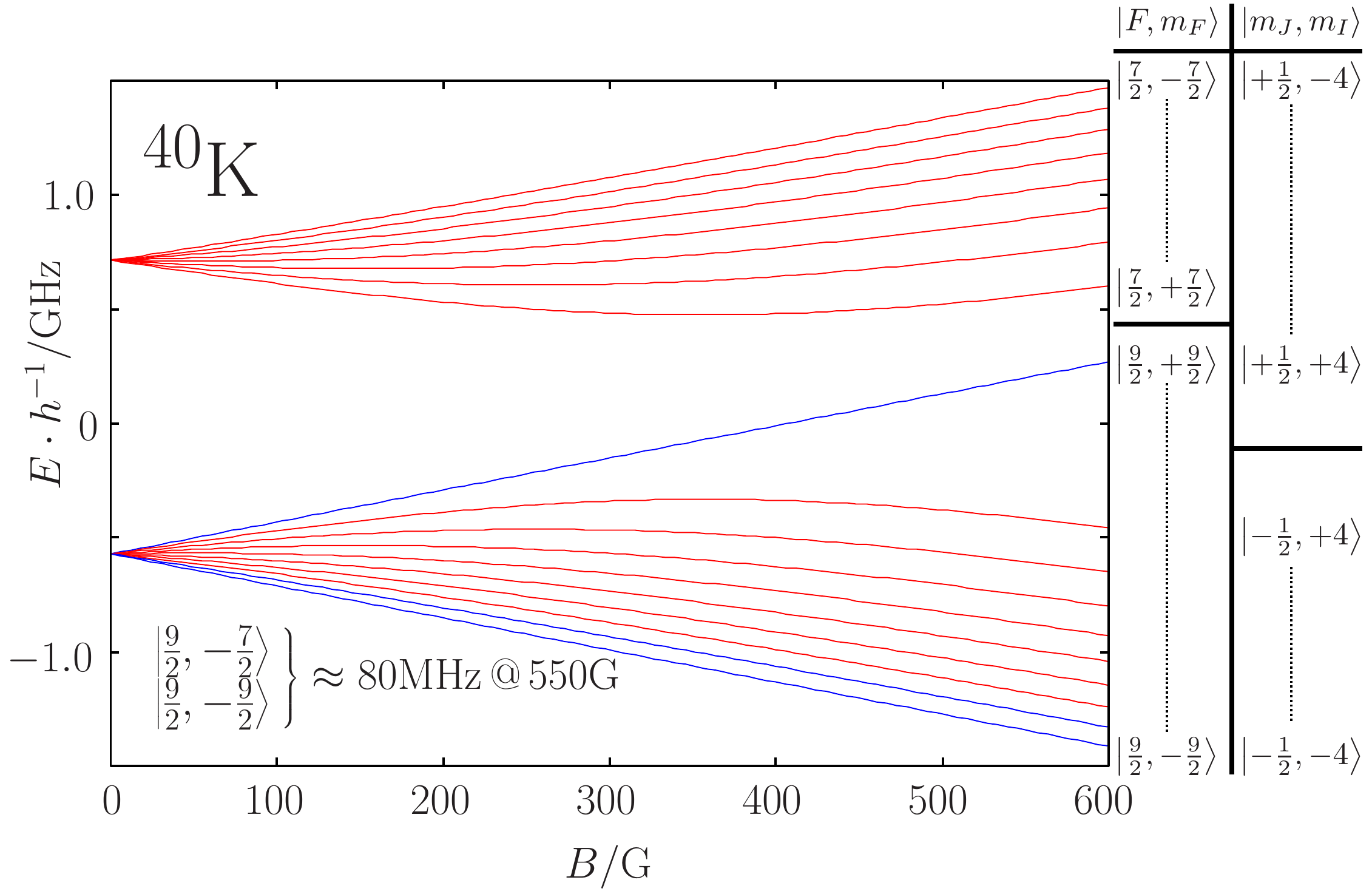}
  \caption{\pot\ $4^2S_{1/2}$ energy levels in the presence of an external magnetic field. The magnetically trapped $\left|\frac92,\frac92\right>$ state and the $\left|\frac92,-\frac72\right>$ / $\left|\frac92,-\frac92\right>$ states used for rf spectroscopy are highlighted in blue.}
  \label{fig:br_k40}
\end{figure}

In order to understand rf spectroscopy as used for molecule creation in this work, let us first look at the energy levels of \rub\ and \pot\ in the presence of a homogeneous external magnetic field. For $J=I\pm1/2$, the well-known Breit-Rabi formula yields the energy levels in the presence of an external magnetic field. Figure~\ref{fig:br_k40} shows energy levels of \pot\ for fields of experimental interest. Due to the low hyperfine splitting, relatively low fields are required to drive \pot\ into the Paschen-Back regime (compared to, for example, \rub). The resulting magnetic field insensitivity of transitions between neighboring levels at experimental fields is one of the reasons for the success of rf spectroscopy in this system. States are labelled both by their low-field ($\vec{F}=\vec{I}+\vec{J}$) limit $F$ and $m_F$ quantum numbers and by the independent $m_I$ and $m_J$ quantum numbers in the Paschen-Back regime.

The Feshbach resonance that was used for molecule production in this work occurs between \rub\ atoms in the $\left|1,1\right>$ and \pot\ atoms in the $\left|9/2,-9/2\right>$ absolute ground states at a magnetic field of 547\,G. The idea for determining the energy spectrum of the $\left|9/2,-9/2\right>\otimes\left|1,1\right>$ state is to drive a transition between the \pot\ $\left|9/2,-7/2\right>$ state and the Feshbach-resonant $\left|9/2,-9/2\right>$ state. The former features a scattering length which is independent of $B$ over the magnetic field range studied in the experiment and small (given approximately by the background scattering length)\footnote{The closest known heteronuclear Feshbach resonances in this state are located at 522\,G and 584\,G~\cite{Ferlaino2006b} and have theoretical widths of below 1\,G.}; the energy of the latter strongly varies as a function of $B$ across the resonance compared to the undisturbed $\left|9/2,-9/2\right>$ energy level. The interaction energy (binding energy) of the Feshbach-resonant state $\left|9/2,-9/2\right>$ is thus given as the difference of the observed $\left|9/2,-7/2\right>$ $\rightarrow$ $\left|9/2,-9/2\right>$ transition frequency from the undisturbed transition frequency which can be calculated from the Breit-Rabi formula or extracted from a measurement where no bosons causing the collisional shift are present (see inset in fig.~\ref{fig:sample_rf_spectrum}). Note that this measurement of the collisional shift is up to the small constant offset caused by the non-resonant $\left|9/2,-7/2\right>\otimes\left|1,1\right>$ interactions.

\subsection{Experimental protocol}
The experimental implementation of this scheme is as follows: A quantum degenerate mixture of \pot\ and \rub\ in the ``magic'' optical dipole trap is prepared in the non Feshbach-resonant $\left|9/2,-7/2\right>\otimes\left|1,1\right>$ state using the exact same procedure as in our experiments on tuning of interactions. The magnetic field is set to a value close to the Feshbach resonance occurring at 547\,G. In total, we allow 150\,ms for the field to stabilize close to its final value. During the last 100\,ms, the 3D optical lattice intensity is increased from zero to its final value using a linear ramp. At magnetic field values of 547\,G, the \pot\ $\left|9/2,-7/2\right>$ $\rightarrow$ $\left|9/2,-9/2\right>$ transition occurs at about 80\,MHz. In order to probe the energy spectrum for a given magnetic field, an rf pulse with a Gaussian amplitude envelope ($1/e^2$ full linewidth of 400\,$\mu$s) and total duration of 800~$\mu$s is irradiated using the standard evaporation antenna\footnote{The pulse shape is chosen such as to avoid the side lobes in spectroscopy that usually result from probing a two-level system with square pulses. A detailed discussion of the observed pulse shapes can be found in~\cite{OspelkausC2006a}.}. For a description of the versatile rf/microwave manipulation system used to generate these pulses, see~\ref{sec:rf_setup} and~\cite{OspelkausC2006a,SuccoThesis}.

About 1\,ms after the end of the pulse, the lattice intensity is ramped down using a linear ramp within 1\,ms in order to reduce the kinetic energy, and the dipole trap is then suddenly switched off, marking the beginning of the time of flight. During time of flight, we detect both remaining \pot\ $\left|9/2,-7/2\right>$ atoms and \rub\ $\left|1,1\right>$ atoms as well as weakly bound molecules and \pot\ atoms in the $\left|9/2,-9/2\right>$ channel.

\begin{figure}
  \centering
  \includegraphics[width=0.7\columnwidth]{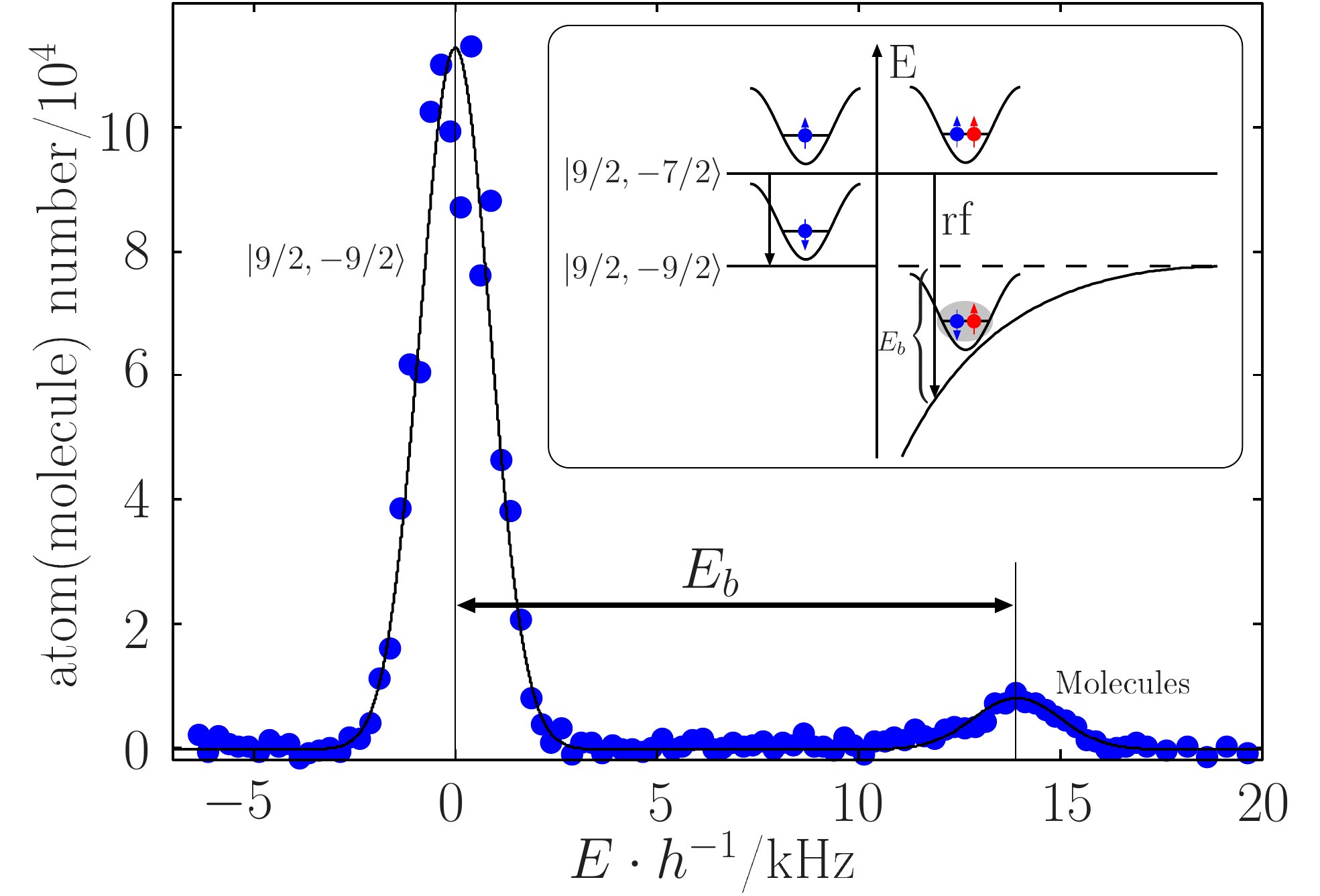}
  \caption{Rf spectroscopy of \pot\---\rub\ at 547.13\,G in a 40.5\,$E_r^{\mathrm{Rb}}$ deep optical lattice~\cite{Ospelkaus2006d}. {\it Copyright (2006) by the American Physical Society.}}
  \label{fig:sample_rf_spectrum}
\end{figure}

At fixed magnetic field, the rf pulse frequency is varied in the vicinity of the undisturbed \pot\ $\left|9/2,-7/2\right>$ $\rightarrow$ $\left|9/2,-9/2\right>$ rf transition to obtain rf spectra as in fig.~\ref{fig:sample_rf_spectrum} for $B$=547.13\,G. The spectrum shows the number of \pot\ atoms in the $\left|9/2,-9/2\right>$ state and weakly bound Feshbach molecules as a function of rf detuning from the pure \pot\ $\left|9/2,-7/2\right>$ $\rightarrow$ $\left|9/2,-9/2\right>$ transition at the set value of the magnetic field. The large peak at zero detuning occurs whether bosons are present in the optical lattice or not. It stems from lattice sites where only one fermion and no boson is present and is used as a precise magnetic field calibration in connection with the Breit-Rabi formula\footnote{Note that the observed $1/e^2$ half linewidth of 1.7\,kHz is in excellent agreement with what we expect based on our Gaussian rf pulse~\cite{OspelkausC2006a}.}.

The second peak occurring about 14\,kHz above the undisturbed (atomic) peak is due to the existence of a Feshbach-resonant shift at lattice sites where one boson and one fermion are present. The strong interaction between the two atoms at the Feshbach resonance introduces a strong differential energy shift between the $\left|9/2,-7/2\right>$ and the $\left|9/2,-9/2\right>$ levels. In this case, additional energy is required to drive the rf transition. This implies that in the presence of \rub, the energy of the target state $\left|9/2,-9/2\right>$ is lower than in the reference level, which shows that the target state has a negative binding energy and that heteronuclear molecules have been formed through rf association. Forming heteronuclear molecules through rf association thus measures the binding energy of these molecules at the same time (about 14\,kHz in this case). In the particular case of fig.~\ref{fig:sample_rf_spectrum} recorded at a magnetic field of 547.13\,G, the interaction between \pot\ and \rub\ is attractive which, in free space, would not allow molecules to be created because the bound molecular state is above threshold. As we shall see, the negative binding energy of the molecules in fig.~\ref{fig:sample_rf_spectrum} is due to the presence of the optical lattice which admits the presence of a two-body bound state even for attractive interactions. The resulting  pair close to threshold will decay into free particles when the lattice potential is removed. In order to avoid potential confusion, we will  use the term `attractively interacting atoms' for this type of atom pair in the following discussion.

\subsection{Energy spectrum}

\begin{figure}
  \centering
  \includegraphics[width=0.6\columnwidth]{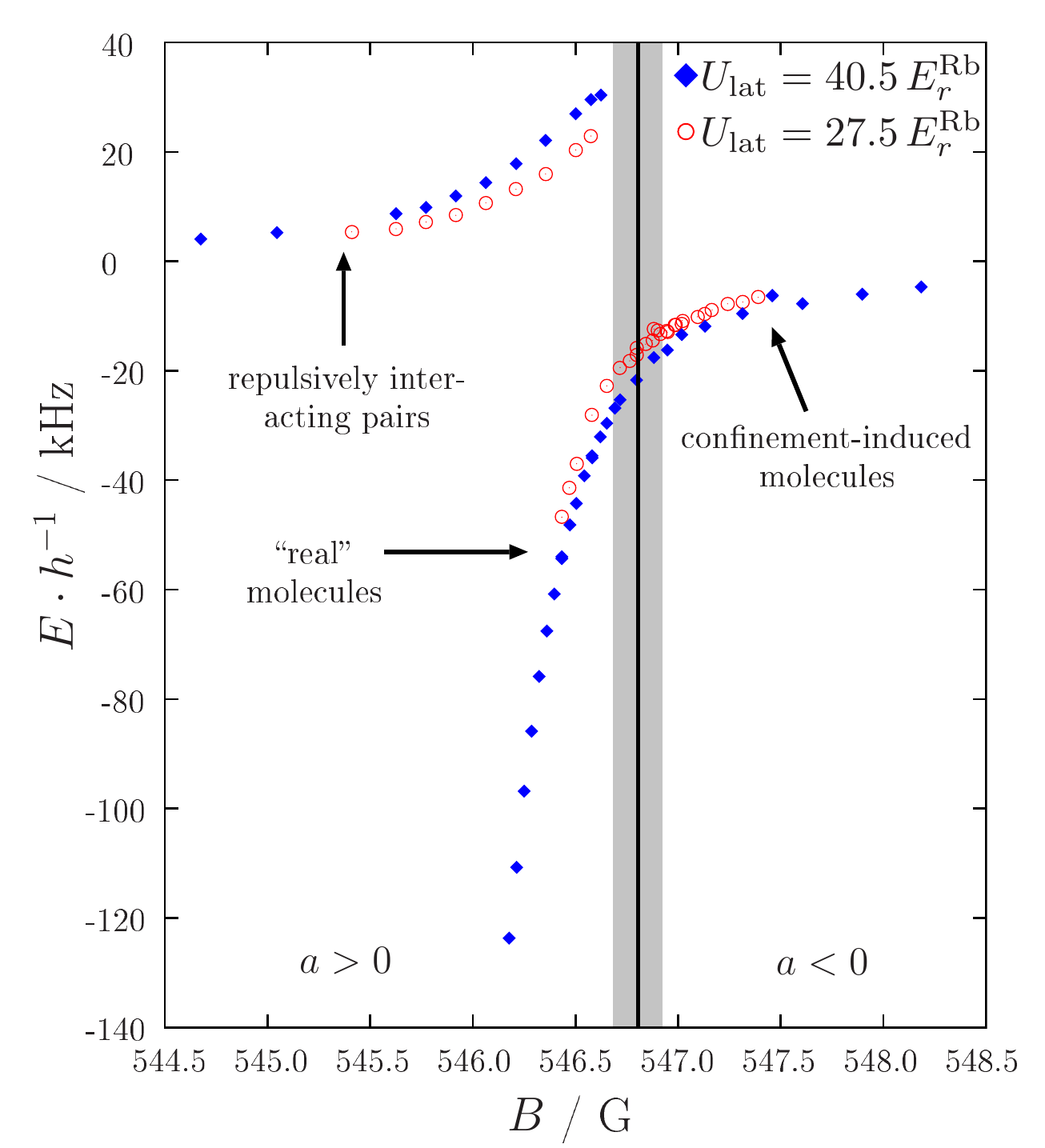}
  \caption{Observed energy spectrum of \pot\ \rub\ heteronuclear molecules and atom pairs at an individual lattice site for two different values of the lattice depth~\cite{Ospelkaus2006d}. {\it Copyright (2006) by the American Physical Society.}}
  \label{fig:binding_energy}
\end{figure}

In order to extract information about the energy spectrum over the whole magnetic field range of the Feshbach resonance, spectra as in fig.~\ref{fig:sample_rf_spectrum} have been recorded for a number of magnetic fields around 547\,G. For each of the spectra, the molecular binding energy has been extracted by fitting both atomic and molecular peaks and taking the difference in position between the atomic and the molecular peak. The resulting spectrum is shown in fig.~\ref{fig:binding_energy}, recorded for two different values of the lattice depth, 40.5\,$E_r^{\mathrm{Rb}}$  and 27.5\,$E_r^{\mathrm{Rb}}$.

The vertical bar in fig.~\ref{fig:binding_energy} represents the center position $B_0$ of the Feshbach resonance as determined from \pot-\rub\ interactions in an optical dipole trap (see our discussion on tuning of interactions). In the magnetic field range considered here, the interaction is attractive above $B_0$ and repulsive below $B_0$.

The energy spectrum shows two basic branches. The lower branch or molecular branch is characterized by the presence of a negative binding energy, and the upper branch, which is only observed for repulsive interactions, is characterized by a positive ``binding energy''. In a system of pure bosons confined in an optical lattice, the behavior of these repulsively interacting pairs has been studied in recent experiments in Innsbruck~\cite{innsbruck_repulsive_pairs}.

In the molecular branch, one can distinguish between attractive and repulsive interactions. Above $B_0$, we produce attractively interacting atom pairs~\cite{eth_confinement_molecules,Stoeferle2006a} which exhibit a negative binding energy due to the presence of the lattice potential. The sample rf spectrum in fig.~\ref{fig:sample_rf_spectrum} has been recorded in this regime. As the resonance center is crossed, there is a smooth transition into ``real'' molecules which are stable even in the absence of an external potential. As the scattering length decreases from infinity (at the resonance) towards zero, these molecules become  more and more deeply bound. The maximum binding energy observed in the experiment is about -130\,kHz.

Thus, using rf association, we have created heteronuclear molecules for the first time. The rf association process allows us to precisely determine the binding energy of the molecules while we create them. A key advantage of this scheme is that it provides a lot of information about the system without any potentially cumbersome magnetic field ramps. 

Several aspects of this experiment have been crucial in obtaining these results. One is that when the rf is detuned from any resonance, the signal recorded in the target state $\left|9/2,-9/2\right>$ is effectively zero. So the molecule signal which we measure in that channel is a signal {\it on zero}, and not a small dip in some fluctuating atom number. This was particularly helpful in the first stages of the experiment, where the molecule fraction was very small due to a small number of lattice sites occupied by one \rub\ and one \pot\ atom.

A second aspect is imaging of the atoms and molecules. While in principle we could have used the Stern-Gerlach technique for state-selective detection in \pot, this technique limits the time of flight to a certain minimum value given by the necessity to apply the field gradient between release from the trap and the actual imaging flash. This minimum time of flight can make it difficult to detect small numbers of molecules, in particular when in the optimization phase of the experiment, since after a large time of flight, the gas becomes very dilute. More importantly, with Stern-Gerlach imaging, the molecules cannot be imaged directly and need to be dissociated prior to imaging.

Instead, we have chosen high field imaging in order to state-selectively image both \pot\ atoms and molecules. High field imaging relies on the fact that at the high magnetic field close to the Feshbach resonance where the \pot\ atoms are mostly in the Paschen-Back regime (Hyperfine structure of the Zeeman effect), it is possible to construct a cycling transition starting from any arbitrary magnetic sublevel of the ground state. In this case, the total shell angular momentum $J$, the nuclear spin $I$ and the corresponding $z$ projections $m_J$ and $m_I$ are good quantum numbers, and we have cycling transitions ($m_J=\pm 1/2$; $m_I=-I\ldots+I$) $\rightarrow$ ($m_J=\pm 3/2$; $m_I=-I\ldots+I$). For the case of weakly bound Feshbach molecules, one photon of the detection pulse will dissociate the molecule, and then subsequent scattering processes detect atoms in the corresponding spin state. This detection process is spin dependent because in general, the various optical cycling transitions for the different spin states differ by several 10\,MHz at least. The particular scheme used in our experiments is described in~\ref{sec:imaging}.

Crucial to rf spectroscopy and rf association as discussed in this work is the influence of environmental magnetic field noise and magnetic field drifts in combination with the magnetic field sensitivity of the rf transition used in the experiment. At the magnetic fields of interest for this work (547\,G), the rf transition in \pot\ used here has a small magnetic field sensitivity of 67\,kHz/G. This is mainly a result of \pot\ being close to the Paschen-Back regime. The other key to the excellent quality of the rf spectra is suppression of magnetic field noise and drifts. By eliminating sources of AC magnetic noise, line triggering the experiment and using a custom built precision regulator, we have achieved a magnetic field reproducibility of 2.7\,mG at 547\,G (57 measurements on 11 consecutive days, as seen from the atomic peak in rf spectroscopy), corresponding to a magnetic field reproducibility of $\approx5\cdot 10^{-6}$.

\subsection{Properties of heteronuclear Feshbach molecules in 3D optical lattices}

After discussing creation of heteronuclear molecules, let us look at some basic properties of these weakly bound dimers, such as a quantitative understanding of the energy spectrum, and the lifetime and creation efficiency of the molecules. These aspects are particularly relevant in view of possible applications, such as optical transfer into more deeply bound molecular states and ultimately the creation of a polar molecular quantum gas. The quantitative analysis presented here has been the result of a collaboration with F.\ Deuretzbacher, K.\ Plassmeier and D.\ Pfannkuche at Hamburg (1.\ Institut f\"ur Theoretische Physik) and F.\ Werner (ENS, Paris)~\cite{two_particles_hh}.

The simplest possible model for our system is that of two atoms, one \pot\ and one \rub, interacting via $s$-wave scattering where the scattering length as a function of magnetic field is given by equation~(\ref{eq:scattering_length_at_Feshbach}), at an individual lattice site with the shape of the lattice potential approximated by a harmonic oscillator. This basic model has been brought forward by Busch and coworkers in 1998~\cite{TwoColdAtoms}. In their article, the authors specifically consider interaction in the form of a regularized $\delta$-potential between the two atoms and analyze the special case of two {\it identical} atoms with equal masses and trap frequencies. The corresponding Hamiltonian is shown to have an exact solution.

In our case, there are several important differences from this simple model. The most obvious one is that we deal with a system where the two particles are not identical. The second is that for experimentally relevant lattice depths, the harmonic approximation for the lattice potential is only a very rough approximation --- the anharmonicities are significant, as we shall see. Another aspect, which we will not be concerned with here, are deviations from the regularized $\delta$ potential in combination with equation~(\ref{eq:scattering_length_at_Feshbach}). These deviations will be relevant for larger binding energies (see e.\ g.~\cite{Zirbel2008b}), but not for the regime of binding energies considered here.

Our model Hamiltonian thus consists of the kinetic energy of the two atoms, of the trapping potential and the interaction potential:
\begin{eqnarray*}
H & = & -\frac{\hbar^2}{2m_1}\Delta_1 - \frac{\hbar^2}{2m_2}\Delta_2 \\
  &   & +\frac{1}{2}m_1\omega_1^2r_1^2 + \frac{1}{2}m_2\omega_2^2r_2^2 \\
  &   & +\frac{2\pi\hbar^2a}{\mu}\delta(\vec r)\frac{\partial}{\partial r}r
        +V_{anh}(\vec r_1, \vec r_2)
\end{eqnarray*}
where $a$ is the s-wave scattering length between the two atoms, $m_{1,2}$ the respective mass, $\mu=m_1\cdot m_2/(m_1+m_2)$ the reduced mass, $\vec r_{1,2}$ the respective single particle coordinate, $\omega_{1,2}$ the harmonic trap frequency for atom 1 and 2 and $V_{anh}$ contains anharmonic correction to the external trapping potential. Reference \cite{TwoColdAtoms} proceeds by introducing relative and center of mass coordinates  $\vec r=\vec r_1-\vec r_2$ and $\vec R=\frac{m_1\vec r_1+m_2\vec r_2}{M}$ where $M=m_1+m_2$ is the total mass. By introducing frequencies 
\begin{eqnarray*}
\omega_{CM} & := & \sqrt{\frac{m_1\omega_1^2+m_2\omega_2^2}{M}}\\
\omega_{rel} & := & \sqrt{\frac{m_2\omega_1^2+m_1\omega_2^2}{M}}\\
\Delta\omega &:= & \sqrt{\omega_1^2+\omega_2^2}\quad,
\end{eqnarray*}
the resulting Hamiltonian takes the form
\begin{eqnarray}
\label{eq:comrelham}
H &
= &
-\frac{\hbar^2}{2M}\Delta_{CM}
+\frac{1}{2}M\omega_{CM}^2R^2\\
\nonumber
&
&
-\frac{\hbar^2}{2\mu}\Delta_{rel}
+\frac{1}{2}M\omega_{rel}^2r^2
+\frac{2\pi\hbar^2a}{\mu}\delta(r)\frac{\partial}{\partial r}r\\
\nonumber
&
&
+\mu\Delta\omega^2\vec R\cdot\vec r
+V_{anh}(\vec R, \vec r)\\
\nonumber
&
= &
H_{CM}+H_{rel}+H_{co}
\end{eqnarray}
with one part $H_{CM}$ involving only center of mass coordinates, one term $H_{rel}$ containing only relative coordinates and a final term $H_{co}$ potential coupling the two coordinates. The crucial point here is that in the absence of anharmonicities and for equal trap frequencies, the coupling term vanishes and the center of mass and relative motion completely decouple, allowing the identification of exact solutions. This is the special case treated by ref.~\cite{TwoColdAtoms}.

This decoupling is absent in our general case of unidentical atoms trapped in an anharmonic lattice well and prevents us from directly using these exact solutions. Instead, given ref.~\cite{TwoColdAtoms}, we can write down exact solutions for the first two groups of terms $H_{CM}$ and $H_{rel}$. Using a numerical procedure, the coupling term $H_{co}$ is then exactly diagonalized in the basis set given by the eigenfunctions of the first two groups of terms, and we can obtain both wave functions and energy structure of the full Hamiltonian.

\begin{figure}
  \raggedleft
    \begin{minipage}{0mm}{(a)}\end{minipage}%
    \begin{minipage}[t]{70.5mm}%
      \includegraphics[width=1.0\columnwidth]{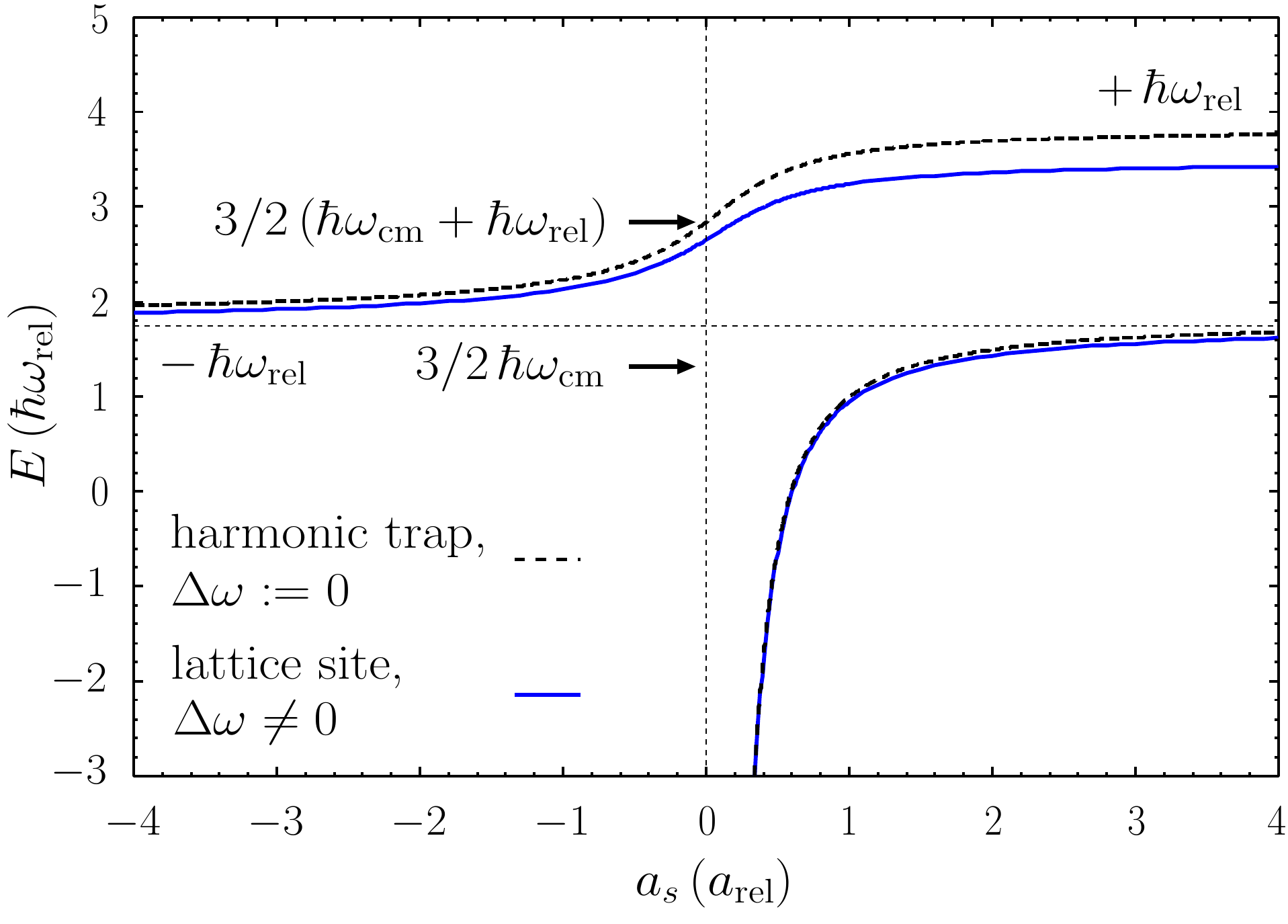}%
    \end{minipage}%
    \begin{minipage}[t]{0mm}{(b)}\end{minipage}%
    \begin{minipage}[t]{75mm}%
      \includegraphics[width=1.0\columnwidth]{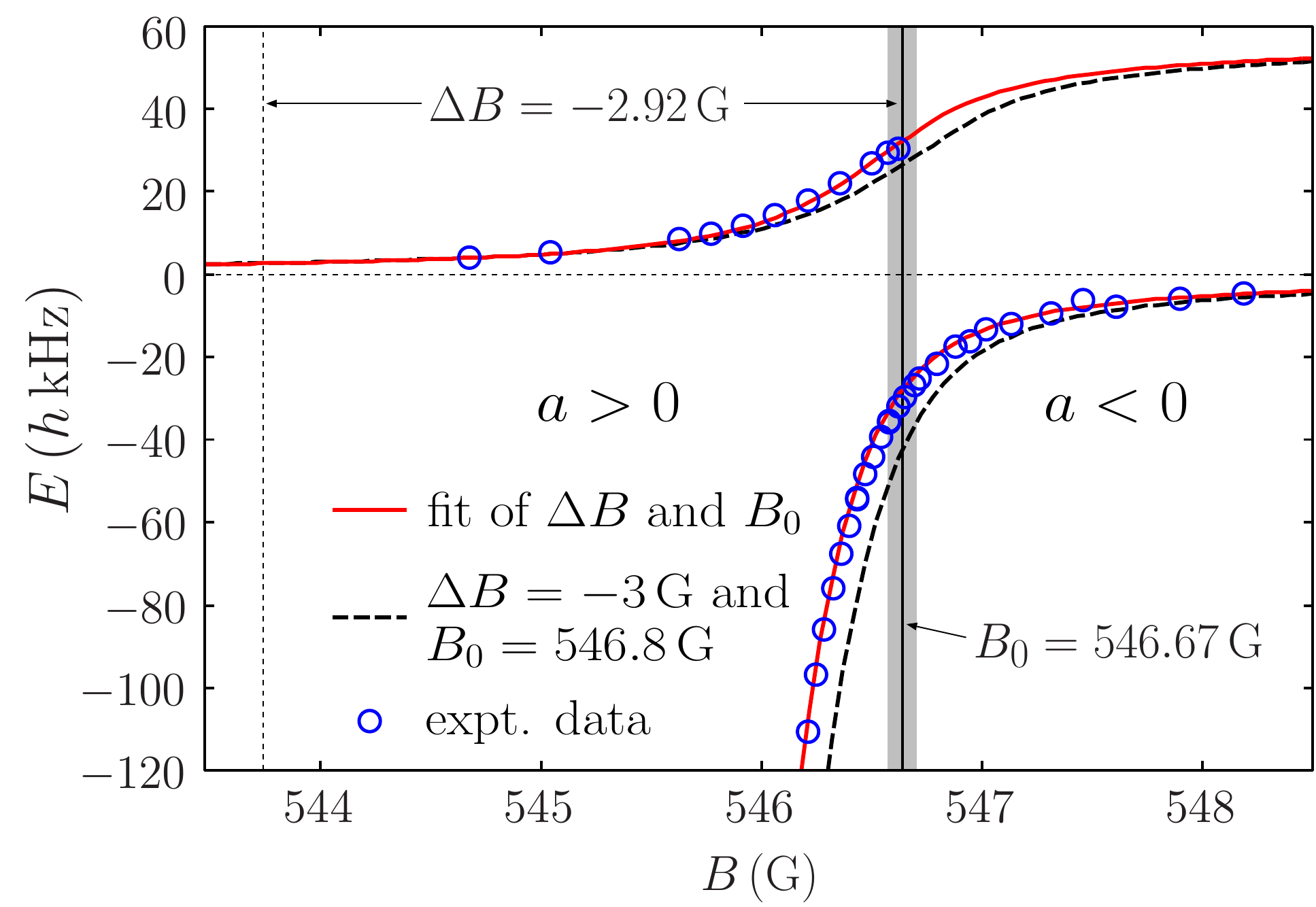}%
    \end{minipage}%
  \caption{ (a) Energy spectrum of a heteronuclear system (\pot--\rub) for the pseudopotential model at a lattice depth of 40.5\,$E_r^{\mathrm{Rb}}$~\cite{two_particles_hh}.
            (b) Experimentally measured binding energy (dots) compared to (dashed line) theory with literature parameters for width $w$ and center position $B_0$ of the Feshbach resonance and (red solid line) fitted curve resulting in $B_0=546.669(24)_{syst}(2)_{stat}$\,G~\cite{two_particles_hh}.
            {\it Copyright (2008) by the American Physical Society.}
          }
  \label{fig:energy_spectrum_pseudo}
\end{figure}

Figure~\ref{fig:energy_spectrum_pseudo}(a) illustrates the results for the experimentally relevant case of a lattice depth of $40.5\,E_r^{\mathrm{Rb}}$. The blue solid line shows the solution to the full Hamiltonian as obtained by the exact diagonalization procedure, while the dashed line shows the energy spectrum resulting from the non-coupling terms of ref.~\cite{TwoColdAtoms} alone. The difference between the two curves is most pronounced in the repulsively interacting atom branch of the spectrum, which is a result of the relative wave function extending over a significant part of the external potential, and the coupling between center of mass and relative motion is important. As the molecule size becomes smaller and smaller for larger negative binding energies, the coupling term $H_{co}$ becomes less important.

Figure~\ref{fig:energy_spectrum_pseudo}(b) compares the solution to the full coupled problem to the experimentally measured binding energies. As can be seen, there is a good agreement between our experimentally measured binding energy and theory; however, there seems to be a slight shift between experiment and theory which is most likely due to insufficient prior knowledge in the determination of the parameters of eq.~(\ref{eq:scattering_length_at_Feshbach}) connecting the scattering length and the magnetic field. By fitting the theoretical result for the binding energy to the experimental data (red solid line), we can thus obtain a new value for the Feshbach resonance center position $B_0=546.669(24)_{syst}(2)_{stat}$\,G. Details of the fitting procedure and the error analysis can be found in ref.~\cite{two_particles_hh}.

In view of possible applications, the lifetime of weakly bound Feshbach dimers has been a particular concern since their first demonstration~\cite{Donley2002a}. In general, these molecules are very fragile objects due to their highly excited internal state, leading to large collisional losses, possibly both molecule-molecule and atom-molecule collisions. Initial experiments with molecules created from bosonic atoms~\cite{Donley2002a,stanford_cesium_molecules,Xu2003a,Herbig2003a,Duerr2004a} correspondingly showed a very short lifetime, meaning that degeneracy could be achieved, but not thermal equilibrium. This limitation has been overcome in the presence of a deep 3D optical lattice~\cite{Thalhammer2006a,mpq_lattice_molecules}. Molecules created from two fermionic atoms have shown a long lifetime close to Feshbach resonances due to Pauli-forbidden inelastic decay~\cite{petrov_dimer_stability}. For heteronuclear molecules composed of a bosonic and a fermionic atom, the situation is somewhat more complicated; suppression of collisions due to their fermionic character is expected for more deeply bound molecules~\cite{Stan2004a}. In addition, collisions with the remaining free atoms are expected to strongly limit the lifetime of the molecules, and this effect will tend to become more important the deeper the molecule is bound.

\begin{figure}
  \raggedleft
  \begin{minipage}[t]{5mm}{(a)}\end{minipage}%
  \begin{minipage}[t]{70mm}%
    \includegraphics[width=1.0\columnwidth]{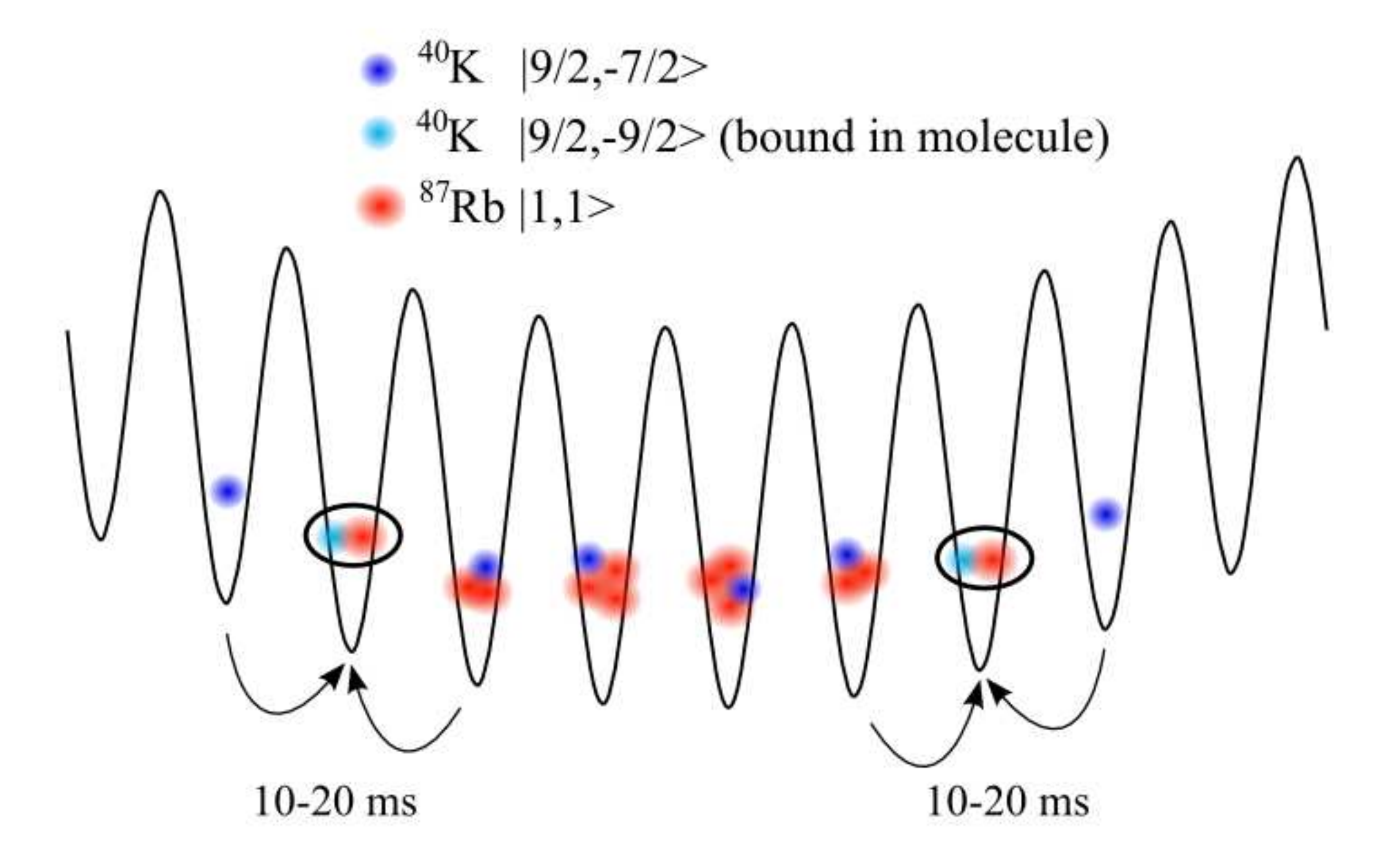}%
  \end{minipage}%
  \begin{minipage}[t]{0mm}{(b)}\end{minipage}%
  \begin{minipage}[t]{75mm}%
    \includegraphics[width=1.0\columnwidth]{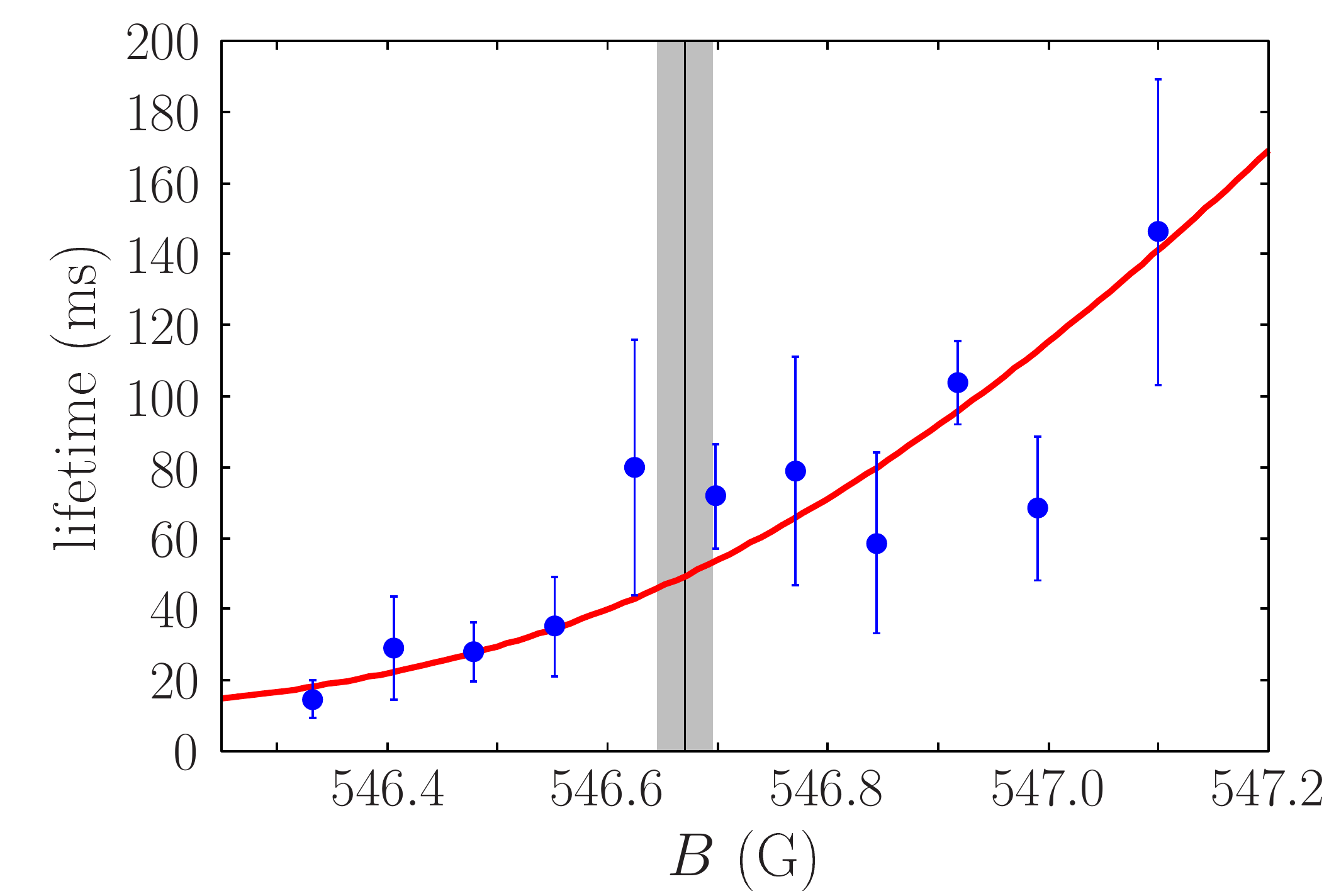}%
  \end{minipage}%
  \caption{ (a) Cartoon picture of expected lattice occupation~\cite{two_particles_hh}.
            (b) Lifetime of the molecular sample as a function of magnetic field~\cite{two_particles_hh}.
            {\it Copyright (2008) by the American Physical Society.}
          }
  \label{fig:molecule_lifetime}
\end{figure}

In the experiment described here, heteronuclear molecules are produced within 3D optical lattices where the large tunneling time strongly reduces inelastic collisional losses. Figure~\ref{fig:molecule_lifetime}(b) shows a measurement of the lifetime of the heteronuclear molecular sample in an optical lattice with a depth of 40.5\,$E_r^{\mathrm{Rb}}$ as a function of magnetic field. The lifetimes are the $1/e$ times obtained from exponential fits to the molecule number decay curves. We find a lifetime of about 120\,ms for weakly bound attractively interacting atom pairs. In the vicinity of the resonance, the lifetime is about 80\,ms, and drops to 20 to 40\,ms for more deeply bound molecules.

So far, we have excluded tunneling between neighboring lattice sites from our discussion. Tunneling is not relevant for the short timescales involved in the molecule creation process. The presence of the optical lattice results in a very long tunneling time between 10 and 20\,ms for \pot. The tunneling time for \rub\ is larger as a result of the larger mass. Ultimately, however, tunneling becomes important, as we shall see. For a cartoon picture of our expected lattice occupation, see fig.~\ref{fig:molecule_lifetime}(a). In the optical lattice, prior to molecule creation, we expect a global occupation of at most one \pot\ atom in the $F=9/2,m_F=-7/2$ state per lattice site (Pauli exclusion principle). As far as \rub\ is concerned, we expect a central occupation number of three to five bosons per lattice site, dropping to one in the outer regions of the optical lattice. It is in this outer shell, at lattice sites with one \pot\ and one \rub\ atom, that the rf association creates molecules. Once the molecular association has happened, the sites with molecules are surrounded by other sites with atoms that are quantum-mechanically distinguishable from the molecules (\pot\ in the $F=9/2,m_F=-7/2$ state and \rub). There is some rate for tunneling of these remaining atoms to the lattice sites occupied by the molecules. Once the tunneling process has happened, the molecule and the atom can undergo three-body recombination, and the probability for this process can be determined from our simple theoretical model. Following Petrov {\it et al.}~\cite{petrov_dimer_stability}, the decay probability (after tunneling has happened) can be calculated as the probability of finding all three particles (the two constituents of the molecule, plus the atom that has tunneled to the molecular site) in a very small volume. For the tunneling atom, we use the wave function of a non-interacting \pot\ atom trapped at a lattice site. For the molecule, we use the wave function obtained from our solution to the coupled Hamiltonian~(\ref{eq:comrelham}). The resulting theory curve, scaled by a global factor to allow comparison to the experiment, is shown in fig.~\ref{fig:molecule_lifetime}(a) and is in good agreement with the experimentally observed magnetic field dependence of the lifetime. Ultimately, this means that while the molecular lifetime in the optical lattice is already excellent for applications (such as transfer to lower lying molecular states and ultimately the absolute ground state), the lifetime can be further increased by removing any unpaired atoms from the optical lattice. At high magnetic fields, this can be achieved by a combination of rf pulses and resonant light beams.  After completion of this work, one of us (S.\ O.) has joined a project aimed at transferring heteronuclear molecules created in {\it dipole traps} into the absolute internal molecular ground state. This project has published an extensive study of stability of molecules in the dipole trap when they collide with various atomic collision partners~\cite{Zirbel2008a}.

\begin{figure}
  \raggedleft
  \begin{minipage}[t]{6mm}{(a)}\end{minipage}%
  \begin{minipage}[t]{73mm}%
    \includegraphics[width=1.0\columnwidth,clip]{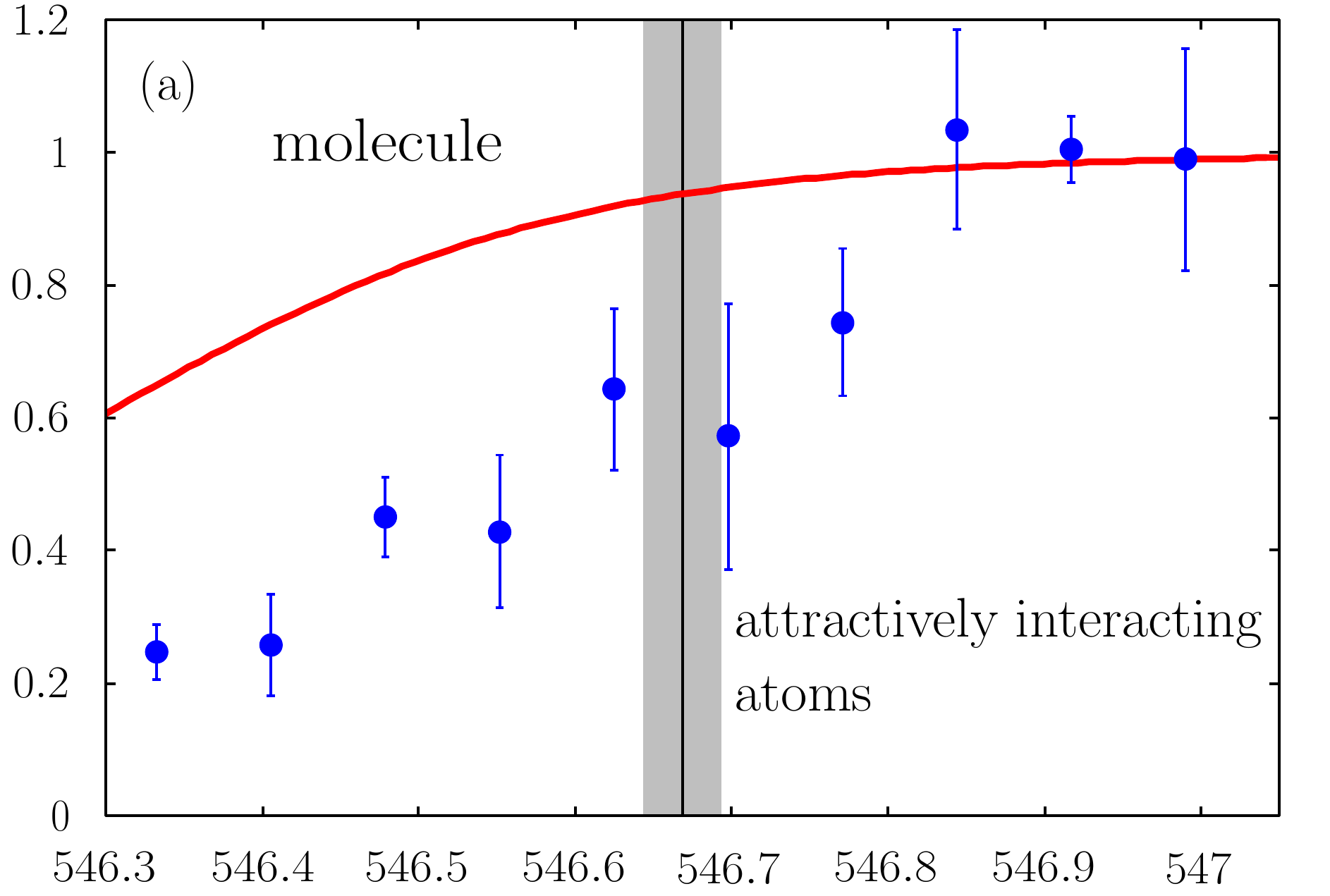}\\%
    \centerline{$B$ / G}%
  \end{minipage}%
  \begin{minipage}[t]{6mm}{(b)}\end{minipage}%
  \begin{minipage}[t]{73mm}%
    \includegraphics[width=1.0\columnwidth,clip]{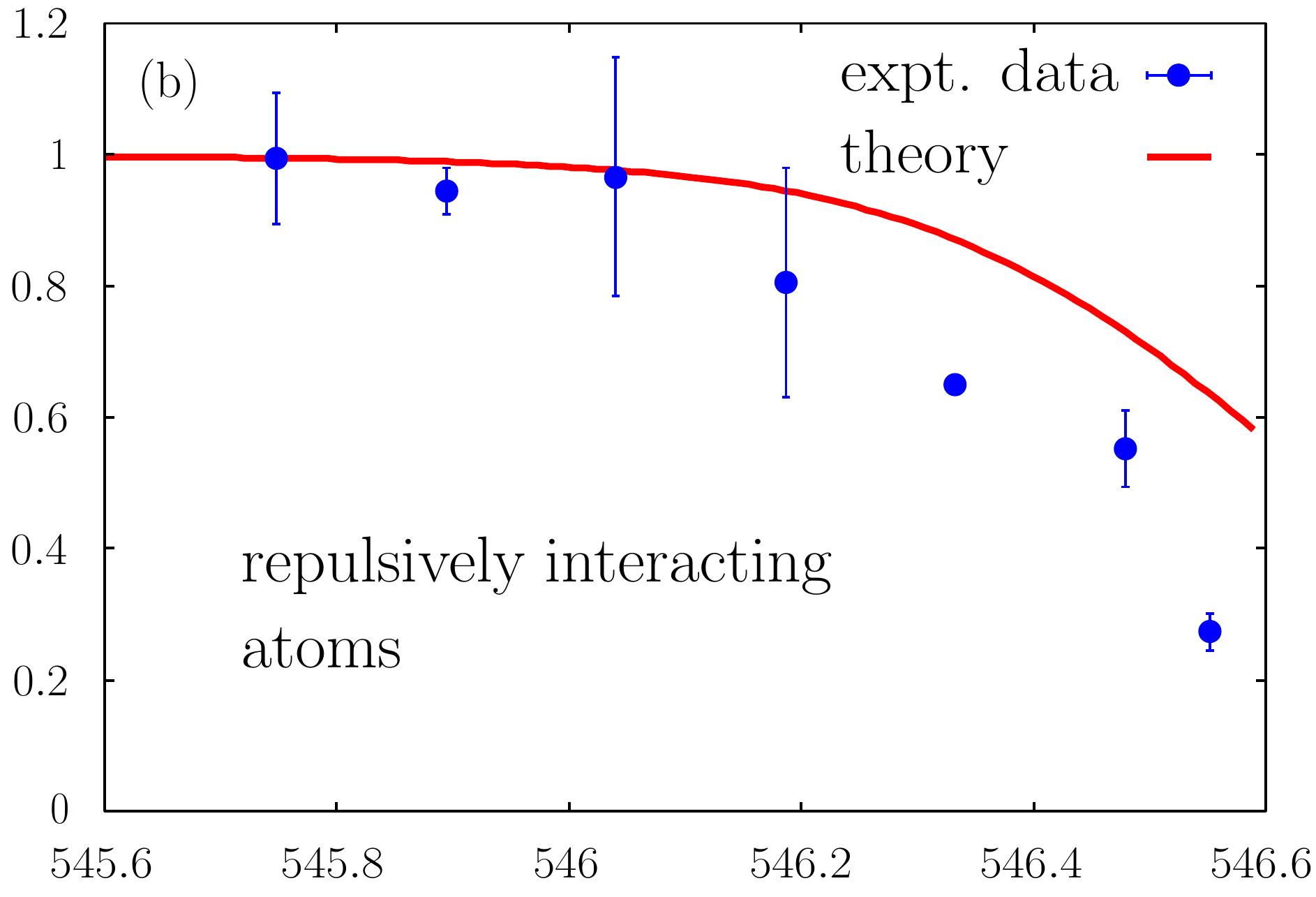}\\%
    \centerline{$B$ / G}%
  \end{minipage}%
  \caption{Rf association efficiency for both the molecular (a) and the repulsively interacting atom pair branch (b). {\it Copyright (2008) by the American Physical Society.}}
  \label{fig:efficiency}
\end{figure}

An important aspect of molecule production as demonstrated in this experiment is the efficiency of rf association, in particular as a function of magnetic field. There are two different aspects to the overall efficiency. A fundamental limit on the achievable {\it absolute} efficiency in any current molecule creation experiment is imposed by the number of doubly occupied lattice sites as compared to the overall atom number. This relation can be somewhat optimized experimentally by playing with temperature and overall particle numbers. 

A second aspect is the {\it relative} efficiency as a function of magnetic field and rf pulse power (i.\ e.\ at which fraction of the lattice sites occupied by one \rub\ and one \pot\ atom does the rf pulse create molecules?). In our experiments, in order to find suitable parameters for the rf pulses, we have driven Rabi oscillations on the atomic transition and then set the parameters such as to always obtain a $\pi$ pulse on the atomic transition. For all of our experiments, we have preserved this choice of parameters. Figure~\ref{fig:efficiency} shows the resulting relative transfer efficiency as a function of magnetic field. We expect that for heteronuclear scattering lengths in the final state which are close to the initial state ($F=9/2,m_F=-7/2$), i.\ e.\ at magnetic fields significantly above the resonance center or otherwise far detuned, we should create attractively interacting atom pairs at all available lattice sites occupied by one \rub\ and one \pot\ atom. The experimental transfer efficiency as determined from the ratio between the atomic and molecular peak in the rf spectroscopy has therefore been normalized to one in this region.

In~\cite{two_particles_hh}, we have established a model for the association process based on a Rabi model. After the rotating wave approximation and after integrating out spatial degrees of freedom, we obtain Rabi flopping on the molecular association transition with a Rabi frequency which is reduced by a factor of $\left< \Phi_i | \Phi_f \right>$ compared to the atomic Rabi frequency. Here, $\left< \Phi_i | \Phi_f \right>$ is the overlap integral between the initial and final motional states of the rf association process as determined from the pseudopotential model.

Since we have chosen constant pulse parameters in the experiment, we expect the transfer efficiency to drop as the initial and final states become more and more dissimilar. This can be seen very clearly in fig.~\ref{fig:efficiency}, both in the experimental data and in the prediction of the Rabi model. The quantitative agreement, however, is not complete. A further possibility of testing the hypothesis of the Rabi model is to look for Rabi oscillations between atoms and molecules on the molecular rf transition. This could provide further insight into the association process.

\section{Conclusions and Outlook}

During our two theses~\cite{SilkeThesis,OspelkausC2006a}, we have realized an experimental apparatus for the preparation of quantum degenerate Fermi-Bose mixtures in 3D optical lattices with tunable interactions. We have presented important milestones towards a mixed statistics many-body system as a versatile model system for the simulation of quantum many-body Hamiltonians and demonstrated important steps on the way towards quantum degenerate polar molecular gases with long-range dipolar interactions.

Starting from the realization of the so far largest particle numbers in a magnetically trapped degenerate \pot--\rub\ system, we have investigated interaction effects due to the large and attractive background interaction between the components. Tuning of heteronuclear interactions in the vicinity of a  Feshbach resonance has been demonstrated for the first time within this work and has allowed studies of arbitrary interactions between bosons and fermions. The complete phase diagram of harmonically trapped mixtures has been accessed. Starting with experiments on stable density distributions for attractively and repulsively interacting mixtures, for strong  repulsive interactions, phase separation has been observed and a Feshbach-induced collapse for strong attractive interactions. A heteronuclear $p$-wave resonance has been identified as described in~\cite{Ospelkaus2006c}, confirming a theoretical assignment of Feshbach resonances in the \pot--\rub\ system and opening intriguing perspectives for tuning of the anisotropy of the interaction. These studies on heteronuclear Feshbach resonances pave the way for studies with strongly interacting Fermi-Bose mixtures. 

From the very beginning, the experimental setup described here has been designed for studies in 3D optical lattices. Within this work, Fermi-Bose mixtures have been loaded into a three-dimensional optical lattice for the first time, demonstrating a large fermion-induced loss of coherence of the bosonic sample. The results are currently the subject of intense theoretical analysis, and explanations ranging from disorder-related localization phenomena over adiabatic thermodynamic effects and a mean-field induced shift of the  superfluid to Mott-insulator transition are being discussed. Further insight into these observations may be gained by implementing additional diagnostics. This may include Bragg spectroscopy, excitation spectroscopy to probe possible excitation gaps in deep optical lattices, probing of lattice occupation through rf or microwave spectroscopy and noise correlation analysis. The studies may also be extended to mixtures with varying heteronuclear interaction strength (see most recent results by Best {\it et al.}, \cite{Best2008a}).

In 1D or 2D optical lattices, one or several spatial degrees of freedom can experimentally  be frozen out, resulting in the preparation of low-dimensional systems. Depending on interaction strength and lattice occupation, charge-density wave or pairing phases may be observable. Low dimensional systems are of particular interest for direct comparison of experimental results to theory. For 1-dimensional situations (2D optical lattice) exactly solvable models exist and direct comparison to quasi-exact DMRG calculations is possible.

The combination of tunable interactions with 3-dimensional optical lattices has resulted in the first demonstration of ultracold heteronuclear Feshbach molecule formation. Long-lived heteronuclear Feshbach molecules have been created in a controlled fashion through rf association of atoms to molecules. This has allowed a precise determination of the binding energy of the molecular sample as a function of the magnetic field. Lifetimes between 20\,ms and 120\,ms have been experimentally observed, and excellent agreement has been found between experimentally determined properties of the molecules (binding energy, lifetime) and a a simple universal theoretical model.

Molecule formation may be the starting point for studies of pairing phases in the optical lattice, including studies of 3-body bound states. After molecule association and removal of the left-over atomic fraction, the lattice potential may  be ramped down. The molecule formed from a bosonic \rub\ and a fermionic \pot\ atom is again a fermion, and the observation of a molecular Fermi sea may be demonstrated.

In the optical lattice, Feshbach molecules  are  created in their absolute external ground state at a lattice site, but in a highly excited rovibrational internal state. Pulsed two-color photoassociation may be used to transfer these molecules into their internal ground state. The resulting molecule would exhibit a permanent electric dipole moment (polar molecule). The dipole moment gives rise to a long-range, anisotropic interaction. In the optical lattice, this long-range interaction may be used to implement quantum computation schemes. In the many-body limit, these polar molecules give access to novel quantum gases with long-range anisotropic interactions. 

After completion of this work, one of us (S. O.) has joined a project~\cite{Zirbel2008a,Zirbel2008b} aimed at optically transferring~\cite{Ospelkaus2008a} weakly bound Feshbach molecules into the absolute ro-vibrational and electronic ground state, and most recently realized a dense ultracold all ground state polar molecular gas~\cite{Ni2008a}.

\appendix
\section{Rf / microwave manipulation}
\label{sec:rf_setup}

In our experiments, we have used a novel rf / microwave manipulation scheme~\cite{SuccoThesis,OspelkausC2006a}, where all timinig, frequency and amplitude control is derived from one fast arbitrary frequency generator. The versatility and simplicity of operation as well as the precise timing down to the nanosecond level has greatly enhanced the experimental flexibility. The setup was developed based on the observation that many of the commercially available frequency generators provide a plethora of modulation options, whereas they are usually unable to rapidly and accurately perform for example simple linear frequency sweeps or AM modulation with a given envelope. The usual Adiabatic Rapid Passage (ARP) sequences for Zeeman state preparation require sweep times on the order of 10\,ms, whereas the dwell time of many commercial synthesizers, i.\ e.\ the time duration of each individual ``slot'' in a linear sweep, is limited to precisely 10\,ms; the usual workaround is to apply an analog voltage sweep to the FM modulation input. Also, many synthesizers need a lot of time for frequency switching both between frequencies and from fixed frequency operation to FM, sometimes as long as a second. Often, each operation requires its own synthesizer, resulting in an expensive and complex `synthesizer stack'.

\begin{figure}
  \centering
  \includegraphics[width=0.8\columnwidth]{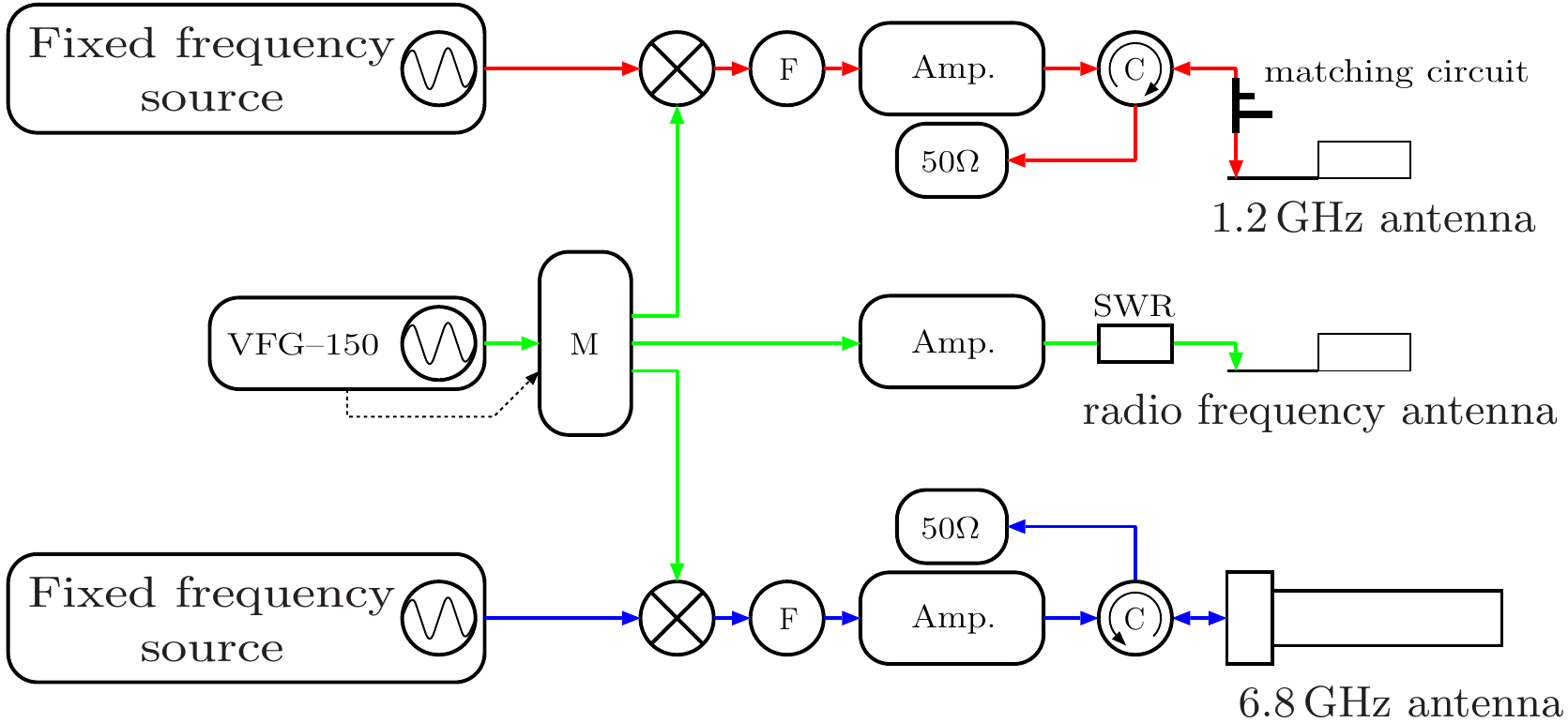}
  \caption{rf/microwave manipulation setup. Amplitude, frequency and timing control is provided by the VFG-150. Its output is connected to an rf multiplexer (M), which either directly feeds the signal to an amplifier and a coil-shaped antenna (up to 150\,MHz), or to two mixers ($\otimes$), summing the output with two fixed frequencies. Undesired sidebands in the resulting signal are eliminated with filters (F), and the signal is amplified and sent through a circulator to eliminate back reflections from the antenna. The resulting power is delivered to the atoms using a coil shaped device (for \pot\ hyperfine manipulation close to 1.2\,GHz) or using an open end waveguide (for \rub\ hyperfine manipulation at 6.8\,GHz). {\it Adapted with permission from~\cite{SuccoThesis}.}}
  \label{fig:rf_setup}
\end{figure}

In this experiment, we have implemented a solution which relies on just one synthesizer (VFG-150, available through Toptica Photonics, and developped by T. Hannemann in the group of C. Wunderlich), offering precise control of amplitude, frequency and phase every 5\,ns. The synthesizer covers an output frequency range of several 100\,kHz to 150\,MHz. The device (equipped with a USB2.0 input) is remote controlled from a small server application running on a PC. We have made the source code of this application available at \url{http://vfg-control.sourceforge.net/}. Many of the required rf manipulations can be performed directly using the output of the VFG-150 (evaporation, state preparation at low magnetic field and rf spectroscopy of \pot\ at high magnetic fields). For other applications, such as transfer from one hyperfine manifold to the other, the required frequencies exceed the direct output range of the VFG-150. To cover all these frequencies, we have connected the output of the VFG-150 to an rf multiplexer which either directs it directly to an amplifier and antenna or to a series of mixers which mix it with fixed frequency microwave sources. We have used mixing with a 1.2\,GHz fixed frequency to generate sweeps and pulses for \pot\ hyperfine manipulation and with a 6.8\,GHz fixed frequency to transfer \rub\ from $F=2$ to $F=1$ (see fig.~\ref{fig:rf_setup}).

The advantage of this approach is that just one synthesizer is being used. The overall setup is versatile to operate; it allows all kinds of shaped pulses (e.\ g.\ the Gaussian amplitude envelope pulses used for molecule creation) and offers a 5\,ns timing accuracy. At the same time, this scheme is highly cost-effective because it requires only one versatile source (the VFG-150) and inexpensive fixed frequency sources.

\section{`Magic' optical dipole trap}
\label{sec:magic_trap}

Compared to magnetic trapping, optical dipole traps offer the advantage of confining an arbitrary spin state. Furthermore, the value of the magnetic field becomes an experimental degree of freedeom. This capability is important for experiments with Feshbach resonaces as described in this article. A special interest is in relatively weak confinement, allowing optical lattice geometries which are as close as possible to a homogeneous situation.

However, when working with shallow traps and mixtures, the differential gravitational sag of the two components becomes an important issue. An atom with mass $m$ confined in a harmonic trap with trap frequency $\omega$ in the presence of gravity experiences a gravitational sag given by $z_0=-(g/\omega^2)$. This expression is found by calculating the local minimum of the combined gravitational and trapping potential. Different atomic species and even the different spin states of the same atomic species will in general experience a different trapping frequency when confined in the same magnetic or optical dipole trap, giving rise to a difference in gravitational sag. The importance of this issue depends on the value of the trap frequency. It is generally a non-issue for the compressed magnetic trap or for strong dipole traps. It does however play a significant role in the weak isotropic magnetic trap used for transfer of laser precooled atoms and for experiments in shallow and weak dipole traps such as the ones used in experiments with optical lattices aiming at a situation close to homogeneous systems. The difference in gravitational sag can then lead to a significantly reduced overlap of the two species. For example, for a far detuned optical dipole trap (equal trapping potential for both species) with a vertical trap frequency of 50\,Hz, the differential gravitational sag is 49\,$\mu$m!

The ``magic'' crossed dipole trap discussed below has allowed us to successfully overcome this limitation: if we can make the trap frequency in the direction of gravity the same for both species, the gravitational sag will automatically be the same for \rub\ and \pot. In an optical dipole trap, this can be done by a suitable choice of the dipole laser wavelength for confinement in the direction of gravity. This so-called ``magic wavelength'' can be calculated to be 806.7\,nm taking into account contributions from the D1 and D2 lines.

\begin{figure}
  \centering
  \includegraphics[width=0.5\columnwidth]{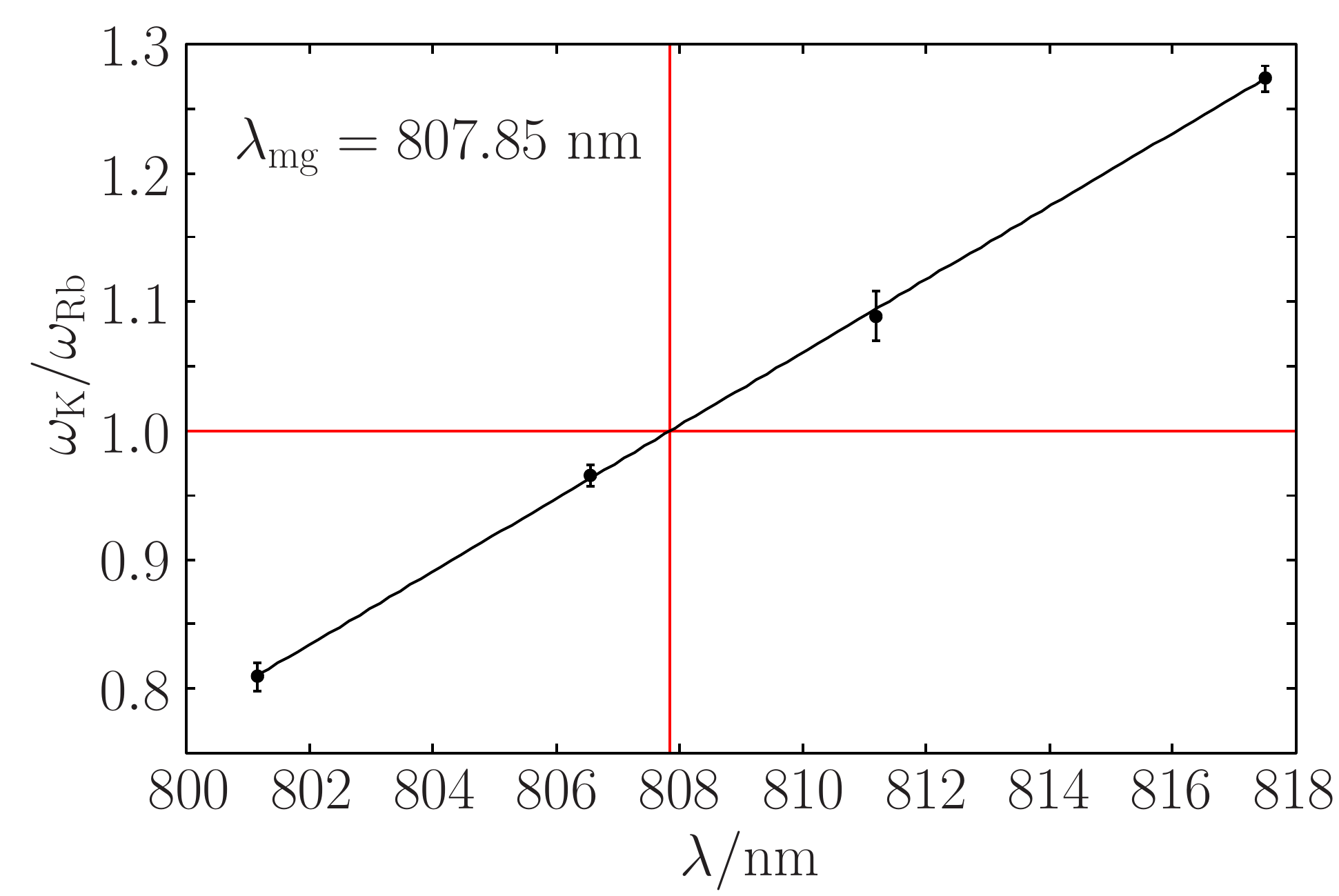}
  \caption{Measurement of the magic wavelength for our dipole trap.}
  \label{fig:frequency_measurement_magic_wavelength}
\end{figure}

With this idea in mind, we first performed a measurement of the magic wavelength by measuring the ratio of \pot\ and \rub\ trap frequencies in a dipole trap as a function of wavelength and found the magic wavelength to be 807.9\,nm (see fig.~\ref{fig:frequency_measurement_magic_wavelength}). Based on this information, a crossed beam optical dipole trap was designed with one beam at the magic wavelength providing confinement in the $z$ (vertical) and $x$ direction. The potential minimum of the atoms in this beam is made to overlap with the position in the magnetic trap. We tend to hold the intensity of this beam constant. A second beam is derived from our Yb:YAG lattice laser ($\lambda=1030$\,nm) and provides confinement in the $y$ and $z$ direction. This beam is overlapped with the atomic position in the ``magic'' beam. This is also the beam that we use for optical evaporation in the crossed dipole trap by lowering its intensity after transfer of atoms into the optical trap. This optical dipole trap allows us to go to weak optical dipole traps ($\omega=2\pi\cdot 50$\,Hz) without having issues with reduced overlap of the two clouds. It has been the basis for our experiments on tuning of interactions, mixtures in lattices and heteronuclear molecules. A detailed discussion of this novel dipole trap can be found in~\cite{SilkeThesis}.

\section{Optical lattice}
\label{sec:lattice_setup}

Part of our experiment is a 3D optical lattice setup. The setup uses an Yb:YAG disc laser with an output power of 20\,W as a laser source. In order to avoid motional heating in the optical lattice due to a `shaking' interference pattern, both frequency and intensity of the laser have been stabilized. The short term frequency stabilization locks the laser frequency to a reference cavity and has required the custom development of frequency control elements for the laser cavity. On longer time scales, the reference cavity follows the laser frequency. The achieved short term linewidth is 20\,kHz relative to the reference cavity. Details can be found in~\cite{WilleThesis,OspelkausC2006a}.

\begin{figure}
  \centering
  \includegraphics[width=0.6\columnwidth]{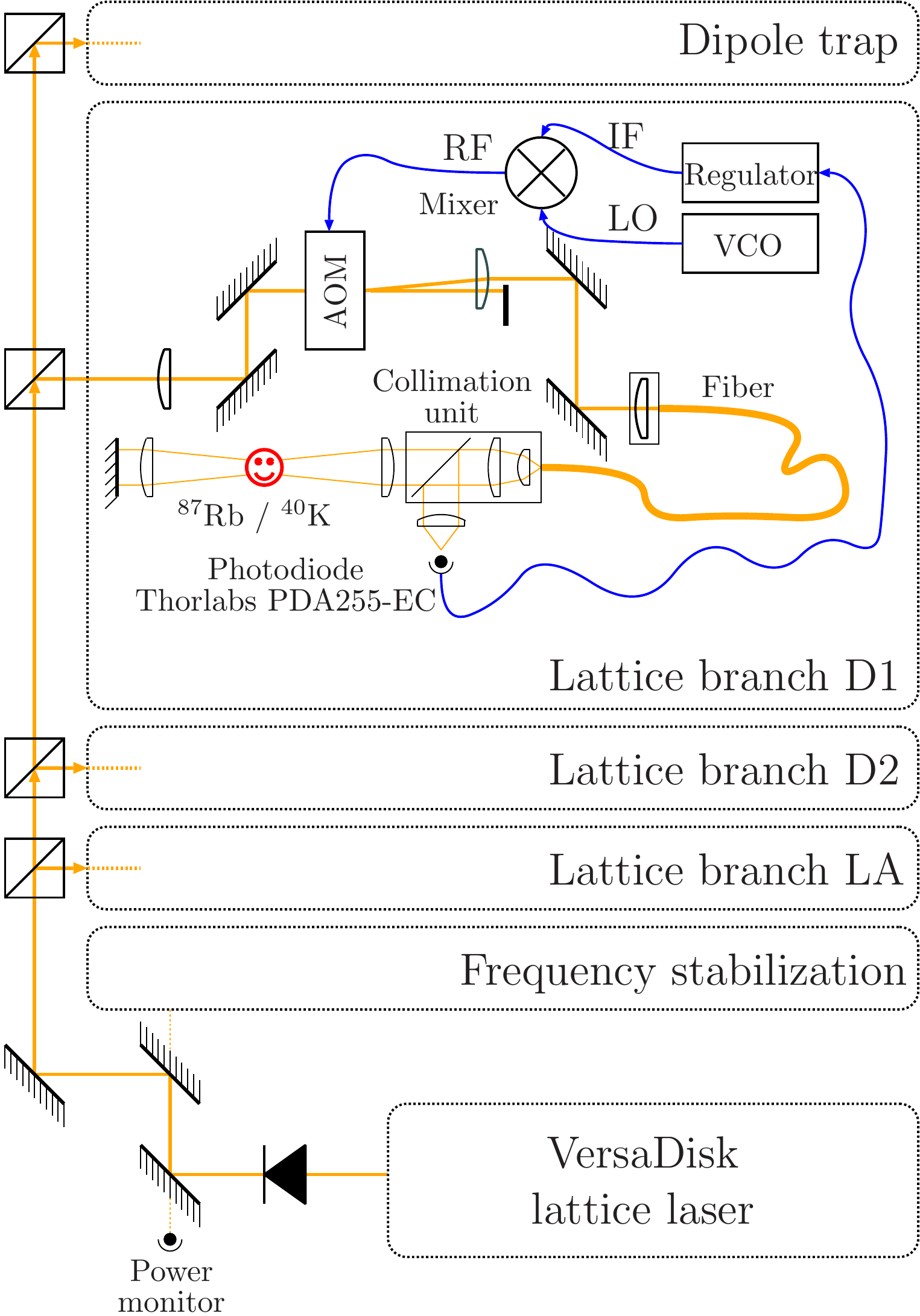}
  \caption{Optical lattice setup with beam preparation, intensity stabilization, collimation unit and retroreflection. Not shown: Imaging along lattice axes and edge filters for overlapping with MOT beams.}
  \label{fig:lattice_branch}
\end{figure}

Figure~\ref{fig:lattice_branch} shows a general view of the beam preparation scheme for the optical lattice setup. Behind an optical diode and pickups for laser power monitoring and frequency stabilization, the light from the lattice laser is split up into 4 independent beam paths using polarization optics. Each of these beams passes through an Acousto-Optical Modulator (AOM) in single-pass for intensity control and is coupled into an optical fiber. All of the AOM frequencies are detuned by roughly 10\,MHz with respect to each other in order to make the interference terms between the individual beams so fast that they are averaged out on any experimentally relevant time scale. Three of the beam paths are used for the three orthogonal retroreflected lattice beams with mutually orthogonal polarizations, and one of them is used for one of the two beams of the ``magic'' crossed optical dipole trap. 

After passing through the optical fiber, each of the lattice beams is collimated and focused. A pickup is used for the intensity stabilization. The focussing lens has a focal length of 400\,mm. Having passed this lens, the lattice beam is overlapped with the 3D MOT beams using dichroic mirrors and delivered to the atoms. The standing wave configuration for the optical lattice is produced by separating the MOT and lattice beams again after passing the glass cell using an identical ``disentangling'' dichroic mirror, recollimating the beam using a second lens identical to the focussing lens and retroreflecting it. Not shown in fig.~\ref{fig:lattice_branch} are the dipole trap beams and imaging systems along all of the three lattice axes. These allow us to overlap the lattice beam focus with the atoms and to optimize the lattice alignment. This is done by nulling out oscillations that result from a quick switch-on of the lattice in the presence of lateral misalignment.

\section{Imaging}
\label{sec:imaging}

In our experiment, we detect atoms using resonant absorption imaging. The standard absorption imaging procedure is to record three images, one where the optical density of the atoms attenuates the beam falling onto the CCD (absorption image), one where the laser beam alone is imaged (reference image) and one image without any light (dark image). From these three images, the optical column density of the cloud can be calculated (see e.\ g.~\cite{making_probing}). We have used several variations of this scheme in our experiments. All of these rely on properties of modern CCD chips allowing the user to record two images in very rapid succession (interline transfer CCD chips, such as the SONY ICX205AL). Typically, the interframing time is on the order of $\mu$s, if not less, and after two images, the readout takes on the order of hundreds of ms, larger than a useful time of flight in absorption imaging. We have made use of this capability to take two images in very rapid succession during time of flight in two different ways:

\subsection*{Two-species imaging}
In a first scenario (two-species imaging), we have used the first double image to store absorption images of \pot\ and \rub, respectively. The second double image contains the reference image for both species, and the last double image contains the dark image for both. This allows us to record both species on one camera and on one imaging axis in the same experimental run. This capability has been crucial in the initial optimization of the experiment and in the exeriments on collapse dynamics.

\subsection*{Fast imaging}
A second scenario makes use of the capability to take two images in rapid succession to eliminate the typical interference fringes from the images. Both the absorption and the reference image typically contain interference fringes caused by the coherent nature of the imaging light and etaloning at various surfaces in the optical path. These fringes are removed in the standard normalization procedure as long as they are stationary over experimental time scales. Usually, reference and absorption image are separated by the readout time of the CCD chip, thereby making the result sensitive to drifts of the interference pattern on acoustic time scales. Using the two images taken with $\mu$s separation as absorption and reference image, the interference patterns are stationary in between the two images, and we obtain a much better image quality and sensitivity to small atom numbers. Since the atoms do not fall very far in the short interframing time, we need to make sure the imaging beam does not `see' them when the reference image is taken. For \rub, which we have typically detected out of the $F=1$ state when this imaging procedure was used, we take the reference image first, when the imaging beam resonant with the $F=2\rightarrow F=3$ transition on the D2 line does not see the atoms. We then shine in a repumping pulse which transfers the atoms to $F=2$ and take the absorption image after the reference image.

For \pot, we have taken advantage of the fact that we have usually detected at high magnetic fields (see below). This allows us to take the absorption image first at high magnetic fields, then quickly switch the 547\,G magnetic field off in a couple of 10\,$\mu$s, which drives the atoms out of resonance with the imaging beam and lets us take the reference image. The improved image quality and sensitivity to small atom numbers has been crucial to the measurements on tuning of interactions, mixtures in 3D lattices and heteronuclear molecules.

\subsection*{State-selective imaging}
We have used two different techniques of state-selective imaging to identify the distribution of the atoms over the various sublevels of the hyperfine ground state manifold. The first and more common technique is Stern-Gerlach imaging, which uses a strong magnetic field gradient applied during time of flight to spatially separate the spin states. Stern-Gerlach imaging has the advantage of simultaneously providing information over all spin states. However, it requires the application of the field gradient for several milliseconds, thereby introducing a minimum time of flight which can impede the detection of small atom numbers. A different technique of state selective imaging, in particular when working at high magnetic fields, e.\ g.\ when using Feshbach resonances, is to take advantage of the transition from the Zeeman to the Paschen-Back regime. In this regime, the detection transitions $m_J=\pm 1/2$ to $m_J=\pm 3/2$, with $\Delta m_I=0$, are cycling transitions, and we can build a cycling transition from {\it any} sublevel in the ground state hyperfine manifold. Detection at high magnetic fields has the advantage of allowing a short time of flight, and allowing the detection of both atoms and heteronuclear molecules. However, it requires the atoms to be sufficiently far in the Paschen-Back regime, and for each spin state to be imaged, a separate detection laser and CCD camera is required. This technique has been crucial to our measurements on heteronuclear molecule formation.

In our particular system, the detection transition for the Feshbach-resonant state $F=9/2,m_F=-9/2$ ($m_J=-1/2,m_I=-4$ $\rightarrow$ $m_J=-3/2,m_I=-4$) and for weakly bound molecules is detuned by -764\,MHz from the standard zero field cycling transition ($F=9/2,m_F=9/2$ $\rightarrow$ $F=11/2,m_F=11/2$). The closest other state, which is $F=9/2,m_F=-7/2$ at low field and $m_J=-1/2,m_I=-3$ at high field, has a cycling transition which is detuned by -833\,MHz from the aforementioned standard zero field transition. Hence, we have a frequency difference of 69\,MHz between the detection transitions for the two states used in the rf spectroscopy, allowing us to state-selectively image both weakly interacting atoms in the $F=9/2,m_F=-7/2$ channel and Feshbach molecules and strongly interacting atom pairs in the $F=9/2,m_F=-9/2$ channel. Note that in our particular system, high field detection is only possible with the \pot\ component because at the magnetic fields of interest, \rub\ is closer to the low field (Zeeman) regime than to the Paschen-Back regime, and the corresponding cycling transitions are not available.

In order to perform the detection, we leave the Feshbach magnetic field on during time of flight and detection. For imaging, we use two beams derived from our cooling and detection laser and differing by 69\,MHz in frequency. After laser cooling and during the evaporative cooling, we detune our imaging laser by about 700\,MHz~\cite{LeifThesis} to make it resonant with the high field detection transition. This high field detection technique has allowed us to state-selectively image both atoms and weakly bound molecules with very short time of flight, high sensitivity and unparalleled simplicity (no dissociation of molecules or ramping of fields). Another beneficial aspect of this detection scheme is that in combination with fast imaging (see above), it has allowed us to drastically reduce the typical interference fringes in absorption imaging, again improving on the signal to noise. Details of this scheme can be found in~\cite{OspelkausC2006a}.

\section*{References}
\bibliographystyle{cobib}
\bibliography{co,sos}

\end{document}